\documentclass[10pt]{article}

\usepackage{amsmath}
\usepackage{amssymb}

\usepackage{graphicx}

\usepackage{cite}

\usepackage{color}

\usepackage{setspace} 
\usepackage[labelfont=bf,labelsep=period,justification=raggedright]{caption}

\makeatletter
\renewcommand{\@biblabel}[1]{\quad#1.}
\makeatother

\date{}

\newcommand{\dotp}{\! \cdot \!\!}
\begin{document}

% Title must be 150 characters or less
\begin{flushleft}
{\Large
\textbf{Coordinated optimization of visual cortical maps\\
(I) Symmetry-based analysis }
}
% Insert Author names, affiliations and corresponding author email.
\\
Lars Reichl$^{1,2,3,4,\ast}$, 
Dominik Heide$^{1,5}$, 
Siegrid L\"owel$^{3,6}$,
Justin C. Crowley$^{7}$, 
Matthias Kaschube$^{8}$,
Fred Wolf$^{1,2,3,4\ast}$ 
\\
\bf{1} Max-Planck-Institute for Dynamics and Self-Organization, G\"ottingen, Germany
\\
\bf{2} Bernstein Center for Computational Neuroscience,  G\"ottingen, Germany 
\\
\bf{3} Bernstein Focus Neurotechnology, G\"ottingen, Germany 
\\
\bf{4} Faculty of Physics, Georg-August University, G\"ottingen, Germany
\\
\bf{5} Frankfurt Institute of Advanced Studies, Frankfurt, Germany
\\
\bf{6} School of Biology, Georg-August University, G\"ottingen, Germany
%Institute of General Zoology and Animal Physiology, University Jena, Jena, Germany
\\
\bf{7} Carnegie Mellon University, Department of Biological Sciences, Pittsburgh PA, USA
\\
\bf{8} Physics Department and Lewis-Sigler Institute, Princeton University, Princeton NJ, USA
\\
$\ast$ E-mail: reichl@nld.ds.mpg.de; fred@nld.ds.mpg.de
\end{flushleft}

% Please keep the abstract between 250 and 300 words
\section*{Abstract}
In the primary visual cortex of primates and carnivores, functional architecture
can be characterized by maps of various stimulus features such as orientation preference (OP), ocular 
dominance (OD), and spatial frequency.
It is a long-standing question in theoretical neuroscience whether the observed maps
should be interpreted as optima of a specific energy functional that
summarizes the design principles of cortical functional architecture.
A rigorous evaluation of this optimization hypothesis is particularly demanded
by recent evidence that the functional architecture of orientation columns precisely follows species invariant 
quantitative laws \cite{Kaschube6}.
Because it would be desirable to infer the form of such an optimization principle 
from the biological data, the optimization approach to explain cortical functional architecture 
raises the following questions:
i) What are the genuine ground states of candidate energy functionals and how can 
they be calculated with precision and rigor?
ii) How do differences in candidate optimization principles impact on the predicted 
map structure and conversely what can be learned about an hypothetical underlying optimization
principle from observations on map structure?
iii) Is there a way to analyze the coordinated organization  
of cortical maps predicted by optimization principles in general? 
To answer these questions we developed a general dynamical systems approach to the combined 
optimization of visual cortical maps of OP and another scalar feature such
as OD or spatial frequency preference.
From basic symmetry assumptions we obtain a comprehensive phenomenological classification of possible 
inter-map coupling energies and examine representative examples.
We show that each individual coupling energy leads to a different class of OP solutions with different 
correlations among the maps such that inferences about the optimization principle
from map layout appear viable.
We systematically assess whether quantitative laws resembling experimental observations can result 
from the coordinated optimization 
of orientation columns with other feature maps. 
%A low and a high order gradient-type coupling energy as well as a low and high order 
%product-type coupling energy.
%OP maps are characterized by singularities, called pinwheel centers \cite{Swindale2}.
%We show that all coupling energies above a critical inter-map coupling strength predict  
%spatially periodic and pinwheel-rich OP maps which represent energetic ground states i.e. global optima of the 
%system.

%We explore whether these results also apply to optimization with higher dimensional feature spaces.
%OP maps in the brain are characterized by pinwheels, regions in which columns preferring all 
%possible orientations are organized around a common center in a radial fashion \cite{Swindale2}.

% Please keep the Author Summary between 150 and 200 words
% Use first person. PLoS ONE authors please skip this step. 
% Author Summary not valid for PLoS ONE submissions.   
\section*{Author Summary}
%We ask that all authors of research articles include a 150-200 word non-technical summary 
%of the work as part of the manuscript to immediately follow the abstract. 
%This text is subject to editorial change, should be written in the first-person voice, and should be 
%distinct from the scientific abstract. 
%Aim to highlight where your work fits within a broader context; present the 
%significance or possible implications of your work simply and objectively; and avoid the use of 
%acronyms and complex terminology wherever possible. The goal is to make your findings 
%accessible to a wide audience that includes both scientists and non-scientists. 
%Authors may benefit from consulting with a science writer or press officer to ensure they 
%effectively communicate their findings to a general audience.
Neurons in the visual cortex form spatial representations or maps of several stimulus features.
How are different spatial representations of visual information coordinated in the brain?
In this paper, we study the hypothesis that the coordinated organization of several visual cortical 
maps can be explained by joint optimization.
Previous attempts to explain the spatial layout of functional maps in the visual cortex proposed 
specific optimization principles ad hoc.
Here, we systematically analyze how optimization principles in a general class of models 
impact on the spatial layout 
of visual cortical maps.
For each considered optimization principle we identify the corresponding optima and
analyze their spatial layout.
This directly demonstrates that by studying map layout and geometric inter-map correlations one
can substantially constrain the underlying optimization principle. 
In particular, we study whether such optimization principles can lead to spatially complex patterns and 
to geometric correlations among cortical maps as observed in imaging experiments.
%%%%
\section*{Introduction}
%0: Intro
Neurons in the primary visual cortex are selective to a multidimensional set of visual stimulus features, 
including visual field position, contour orientation, ocular dominance, direction 
of motion, and spatial frequency \cite{HubelWiesel2,LoewelBuch}.
In many mammals, these response properties form spatially complex, two-dimensional patterns
called visual cortical 
maps 
\cite{Grinwald,Blasdel3,Swindale3,Grinwald1,Bartfeld,Blasdel4,Blasdel,Grinwald5,Bosking3,Grinwald_Dir,
Chapman,Rao,Bosking,Das2,Loewel,Stryker2,White,Galuske,Casagrande,White_Dir,White3}.
The functional advantage of a two dimensional mapping of stimulus selectivities is 
currently unknown \cite{Horton4,Koulakov2,Horton5}.
%%%%%%%%%%%%%%%%%%%%%%%%%%%%%%%%%%%%%%%%%%%%%%%%%%%%%%%%
%1: Optimization ( Evolution, Development)
What determines the precise spatial organization of these maps?
%1a) Maps as optima 
%1b) Evolution
It is a plausible hypothesis that natural selection should shape visual cortical maps to build  
efficient representations of visual information improving the 'fitness' of the organism.
Cortical maps are therefore often viewed as optima of some cost function.
For instance, it has been proposed that cortical maps optimize the cortical 
wiring length \cite{Koulakov,Koulakov4} 
or represent an optimal compromise between stimulus coverage and map continuity 
\cite{Durbin,Obermayer2,Swindale11,Schulten,Swindale6,Goodhill7,Goodhill,Swindale5,
Goodhill5,Swindale1,Goodhill2,Sur5,Goodhill9,Sur2}.
If map structure was largely genetically determined map structure might be optimized through 
genetic variation and Darwinian selection on an evolutionary timescale.
%1c) Development :
Optimization may, however, also occur during the ontogenetic maturation of the individual organism 
for instance by the activity-dependent refinement of neuronal circuits.
If such an activity-dependent refinement of cortical architecture realizes an optimization strategy 
its outcome should be interpreted as the convergence towards a ground state of a specific energy
functional.
%1d) E is unknown
This hypothesized optimized functional, however, remains currently unknown.
%%%%%%%%%%%%%%%%%%%%%%%%%%%%%%%%%%%%%%%%%%%%%%%%%%%%%
%2: Single map optimization (Koulakov)
%%%%%%%%%%%%%%%%%%%%%%%%%%%%%%%%%%%%%%%%%%%%%%%%%%%
%3: Multi map optimization 
As several different functional maps coexist in the visual cortex candidate energy 
functionals are expected to reflect the multiple response properties of neurons in
the visual cortex. 
%3a) EN, Kohonen, Coverage&Uniformity
In fact, consistent with the idea of joint optimization of different feature maps
cortical maps are not independent of each 
other \cite{Bartfeld,Blasdel,Bonhoeffer,Crair2,Loewel,Loewel7,Matsuda,Sur5,Casagrande}. 
Various studies 
proposed a coordinated optimization of different feature maps 
\cite{Durbin,Goodhill,Goodhill2,Sur5,Schulten,Obermayer,Sur2,Swindale1,
Swindale5,Bednar,Swindale11,Wolf3}.
%3b) Spatial Rel. <- indication of optimization principle
Coordinated optimization appears consistent with the observed distinct spatial relationships between 
different maps such as the tendency of iso-orientation lines to 
intersect OD borders perpendicularly or the preferential positioning of orientation pinwheels at locations
of maximal eye dominance \cite{Bonhoeffer,Loewel,Blasdel,Bartfeld,Loewel7,Sur5,Casagrande}.
Specifically these geometric correlations have thus been proposed to indicate 
the optimization of a cost function given by a compromise between stimulus 
coverage and continuity \cite{Swindale11,Swindale5,Swindale6,Swindale1,Sur2,Sur5}, 
a conclusion that was questioned by Carreira-Perpinan and Goodhill \cite{Goodhill4}.\\ 
%%%%%%%%%%%%%%%%%%%%%%%%%%%%%%%%%%%%%%%%%%%%%%%%%%
%4: Maps are complex spatial patterns (May result from frustration, conflicting demands)
Visual cortical maps are often spatially complex patterns that 
contain defect structures such as point 
singularities (pinwheels) \cite{Swindale2,Swindale3,Grinwald3,Grinwald5,Grinwald3,Ohki} or line 
discontinuities (fractures) \cite{Ohki2,Bosking3} and that never 
exactly repeat 
\cite{Grinwald,Blasdel3,Swindale3,Grinwald1,Bartfeld,Blasdel4,Blasdel,Grinwald5,Bosking3,Grinwald_Dir,
Chapman,Rao,Bosking,Das2,Loewel,Stryker2,White,Galuske,Casagrande,White_Dir,White3,Kaschube1,Kaschube6}.
%Notably, the occurrence of point singularities was predicted from models of spontaneous symmetry
%breaking in cortical development \cite{Swindale2} almost two decades before their 
%experimental demonstration \cite{}.
It is conceivable that this spatial complexity arises from geometric frustration due to a coordinated
optimization of multiple feature maps in which not all inter-map interactions can be simultaneously 
satisfied \cite{Wolf3,Cho2,Hoffsummer,Hoffsummer2,Hoffsummer3}.
In many optimization models, however, the resulting map layout is spatially not complex
or lacks some of the basic features such as topological defects \cite{Wolf3,Koulakov,Kadar,Cho2,Reichl4}.
In other studies coordinated optimization was reported to preserve defects 
that would otherwise decay \cite{Wolf3,Cho2}.
%%%%%%%%%%%%%%%%%%%%%%%%%%%%%%%%%%%%%%%%%%%%%%%%%%
%6: Questions
An attempt to rigorously study the 
hypothesis that the structure of cortical maps is explained by an optimization process
thus raises a number of questions:
i) What are the genuine ground states of candidate energy functionals and how can 
they be calculated with precision and rigor?
ii) How do differences in candidate optimization principles impact on the predicted 
map structure and conversely what can be learned about an hypothetical underlying optimization
principle from observations on map structure?
iii) Is there a way to analyze the coordinated organization  
of cortical maps predicted by optimization principles in general? 
If theoretical neuroscience was able to answer these questions with greater confidence,
the interpretation and explanation of visual cortical architecture could build
on a more solid foundation than currently available.
%%%%%%%%%%%%%%%%%%%%%%%%%%%%%%%%%%%%%%%%%%%%%%%%%%
%5: Symmetries to determine E
%50) No specific optimization principle
To start laying such a foundation, we examined how symmetry principles in general constrain 
the form of optimization models
and developed a formalism for analyzing map optimization independent of the specific
energy functional assumed.\\
%5a) Gradient descent : dynamical system
Minima of a given energy functional can be found by gradient descent which is 
naturally represented by a dynamical system describing a formal time evolution of the maps.
%5b) Theory of pattern formation: known that symmetries are crucial
Response properties in visual cortical maps are arranged in repetitive modules of a typical spatial length 
called hypercolumn.
Optimization models that reproduce this typical length scale are therefore effectively pattern forming systems
with a so-called 'cellular' or finite wavelength instability, see \cite{Manneville,Cross,Greenside}.
%In such an approach, map formation is described on a mesoscopic level by the formal time 
%evolution of order parameter fields.
In the theory of pattern formation, it is well understood that symmetries play a 
crucial role \cite{Manneville,Cross,Greenside}.
%5c) Our symmetry assmuptions
Some symmetries are widely considered biologically plausible for cortical maps, for 
instance the invariance under spatial translations and rotations or
a global shift of orientation preference \cite{Braitenberg2,Cowan3,Cowan,Wolf1,Wolf2,Wolf3,Reichl4}.
In this paper we argue that such symmetries and an approach
that utilizes the analogy between map optimization and pattern forming systems
can open up a novel and systematic approach to the coordinated optimization
of visual cortical representations.\\
%% Common design
A recent study found strong evidence for a common design in the functional architecture
of orientation columns \cite{Kaschube6}.
Three species, galagos, ferrets, and tree shrews, widely separated in evolution of modern mammals, share 
an apparently universal set of quantitative properties. 
The average pinwheel density as well as the spatial organization of pinwheels within orientation 
hypercolumns, expressed in the
statistics of nearest neighbors as well as the local variability of the pinwheel densities in
cortical subregions ranging from 1 to 30 hypercolumns, 
are found to be virtually identical in the analyzed species.
However, these quantities are different from random maps.
Intriguingly, the average pinwheel density was found to be statistical indistinguishable
from the mathematical constant $\pi$ up to a precision of 2\%.
Such apparently universal laws can be reproduced in relatively simple self-organization 
models if long-range neuronal interactions
are dominant \cite{Wolf1,Wolf2,Kaschube3,Kaschube6}.
As pointed out by Kaschube and coworkers, these findings pose strong constraints 
on models of cortical functional architecture \cite{Kaschube6}. 
Many models exhibiting pinwheel annihilation \cite{Wolf3,Cho2} or pinwheel 
crystallization \cite{Cowan2,Kadar,Reichl4} were found to
violate the experimentally observed layout rules.
In \cite{Kaschube6} it was shown that the common design is correctly predicted
in models that were based only on intrinsic OP properties.
Alternatively, however, it is conceivable that they result
from geometric frustration due to inter-map interactions and joint optimization.
In the current study we therefore in particular examined whether the coordinated optimization of the 
OP map and another feature map
can reproduce the quantitative laws defining the common design.\\ 
% Left-right :Referenzen in Conclusions ?????
%While orientation preference seems to be represented essentially homogeneously throughout the
%visual cortex a bias in response properties has been reported in other feature maps.
%For instance, a bias towards the contralateral eye has been reported in young animals which
%breaks the left-right eye symmetry \cite{Stryker3,Crair,Horton}.\\
%%%%%%%%%%%%%%%%%%%%%%%%%%%%%%%%%%%%%%%%%%%%%%%%%%%
%%%%%%%%%%%%%%%%%%%%%%%%%%%%%%%%%%%%%%%%%%%%%%%%%%%
%7: Outline
The presentation of our results is organized as follows.
%part1
First we introduce a formalism to model the coordinated optimization of 
complex and real valued scalar fields.
Complex valued fields can represent for instance orientation preference (OP) or 
direction preference maps \cite{Grinwald_Dir,White_Dir}.
Real valued fields may represent for instance ocular dominance (OD) \cite{HubelWiesel2}, spatial frequency 
maps \cite{Stryker2,Bonhoeffer} or
ON-OFF segregation \cite{Chapman3}.
We construct several optimization models such that an independent optimization 
of each map in isolation results in a regular OP stripe pattern and, depending on the
relative representations of the two eyes, OD patterns with a regular hexagonal or stripe layout. 
A model-free, symmetry-based analysis of potential optimization principles that couple
the real and complex valued fields provides a comprehensive classification and parametrization 
of conceivable coordinated optimization models and identifies representative forms of coupling energies.
For analytical treatment of the optimization problem
we adapt a perturbation method from pattern formation theory called weakly nonlinear analysis 
\cite{Manneville,Cross,Greenside,Busse,Bodenschatz,Soward,Vinals}.
This method is applicable to models in which the spatial pattern of columns branches off continuously
from an unselective homogeneous state. 
It reduces the dimensionality of the system and
leads to amplitude equations as an approximate description of the system near the symmetry breaking
transition at which the homogeneous state becomes unstable.  
We identify a limit in which inter-map interactions that are formally always bidirectional
become effectively unidirectional. 
In this limit, one can neglect the backreaction of the complex map on the layout of the co-evolving 
scalar feature map.
We show how to treat low and higher order versions of inter-map 
coupling energies which enter at different order 
in the perturbative expansion.\\
% part 2
Second we apply the derived formalism by calculating optima
of two representative low order examples of coordinated optimization models and examine
how they impact on the resulting map layout. 
Two higher order optimization models are analyzed in Text S1.
For concreteness and motivated by recent topical interest \cite{Kaschube6,Miller4,Stevens}, we illustrate the 
coordinated optimization of visual cortical maps for 
the widely studied example of a complex 
OP map and a real feature map such as the OD map.
OP maps are characterized by pinwheels, regions
in which columns preferring all possible orientations are organized around a common center 
in a radial fashion \cite{Swindale2,Ohki,Blasdel5,Bonhoeffer2}. 
In particular, we address the problem of pinwheel stability in OP maps \cite{Wolf2,Wolf3} and
calculate the pinwheel densities predicted by different models.
As shown previously, many theoretical models of visual cortical development and optimization fail 
to predict OP maps possessing stable pinwheels \cite{Wolf3,Koulakov,Kadar,Cho2}. 
%The OP maps we obtained as optima of different optimization principles are therefore
%analyzed for the occurrence of pinwheels.
We show that in case of the low order energies, a strong inter-map coupling will typically lead to OP map 
suppression, causing the orientation selectivity of
all neurons to vanish. 
For all considered optimization models, we identify stationary solutions of the resulting 
dynamics and mathematically demonstrate their stability.
We further calculate phase diagrams as a function of the inter-map coupling strength and the amount
of overrepresentation of certain stimuli of the co-evolving scalar feature map.
We show that the optimization of any of the analyzed coupling energies can lead to 
spatially relatively complex patterns.
Moreover, in case of OP maps, these patterns are typically pinwheel-rich.
The phase diagrams, however, differ for each considered coupling energy, in particular 
leading to coupling energy specific ground states.
We therefore thoroughly analyze the spatial layout of energetic ground states and in particular
their geometric inter-map relationships.
We find that none of the examined models reproduces the experimentally observed pinwheel density
and spatially aperiodic arrangements.
Our analysis identifies a seemingly general condition for interaction induced pinwheel-rich OP optima namely 
a substantial bias in the response properties of the
co-evolving scalar feature map.
%%%%%%%%%%%%%%%%%%%%%%%%%%%%%%%%%%%%%%%%%%%%%%%%%%%%%%
%%%%%%%%%%%%%%%%%%%%%%%%%%%%%%%%%%%%%%%%%%%%%%%%%%%%%%
%%%%%%%%%%%%%%%%%%%%%
%%%%%%%%%%%%%%%%%%%%%
%%%%%%%%%%%%%%%%%%%%%
%%%%%%%%%%%%%%%%%%%%%
% Results and Discussion can be combined.
%\setcounter{section}{1}
\section*{Results}\label{sec:Results} % PAST TENSE
\setcounter{section}{1}
\setcounter{subsection}{0}
\subsection{Modeling the coordinated optimization of multiple maps}
We model the response properties of neuronal populations in the visual cortex by two-dimensional
scalar order parameter fields which are either complex valued or real valued \cite{Swindale2,Swindale7}.
A complex valued field $z(\mathbf{x})$ can for instance describe OP or
direction preference of a neuron located at position $\mathbf{x}$.
A real valued field $o(\mathbf{x})$ can describe for instance OD or the spatial frequency preference. 
Although we introduce a model for the coordinated optimization of general real and complex valued order parameter 
fields we consider $z(\mathbf{x})$ as the field of OP and $o(\mathbf{x})$ as the field of OD throughout this article.
In this case, the pattern of preferred stimulus orientation $\vartheta$ is obtained by
\begin{equation} 
\vartheta(\mathbf{x})=\frac{1}{2} \arg (z).
\end{equation} 
The modulus $|z(\mathbf{x})|$ is a measure of the selectivity at cortical location $\mathbf{x}$.\\
OP maps are characterized by so-called \textit{pinwheels}, regions
in which columns preferring all possible orientations are organized around a common center in a radial fashion.
The centers of pinwheels are point discontinuities of the field $\vartheta(\mathbf{x})$
where the mean orientation preference of nearby columns changes by 90 degrees.
Pinwheels can be characterized by a topological charge $q$ which indicates in particular
whether the orientation preference increases clockwise or counterclockwise around
the pinwheel center,
\begin{equation}
q_i=\frac{1}{2\pi}\oint_{C_i} \nabla \vartheta(\mathbf{x}) d\mathbf{s}\, ,
\end{equation}
where $C_i$ is a closed curve around a single pinwheel center at $\mathbf{x}_i$. 
Since $\vartheta$ is a cyclic variable in the interval $[0,\pi]$ and up to isolated points
is a continuous function of $\mathbf{x}$, $q_i$ can only have values
\begin{equation}
q_i=\frac{n}{2} \, ,
\end{equation}
where $n$ is an integer number \cite{Mermin}. If its absolute value $|q_i|=1/2$, each orientation is 
represented only
once in the vicinity of a pinwheel center.
In experiments, only pinwheels with a topological charge of $\pm 1/2$ are observed, which are 
simple zeros of the field $z(\mathbf{x})$.\\
OD maps can be described by a real valued two-dimensional field $o(\mathbf{x})$, 
where $o(\mathbf{x})<0$ indicates ipsilateral eye dominance and $o(\mathbf{x})>0$ 
contralateral eye dominance of the neuron located at position $\mathbf{x}$. The magnitude
indicates the strength of the eye dominance and thus the zeros of the 
field correspond to the borders of OD.\\
%%%%
In this article, we view visual cortical maps as optima of some energy functional $E$.
The time evolution of these maps can be described by the gradient descent of this energy functional.
The field dynamics thus takes the form
\begin{eqnarray}\label{eq:coupledGeneral}
\partial_t \, z(\mathbf{x},t) &=& F[z(\mathbf{x},t),o(\mathbf{x},t)] \nonumber \\
\partial_t \, o(\mathbf{x},t) &=& G[z(\mathbf{x},t),o(\mathbf{x},t)], 
\end{eqnarray}
where $F[z,o]$ and $G[z,o]$ are nonlinear operators given by
$F[z,o]=-\frac{\delta E}{\delta \overline{z}}$, $G[z,o]=-\frac{\delta E}{\delta o}$.
The system then relaxes towards the minima of the energy $E$.
The convergence of this dynamics towards an attractor is assumed to represent
the process of maturation and optimization of the cortical circuitry.
Various biologically detailed models have been cast to this form \cite{Swindale6,Wolf3,Scherf}.\\
%%%%%%%%%%%%
All visual cortical maps are arranged in repetitive patterns of a typical wavelength $\Lambda$.
We splitted the energy functional $E$ into a part that ensures the emergence of such a typical wavelength
for each map and into a part which describes the coupling among different maps.
A well studied model reproducing the emergence of a typical wavelength by a pattern forming 
instability is of the Swift-Hohenberg type \cite{SwiftHohenberg,Cross}.
Many other pattern forming systems occurring in different physical, chemical, and 
biological contexts (see for instance \cite{Busse,Bodenschatz,Soward,Vinals}) 
have been cast into a dynamics of this type.
Its dynamics in case of the OP map is of the form
\begin{equation}\label{eq:DynamicsSH}
\partial_t \, z(\mathbf{x},t) = \hat{L} z(\mathbf{x},t) -|z|^2z\, , 
\end{equation}
with the linear Swift-Hohenberg operator
\begin{equation}\label{eq:LinearSH}
\hat{L}=r-\left(k_c^2+\Delta \right)^2\, , 
\end{equation}
$k_c=2\pi/\Lambda$, and $N[z(\mathbf{x},t)]$ a nonlinear operator.
The energy functional of this dynamics is given by
\begin{equation}
E=-\int d^2x \left( \overline{z}(\mathbf{x}) \hat{L} z(\mathbf{x})-\frac{1}{2} |z(\mathbf{x})|^4 \right)\, .
\end{equation} 
In Fourier representation, $\hat{L}$ is diagonal with the spectrum
\begin{equation}\label{eq:LinearSHkspace}
\lambda(k)=r-\left(k_c^2-k^2 \right)^2\, . 
\end{equation}
The spectrum exhibits a maximum at $k=k_c$.%, see Fig.~\ref{fig:SwiftHohenberg}(a). 
%\begin{figure}[tb]
%\includegraphics[width=\linewidth]{pics/SwiftHohenberg/SwiftHohenberg_small} 
%\caption{ \textbf{Swift-Hohenberg equation},
%\textbf{\sffamily{A}}
%Cross section through the spectrum $\lambda(k)$ of the
%Swift-Hohenberg operator Eq.~(\ref{eq:LinearSHkspace}), $r=0.1$.
%\textbf{\sffamily{B}} Time evolution of the Power Eq.~(\ref{eq:Power}).
%}
%\label{fig:SwiftHohenberg}
%\end{figure}
For $r<0$, all modes are damped since $\lambda(k)<0 , \, \forall k$ and only the 
homogeneous state $z(\mathbf{x})=0$ is stable.
This is no longer the case for $r>0$ when modes on the \textit{critical circle} $k=k_c$
acquire a positive growth rate and now start to grow, resulting in patterns with a typical 
wavelength $\Lambda=2\pi/k_c$.
Thus, this model exhibits a supercritical bifurcation where the homogeneous state looses 
its stability and spatial modulations start to grow.\\
The coupled dynamics we considered is of the form
\begin{eqnarray}\label{eq:DynamicsSHcoupled_ref}
\partial_t \, z(\mathbf{x},t) &=& \hat{L}_z \, z(\mathbf{x},t) -|z|^2z 
-\frac{\delta U}{\delta \overline{z}} \nonumber \\ 
\partial_t \, o(\mathbf{x},t) &=& \hat{L}_o \, o(\mathbf{x},t) 
-o^3 -\frac{\delta U}{\delta o}+\gamma \, ,
\end{eqnarray}
where $\hat{L}_{\{o,z\}}=r_{\{o,z\}} -\left(k_{c,\{o,z\}}^2+\Delta \right)^2$, and $\gamma$ is a constant.
To account for the species differences in the wavelengths of the pattern we chose two
typical wavelengths $\Lambda_z=2\pi/k_{c,z}$ and $\Lambda_o=2\pi/k_{c,o}$.
The dynamics of $z(\mathbf{x},t)$ and $o(\mathbf{x},t)$ is coupled by interaction terms which
can be derived from a coupling energy $U$.
In the uncoupled case this dynamics leads to pinwheel free OP stripe patterns.
%%%%%%%%%%%%%%%%%%%%%%%%%%%%%%%%%%%%%%%%%%%%%%%%%
\setcounter{section}{1}
\subsection{Symmetries constrain inter-map coupling energies}\label{sec:Phenomeno_part3a}
\begin{figure}
\includegraphics[width=\linewidth]{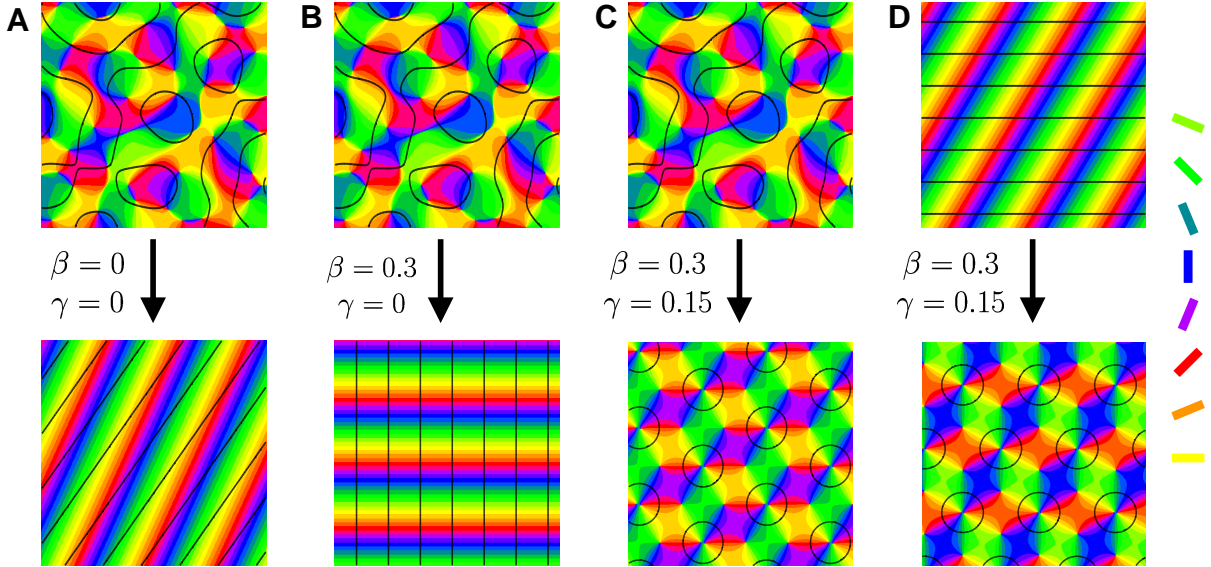}
\caption{
\textbf{Pinwheel annihilation, preservation, and generation} in numerical simulations for different strengths of
inter-map coupling and OD bias, $r_o=0.2, r_z=0.02$.
Color code of OP map with zero contours of OD map superimposed. 
\textbf{\sffamily{A}} $\gamma=0,\beta=0$
\textbf{\sffamily{B}} $\gamma=0,\beta=0.3$
\textbf{\sffamily{C}} and \textbf{\sffamily{D}} $\gamma=0.15,\beta=0.3$.
Initial conditions identical in \textbf{\sffamily{A}} - \textbf{\sffamily{C}}, $T_{f}=10^4, k_{c,o}=k_{c,z}$.}
\label{fig:Numerics}
\end{figure}

How many inter-map coupling energies $U$ exist?
Using a phenomenological approach the inclusion and exclusion of various terms
has to be strictly justified. We did this by symmetry considerations.
The constant $\gamma$ breaks the inversion symmetry $o(\mathbf{x})=-o(\mathbf{x})$
of inputs from the ipsilateral ($o(\mathbf{x})<0$) or contralateral ($o(\mathbf{x})>0$) eye.
Such an inversion symmetry breaking could also arise from quadratic terms such as $o(\mathbf{x})^2$.
In the methods section we detail how a constant shift in the field $o(\mathbf{x})$ can eliminate
the constant term and generate such a quadratic term. Including either a shift or a quadratic term thus
already represents the most general case.
The inter-map coupling energy $U$ was assumed to be invariant under this inversion.
Otherwise orientation selective neurons would, for an equal
representation of the two eyes, develop 
different layouts to inputs from the left or the right eye.
The primary visual cortex shows no anatomical indication that there are any prominent
regions or directions parallel to the cortical layers \cite{Braitenberg2}. 
Besides invariance under translations $\hat{T}_y z(\mathbf{x})=z(\mathbf{x}-\mathbf{y})$ and 
rotations $\hat{R}_\phi z(\mathbf{x})=z(R^{-1}_\phi \mathbf{x})$ of both maps we required that the
dynamics should be invariant under orientation shifts 
$z(\mathbf{x})\rightarrow e^{\imath \vartheta} z(\mathbf{x})$.
Note, that the assumption of shift symmetry is an idealization that uncouples
the OP map from the map of visual space. 
Bressloff and coworkers have presented arguments that Euclidean symmetry that
couples spatial locations to orientation shift represents a more plausible symmetry 
for visual cortical dynamics \cite{Cowan5,Cowan3}, see also \cite{Herrmann2}.
The existence of orientation shift symmetry, however, is not an all or none question.
Recent evidence in fact indicates that shift 
symmetry is only weakly broken in the spatial organization of orientation maps \cite{Schnabel2,Schnabel3}.
%Numerous optimization and developmental models have been proposed that exhibit this symmetry. 
%Note, however, that the proposed set of symmetries has been questioned 
%raising the possibility that the orientation shift symmetry 
%is not a proper symmetry of OP map development \cite{Schnabel2}.
A general coupling energy term can be expressed by integral operators
which can be written as a Volterra series
\begin{equation}\label{eq:Volterra}
E=\sum_{u=u_o+u_z}^{\infty} \int \prod_{i=1}^{u_o} d^2x_i \, o(\mathbf{x}_i) \prod _{j=u_o+1}^{u_o+u_z/2} d^2x_j\,
z(\mathbf{x}_j)\prod_{k=u_o+u_z/2+1}^{u} d^2x_k \, \overline{z}(\mathbf{x}_k)\,
K_u(\mathbf{x}_1,\dots,\mathbf{x}_u) \, ,
\end{equation}
with an $u$-th. order integral kernel $K_u$. 
Inversion symmetry and orientation shift symmetry require $u_o$ to be even and 
that the number of fields $z$ equals the number of fields $\overline{z}$.  
The lowest order term, mediating an interaction between the fields
$o$ and $z$ is given by $u=4, u_o=2$ i.e.
\begin{equation}\label{eq:Volterra4}
 E_4=\int d^2x_1 d^2x_2 d^2x_3 d^2x_4 \, o(\mathbf{x}_1) o(\mathbf{x}_2) z(\mathbf{x}_3) \overline{z}(\mathbf{x}_4)\, K_4(\mathbf{x}_1,\mathbf{x}_2,\mathbf{x}_3,\mathbf{x}_4)\, .
\end{equation}
Next, we rewrite Eq.~(\ref{eq:Volterra4}) as an integral over an energy density $U$.
We use the invariance under translations to introduce new coordinates
\begin{eqnarray}
\mathbf{x}_m &=& (1/4)\sum_j^4 \mathbf{x}_i \nonumber \\
\mathbf{y}_1 &=& \mathbf{x}_1-\mathbf{x}_m \nonumber \\
\mathbf{y}_2 &=&\mathbf{x}_2-\mathbf{x}_m \nonumber \\
\mathbf{y}_3 &=& \mathbf{x}_3-\mathbf{x}_m \, .
\end{eqnarray}
This leads to 
\begin{eqnarray}
 E_4&=&\int d^2 x_m \int d^2y_1 d^2 y_2 d^2 y_3 \, o(\mathbf{y}_1+\mathbf{x}_m)
o(\mathbf{y}_2+\mathbf{x}_m) z(\mathbf{y}_3+\mathbf{x}_m) \nonumber \\ 
&& \overline{z}(\mathbf{x}_m-\sum_i^3(\mathbf{y}_i-\mathbf{x}_m)) K(\mathbf{y}_1,\mathbf{y}_2,\mathbf{y}_3) \nonumber \\
&=&\int d^2x_m \, U_4(\mathbf{x}_m)\, .
\end{eqnarray}
The kernel $K$ may contain local and non-local contributions.
Map interactions were assumed to be local. 
For local interactions the integral kernel is independent of the locations $\mathbf{y}_i$.
We expanded both fields in a Taylor series around $\mathbf{x}_m$
\begin{equation}\label{eq:Taylor_part3a}
z(\mathbf{x}_m+\mathbf{y}_i)=z(\mathbf{x}_m)+\nabla z(\mathbf{x}_m) \mathbf{y}_i +\dots ,\quad  
o(\mathbf{x}_m+\mathbf{y}_i)=o(\mathbf{x}_m)+\nabla o(\mathbf{x}_m) \mathbf{y}_i +\dots
\end{equation}
For a local energy density we could truncate this expansion at the first order in the derivatives. 
The energy density can thus be written
\begin{eqnarray}\label{eq:T4fields_part3a}
U_4(\mathbf{x}_m)&=& \int d^2y_1 d^2 y_2 d^2 y_3
\left(o(\mathbf{x}_m)+\nabla o(\mathbf{x}_m)\mathbf{y}_1 \right)
\left(o(\mathbf{x}_m)+\nabla o(\mathbf{x}_m)\mathbf{y}_2 \right)  \\
&&\left(z(\mathbf{x}_m)+\nabla z(\mathbf{x}_m)\mathbf{y}_3 \right)
\left(\overline{z}(\mathbf{x}_m)-\nabla \overline{z}(\mathbf{x}_m)\sum_i(\mathbf{y}_i-\mathbf{x}_m) \right)
K(\mathbf{y}_1,\mathbf{y}_2,\mathbf{y}_3)\, . \nonumber 
\end{eqnarray}
Due to rotation symmetry this energy density should be invariant under a simultaneous rotation of both fields. 
From all possible combinations of Eq.~(\ref{eq:T4fields_part3a}) only those are invariant
in which the gradients of the fields appear as scalar products.
The energy density can thus be written as
\begin{eqnarray}
U_4&=&f(c_1,c_2,\dots, c_8) \nonumber \\
&=& f(o^2,z^2,z\overline{z},oz,\nabla o  \dotp \nabla o,\nabla z \dotp \nabla z,\nabla z \dotp \nabla \overline{z},\nabla o \dotp \nabla z) \, ,
\end{eqnarray}
where we suppress the argument $\mathbf{x}_m$.
All combinations $c_j$ can also enter via their complex conjugate.
The general expression for $U_4$ is therefore
\begin{equation}
 U_4=\sum_{i>j}l^{(1)}_{ij} c_ic_j+\sum_{i>j}l^{(2)}_{ij} \overline{c}_i \overline{c}_j+\sum_{i,j}l^{(3)}_{ij} c_i \overline{c}_j\, .
\end{equation}
From all possible combinations we selected those which are invariant
under orientation shifts and eye inversions.
This leads to
\begin{eqnarray}
 U_4&=& l_1o^4 + l_2 |z|^4 + l_3 (\nabla o \dotp \nabla o)^2 + l_4 |\nabla z \dotp \nabla z|^2 \nonumber \\
&& +l_5 (\nabla z \dotp \nabla \overline{z})^2 +l_6 (\nabla o \dotp \nabla o) o^2 + l_7 (\nabla z \dotp \nabla \overline{z}) |z|^2 \nonumber \\
&& +l_8 (\nabla z \dotp \nabla z) \overline{z}^2 + l_9 (\nabla \overline{z} \dotp \nabla \overline{z}) z^2 \nonumber \\
&&+l_{10} (\nabla o \dotp \nabla z)o\overline{z} +l_{11} (\nabla o \dotp \nabla \overline{z} )o z \nonumber \\
&&+l_{12} o^2|z|^2 +l_{13}|\nabla o \dotp \nabla z|^2 +l_{14} (\nabla z \dotp \nabla \overline{z})o^2 \nonumber \\
&& +l_{15} (\nabla o \dotp \nabla o)|z|^2 +l_{16} (\nabla z \dotp \nabla \overline{z})(\nabla o \dotp \nabla o) \, .
\end{eqnarray}
The energy densities with prefactor $l_1$ to $l_9$ do not mediate a coupling between OD and OP fields 
and can be absorbed into the single field energy functionals.
The densities with prefactors $l_{8}$ and $l_{9}$ (also with $l_{10}$ and $l_{11}$) are complex 
and can occur only together with $l_{8}=l_{9}$ ($l_{10}=l_{11}$) to be real. 
These energy densities, however, are not bounded from below as their real and imaginary parts can have arbitrary
positive and negative values.
%%%%
The lowest order terms which are real and positive definite are thus given by
\begin{equation}\label{eq:T4}
U_4=l_{12} o^2|z|^2+l_{13} |\nabla \dotp o\nabla z|^2+l_{14} o^2\nabla z \dotp \nabla 
\overline{z}+l_{15} \nabla o \dotp \nabla o |z|^2+l_{16}\left(\nabla z \dotp \nabla \overline{z}\right)
\left(\nabla o \dotp \nabla o \right)\, .
\end{equation}
The next higher order energy terms are given by 
\begin{equation}\label{eq:T6}
U_6=o^2|z|^4 + |z|^2o^4+ o^4 \nabla z \dotp \nabla \overline{z} +\dots
\end{equation}
Here the fields $o(\mathbf{x})$ and $z(\mathbf{x})$ enter with an unequal power. In the corresponding field
equations these interaction terms enter either in the linear part or 
in the cubic nonlinearity. We will show in this article that 
interaction terms that enter
in the linear part of the dynamics can lead to a suppression of the pattern
and possibly to an instability of the pattern solution.
Therefore we considered also higher order interaction terms.\\
These higher order terms contain combinations of terms in Eq.~(\ref{eq:T4}) and are given by
\begin{eqnarray}\label{eq:T8}
U_8&=& o^4|z|^4 + |\nabla o \dotp \nabla z|^4+o^4 \left(\nabla z \dotp \nabla \overline{z}\right)^2 
+\left(\nabla o \dotp \nabla o\right)^2 |z|^4 \nonumber \\ && 
+\left(\nabla z \dotp \nabla \overline{z}\right)^2
\left(\nabla o \dotp \nabla o \right)^2+o^2|z|^2|\nabla o \dotp \nabla z|^2 +\dots
\end{eqnarray}
As we will show below examples of coupling energies
\begin{equation}\label{eq:Energy_all}
U=\alpha \, o^2|z|^2+\beta \, |\nabla z \dotp \nabla o|^2 +\tau \, o^4 |z|^4 +\epsilon\, |\nabla z \dotp \nabla o|^4 \, ,
\end{equation}
form a representative set that can be expected to reproduce experimentally observed map relationships.
For this choice of energy the corresponding interaction terms in the 
dynamics Eq.~(\ref{eq:DynamicsSHcoupled_ref}) are given by
\begin{eqnarray}\label{eq:InteractionTerm}
 -\frac{\delta U}{\delta \overline{z}}&=&
N_\alpha[o,o,z]+N_\beta[o,o,z]+N_\epsilon[o,o,o,o,z,z,\overline{z}]+N_\tau[o,o,o,o,z,z,\overline{z}]\nonumber \\
&=&-\alpha o^2z+\beta\nabla \left(a\nabla o\right)  
+ \epsilon \, 2\nabla \left(|a|^2a\nabla o\right) -2\tau \,o^4|z|^2z ,\nonumber \\
-\frac{\delta U}{\delta o}&=&
\widetilde{N}_\alpha[o,z,\overline{z}]+\widetilde{N}_\beta[o,z,\overline{z}]+
\widetilde{N}_\epsilon[o,o,o,z,z,\overline{z},\overline{z}]+
\widetilde{N}_\tau[o,o,o,z,z,\overline{z},\overline{z}]\nonumber \\
&=&-\alpha o|z|^2+\beta \nabla \left( \overline{a} \nabla z \right) + \epsilon \, 2 \nabla \left(|a|^2\overline{a} \nabla z \right)-2\tau\, o^3|z|^4+c.c. 
\end{eqnarray}
with $a=\nabla z \dotp \nabla o$ and $c.c.$ denoting the complex conjugate.
In general, all coupling energies in $U_4, U_6$, and $U_8$ can occur in the dynamics and we
restrict to those energies which are expected to reproduce the observed geometric relationships
between OP and OD maps.
It is important to note that with this restriction we did not miss any essential parts of the model.
When using weakly nonlinear analysis the general form of the near threshold dynamics is insensitive to the
used type of coupling energy and we therefore expect similar results
also for the remaining coupling energies. \\
%%%%%%%%%%%%%%%%%%%%%%%%%%%%%%%%%%%%%%%%%%%%%%%%
Numerical simulations of the dynamics Eq.~(\ref{eq:DynamicsSHcoupled_ref}), see 
\cite{Reichl4,Reichl6}, with the coupling 
energy Eq.~(\ref{eq:Energy_all})
and $\alpha=\epsilon=\tau=0$ are shown in Fig.~\ref{fig:Numerics}.
The initial conditions and final states are shown for different bias terms $\gamma$ 
and inter-map coupling strengths $\beta$. 
We observed that for a substantial contralateral bias and above a
critical inter-map coupling pinwheels are preserved from random initial conditions or
are generated if the initial condition is pinwheel free.
Without a contralateral bias the final states were pinwheel free stripe solutions 
irrespective of the strength of the inter-map coupling.
%%%%%%%%%%%%%%%%%%%%%%%
\subsection{Calculating ground states by coupled amplitude equations}\label{sec:CoupledAmplitudeEquations_part3}
We studied Eq.~(\ref{eq:DynamicsSHcoupled_ref}) with the low order inter-map
coupling energies in Eq.~(\ref{eq:Energy_all}) using weakly nonlinear analysis.
We therefore rewrite Eq.~(\ref{eq:DynamicsSHcoupled_ref}) as
\begin{eqnarray}\label{eq:fieldEq_forAmplDeriv}
 \partial_t \, z(\mathbf{x},t) &=& r_z z(\mathbf{x},t) -\hat{L}_z^0\, z(\mathbf{x},t) -N_{3,u}[z,z,\overline{z}] -N_{3,c}[z,o,o] \nonumber \\
 \partial_t \, o(\mathbf{x},t) &=& r_o o(\mathbf{x},t) -\hat{L}_o^0 \, o(\mathbf{x},t) +N_{2,u}[o,o]-N_{3,u}[o,o,o] -\tilde{N}_{3,c}[o,z,\overline{z}]\, ,
\end{eqnarray}
where we shifted both linear operators as $\hat{L}_z=r_z+\hat{L}_z^0$, $\hat{L}_o=r_o+\hat{L}_o^0$.
The constant term $\gamma$ in Eq.~(\ref{eq:DynamicsSHcoupled_ref}) is replaced by a
quadratic interaction term $N_{2,u}[o,o]=\tilde{\gamma} o^2$ with 
$\tilde{\gamma}\propto \sqrt{r_o}$, see Methods.
The uncoupled nonlinearities are given by $N_{3,u}=|z|^2z$, $\tilde{N}_{3,u}=o^3$ while 
$N_{3,c}$ and $\tilde{N}_{3,c}$ are the nonlinearities of the low order inter-map 
coupling energy Eq.~(\ref{eq:InteractionTerm}).
We study Eq.~(\ref{eq:fieldEq_forAmplDeriv}) close to the pattern forming bifurcation where
$r_z$ and $r_o$ are small. 
We therefore expand both control parameters in powers of the small expansion parameter $\mu$
\begin{eqnarray}
 r_z &=&\mu r_{z1}+\mu^2 r_{z2} +\mu^3 r_{z3}+\dots \nonumber \\
 r_o &=& \mu r_{o1} +\mu^2 r_{o2} +\mu^3 r_{o3}+\dots  \, .
\end{eqnarray}
Close to the bifurcation the fileds are small and thus nonlinearities are weak.
We therefore expand both fields as
\begin{align}
o(\mathbf{x},t)&=\mu o_1(\mathbf{x},t)+\mu^2 o_2(\mathbf{x},t)+\mu^3 o_3(\mathbf{x},t)+\dots \nonumber \\
z(\mathbf{x},t)&=\mu z_1(\mathbf{x},t)+\mu^2 z_2(\mathbf{x},t)+\mu^3 z_3(\mathbf{x},t)+\dots
\end{align}
We further introduced a common slow timescale $T=r_z t$ and insert the expansions 
in Eq.~(\ref{eq:fieldEq_forAmplDeriv}) and get
\begin{eqnarray}\label{eq:Expansion_z_coupled}
 0&=&\mu  \hat{L}^0 z_1 \nonumber \\
&&+\mu^2\left(-\hat{L}^0z_2+r_{z1} z_1-r_{z1}\partial_Tz_1 \right)\nonumber\\
&&+\mu^3 \left(-r_{z2} \partial_Tz_1+r_{z2} z_1+r_{z1} z_2-r_{z1} \partial_T z_2-\hat{L}^0z_3-N_{3,u}[z_1,z_1,\overline{z}_1]\right) \nonumber \\
&&+\mu^3\left(- N_{3,c}[z_1,o_1,o_1]\right) \nonumber \\
&&\vdots 
\end{eqnarray} 
and
\begin{eqnarray}\label{eq:Expansion_o_coupled}
 0&=&\mu  \hat{L}^0 o_1 \nonumber \\
&&+\mu^2\left(-\hat{L}^0o_2+r_{o1} o_1- r_{z1} \partial_T o_1 +
\sqrt{\mu r_{o1}+\mu^2  r_{o2}+\dots}\tilde{N}_{2,u}[o_1,o_1] \right)\nonumber\\
&&+\mu^3\left(- r_{z2} \partial_T o_1+r_{o2} o_1+r_{o1} o_2- r_{z1} \partial_T o_2-\hat{L}^0o_3-\tilde{N}_{3,u}[o_1,o_1,o_1]\right) \nonumber \\
&&+\mu^3\left(- \tilde{N}_{3,c}[o_1,z_1,\overline{z}_1]\right) \nonumber \\
&&\vdots 
\end{eqnarray}
%For simplicity, we have written only the nonlinearities
%corresponding to the gradient-type inter-map coupling energies where
%we scaled out the inter-map coupling strength $\beta$ and $\epsilon$.
We consider amplitude equations up to third order as this is the order where the 
nonlinearity of the low order inter-map coupling energy enters first.
For Eq.~(\ref{eq:Expansion_z_coupled}) and Eq.~(\ref{eq:Expansion_o_coupled}) to be fulfilled
each individual order in $\mu$ has to be zero. 
At linear order in $\mu$ we get the two homogeneous equations
\begin{equation}
 \hat{L}_z^0 z_1=0\, , \quad  \hat{L}_o^0 o_1=0 \, .
\end{equation}
Thus $z_1$ and $o_1$ are elements of the kernel of $\hat{L}_z^0$ and $\hat{L}_o^0$.
Both kernels contain linear
combinations of modes with a wavevector on the corresponding critical circle i.e.
\begin{eqnarray}\label{eq_z1o1}
z_1(\mathbf{x},T)&=&\sum\limits_j^{n} \left( A_j(T)e^{\imath \vec{k}_j \cdot \vec{x}} +
 A_{j^-}(T)  e^{-\imath \vec{k}_j \cdot \vec{x}} \right)\nonumber \\
o_1(\mathbf{x},T)&=&\sum\limits_j^{n} \left( B_j(T)e^{\imath \vec{k}'_j \cdot \vec{x}} + 
\overline{B}_{j}(T) e^{-\imath  \vec{k}'_j \cdot \vec{x}} \right) ,
\end{eqnarray}
with the complex amplitudes $A_j=\mathcal{A}_j e^{\imath \phi_j}$,  
$B_j=\mathcal{B}_j e^{\imath \psi_j}$ and
$\vec{k}_j=k_{c,z} \left(\cos (j \pi/n), \sin (j \pi/n) \right)$, 
$\vec{k}'_j=k_{c,o} \left(\cos (j \pi/n), \sin (j \pi/n) \right)$. 
In view of  the hexagonal or stripe layout of the OD pattern shown in Fig.~\ref{fig:Numerics}, 
$n=3$ is an appropriate choice.
Since in cat visual cortex the typical wavelength for OD and OP maps are approximately the 
same \cite{Kaschube1,Kaschube2} i.e. $k_{c,o}=k_{c,z}=k_c$
the Fourier components of the emerging pattern are located on a common critical circle.
To account for species differences we also analyzed models with detuned OP and OD wavelengths
in part (II) of this study.\\
At second order in $\mu$ we get 
\begin{eqnarray}
\hat{L}^0 z_2+r_{z1} z_1-r_{z1} \partial_T z_1 &=& 0 \nonumber \\
\hat{L}^0 o_2+r_{o1} o_1- r_{z1} \partial_T o_1 &=&0 \, .
\end{eqnarray}
As $z_1$ and $o_1$ are elements of the kernel $r_{z1}=r_{o1}=0$.
At third order, when applying the solvability condition (see Methods), we get
\begin{eqnarray}
 r_{z2} \partial_T \, z_1 &=& r_{z2}  z_1 - \hat{P}_c N_{3,u}[z_1,z_1,\overline{z}_1] -
\hat{P}_c N_{3,c}[z_1,o_1,o_1]  \\
r_{z2} \partial_T \, o_1&=& r_{o2} o_1-\sqrt{r_{o2}} \, \hat{P}_c \tilde{N}_{2,u}[o_1,o_1]-
\hat{P}_c \tilde{N}_{3,u}[o_1,o_1,o_1] -  \hat{P}_c \tilde{N}_{3,c}[o_1,z_1,\overline{z}_1]\, . \nonumber
\end{eqnarray}
We insert the leading order fields Eq.~(\ref{eq_z1o1}) and obtain the amplitude equations
\begin{eqnarray}\label{eq:withkappa}
r_{z2} \partial_T A_i &=& r_{z2} A_i
-\sum_j h_{ij} |B_j|^2 A_i-hB_i^2 A_{i^-} 
-\sum_j g_{ij}|A_j|^2A_i 
-\sum_j f_{ij}A_j A_{j^-}\overline{A}_{i^-}-\dots \nonumber \\
 r_{z2} \partial_T B_i &=& r_{o2} B_i -\sum_j h_{ij} |A_j|^2 B_i
-2\sqrt{r_{o2}}\,  \overline{B}_{i+1}\overline{B}_{i+2}-\sum_{j} \widetilde{g}_{ij} |B_j|^2 B_i -\dots \, . 
\end{eqnarray}
For simplicity we have written only the simplest inter-map coupling terms. 
Depending on the configuration of active modes additional contributions may enter the amplitude equations.
In addition, for the product-type coupling energy, there are coupling terms 
which contain the constant $\delta$, see Methods and Eq.~(\ref{eq:Potential_Cubic_part3}).
The coupling coefficients are given by
\begin{align}\label{eq:coupling_atthird_part3}
%\begin{array}{cc}
g_{ij}= & \, e^{-\imath \vec{k}_i \cdot \vec{x}} \left( N_{3,u}[e^{\imath \vec{k}_i \cdot \vec{x}},
e^{\imath \vec{k}_j \cdot \vec{x}},e^{-\imath \vec{k}_j \cdot \vec{x}}] 
+ N_{3,u}[e^{\imath \vec{k}_j \cdot \vec{x}},e^{\imath \vec{k}_i \cdot \vec{x}},
e^{-\imath \vec{k}_j \cdot \vec{x}}] \right)=2 \nonumber \\
g_{ii}= & \, e^{-\imath \vec{k}_i \cdot \vec{x}} \,  N_{3,u}[e^{\imath \vec{k}_i \cdot \vec{x}},
e^{\imath \vec{k}_i \cdot \vec{x}},
e^{-\imath \vec{k}_i \cdot \vec{x}}] =1\nonumber \\
g_{ij^-}= & \, e^{-\imath \vec{k}_i \cdot \vec{x}} \left( N_{3,u}[e^{\imath \vec{k}_i \cdot \vec{x}},
e^{-\imath \vec{k}_j \cdot \vec{x}},e^{\imath \vec{k}_j \cdot \vec{x}}] 
+ N_{3,u}[e^{-\imath \vec{k}_j \cdot \vec{x}},e^{\imath \vec{k}_i \cdot \vec{x}},
e^{\imath \vec{k}_j \cdot \vec{x}}] \right)=2 \nonumber \\
f_{ij}=& \, e^{-\imath \vec{k}_i \cdot \vec{x}} \left( N_{3,u}[e^{\imath \vec{k}_j \cdot \vec{x}},
e^{-\imath \vec{k}_j \cdot \vec{x}},e^{\imath \vec{k}_i \cdot \vec{x}}]+
N_{3,u}[e^{-\imath \vec{k}_j \cdot \vec{x}},
e^{\imath \vec{k}_j \cdot \vec{x}},e^{\imath \vec{k}_i \cdot \vec{x}}] \right)=2 \nonumber \\
f_{ii}=&\, 0 \nonumber \\
h_{ij}= & \, e^{-\imath \vec{k}_i \cdot \vec{x}} 
 N_{3,c}[e^{\imath \vec{k}_i \cdot \vec{x}},e^{\imath \vec{k}_j \cdot \vec{x}},
e^{-\imath \vec{k}_j \cdot \vec{x}}]  \nonumber \\
h_{ii}= & \, e^{-\imath \vec{k}_i \cdot \vec{x}} \,  N_{3,c}[e^{\imath \vec{k}_i \cdot \vec{x}},
e^{\imath \vec{k}_i \cdot \vec{x}},
e^{-\imath \vec{k}_i \cdot \vec{x}}] \nonumber \\
h= & \, e^{-\imath \vec{k}_i \cdot \vec{x}} \,  N_{3,c}[e^{-\imath \vec{k}_i \cdot \vec{x}},
e^{\imath \vec{k}_i \cdot \vec{x}},
e^{\imath \vec{k}_i \cdot \vec{x}}] \, .
%\end{array}
\end{align}
From Eq.~(\ref{eq:withkappa}) we see that inter-map coupling has two effects on the modes
of the OP pattern. First, inter-map coupling shifts the bifurcation point from $r_z$ to
$\left(r_z-\sum_j h^{(1)}_{ij} |B_j|^2 \right)$. This can cause a potential destabilization of
pattern solutions for large inter-map coupling strength.
Second, inter-map coupling introduces additional resonant interactions that for instance couple the modes
$A_i$ and their opposite modes $A_{i^-}$. 
In case of $A\ll B \ll 1$ the inter-map coupling terms in dynamics of the modes $B$ are small.
In this limit the dynamics of the modes $B$ decouples from the modes $A$ and we can use the uncoupled
OD dynamics, see Methods.
%%%%
When we scale back to the fast time variable and set $r_{z2}=r_z$, $r_{o2}=r_o$ we obtain
\begin{eqnarray}
 \partial_t A_i &=& r_z A_i-\sum_j h_{ij} |B_j|^2 A_i-h  B_i^2 A_{i^-}
-\sum_j g_{ij}|A_j|^2A_i -\sum_j f_{ij}A_j A_{j^-}\overline{A}_{i^-}\nonumber \\
 r_z \partial_t B_i &=& r_o B_i - \sum_j h_{ij} |A_j|^2 B_i
 -2\sqrt{r_o}\,  \overline{B}_{i+1}\overline{B}_{i+2}-\sum_{j} \widetilde{g}_{ij} |B_j|^2 B_i  \, . 
\end{eqnarray}
%%%%
The amplitude equations are truncated at third order. If pattern formation takes place
far from threshold fifth order, seventh order, or even higher order corrections have to be considered.
Moreover, at seventh order the nonlinearities of the higher order inter-map coupling
terms enter the expansion.
A derivation of amplitude equation with higher order inter-map coupling energies is presented in
Text S1.
%%%%%%%%%%%%%%%%
%
\subsection{Interpretation of coupling energies}
%%%%%%%%%%%%%%%%%%%%%%%%%%%%%%%%%%%%%%%%%%%%%%%%
% Discussion of coupling energies
Using symmetry considerations we derived inter-map coupling energies up to eighth order in 
the fields, see Eq.~(\ref{eq:T4}), Eq.(\ref{eq:T6}), and Eq.(\ref{eq:T8}).
Which of these various optimization principles could reproduce realistic inter-map relationships
such as a uniform coverage of all stimulus features?
We identified two types of optimization principles that can be 
expected to reproduce realistic inter-map relationships and good stimulus coverage.
%1: a la Coverage/ Continuity
%Product
First, product-type coupling energies of the form $U=o^{2n}|z|^{2n}\, ,\, n=1,2,...$. 
These energies favor configurations in which regions of high gradients avoid each other
and thus leading to high coverage.
%Gradient
Second, gradient-type coupling energies of the form $U=|\nabla o \dotp \nabla z|^{2n}\, ,\, n=1,2,... $.
In experimentally obtained maps, iso-orientation lines show the tendency to intersect the OD 
borders perpendicularly. 
Perpendicular intersection angles lead to high coverage as large changes of the field $z$ in one direction
lead to small changes of the field $o$ in that direction.
To see that the gradient-type coupling energy favors perpendicular intersection angles
we decompose the complex field $z(\mathbf{x})$ into the selectivity $|z|$ and 
the preferred orientation $\vartheta$. We obtain
\begin{equation}\label{eq:EnergyGradientThird}
U= |\nabla z \dotp \nabla o|^{2n}=   |z|^{2n} \left( |\nabla o \dotp \nabla \ln |z| |^2
+4|\nabla o \dotp \nabla \vartheta|^2  \right)^n.
\end{equation}
If the orientation selectivity is locally homogeneous, i.e. $ \nabla \ln |z| \approx 0$,
then the energy is minimized if the direction of the iso-orientation lines ($\nabla \vartheta$) is
perpendicular to the OD borders. 
%
%Mixture terms
In our symmetry-based analysis we further identified terms that are expected to lead to the opposite behavior
for instance mixture terms such as $U=o^2 \nabla z  \dotp \nabla \overline{z}$.\\ 
%
%
%2: Pinwheel position
Pinwheels are prominent features in OP maps.
We therefore also analyze how different optimization principles impact on the pinwheel positions
with respect to the co-evolving feature maps.
The product-type coupling energies are expected to favor pinwheels at OD extrema.
Pinwheels are zeros of $z$ and are thus expected to reduce this energy term.
They will reduce energy the most when $|o|$ is maximal which should repel 
pinwheels from OD borders, where $o(\mathbf{x})$ is zero. 
Also the gradient-type coupling energy is expected to couple the OD pattern with the position of pinwheels.
To see this we decompose the field $z$ into its real and imaginary part
\begin{equation}
U= \left( |\nabla o \dotp \, \nabla \textrm{Re} z|^2 + |\nabla o \dotp \, \nabla \textrm{Im} z|^2 \right) .
\end{equation}
At pinwheel centers the zero contours of $\textrm{Re}\, z$ and $\textrm{Im} \, z$ cross.
Since there $\nabla \textrm{Re} \, z$ and $\nabla \textrm{Im} \, z$ are almost constant and
not parallel the energy can be minimized only if $|\nabla o|$
is small at the pinwheel centers, i.e. the extrema or saddle-points of $o(\mathbf{x})$.\\
From the previous considerations we assume all coupling coefficients of the energies to be positive.
A negative coupling coefficient can be saturated by
higher order inter-map coupling terms.
In the following, we discuss the impact of the low order inter-map coupling energies on the
resulting optima of the system using the derived amplitude equations.  
The corresponding analysis for higher order inter-map coupling
energies is provided in Text S1.
%%%%
%In experimentally obtained maps iso-orientation lines show the tendency to intersect the OD 
%borders perpendicularly 
%but only the second term in Eq.~(\ref{eq:T4}) is expected to be sensitive to these intersection angles.
%In order to be sensitive to intersection angles the energy is expected 
%to contain terms of the form $|\nabla o \nabla \vartheta|$.
%The first term in Eq.~(\ref{eq:T4}), 
%\begin{equation}\label{eq:EnergyNonGradientThird}
%U=l_{12} \, o^2|z|^2 \, ,
%\end{equation}
%is also expected to favor pinwheels at OD extrema.
%So the expectation that terms such as Eq.~(\ref{eq:EnergyGradientThird}) and Eq.~(\ref{eq:EnergyNonGradientThird}) 
%and its higher order variants lead to a repulsion of pinwheels from OD borders has some intuitive appeal.
%%%%%%%%%%%%%%%%%%%%%%%
%\subsection*{Optima of particular optimization principles}
\setcounter{section}{2}
\setcounter{subsection}{0}
\setcounter{subsubsection}{0}
%\subsection{The impact of inter-map coupling: Low order coupling terms}\label{sec:MapCoupling_loworder}
\subsection{Optima of particular optimization principles: Low order coupling terms}\label{sec:MapCoupling_loworder}
As indicated by numerical simulations and weakly nonlinear analysis of 
the uncoupled OD dynamics, see Methods, we discussed the influence of the OD stripe, 
hexagon, and constant solutions on the OP map using
the coupled amplitude equations derived in the previous section.
% Backreaction
A potential backreaction onto the dynamics of the OD map can be neglected if the modes
$A_j$ of the OP map are much smaller than the modes $B_j$ of the OD map. This can be achieved
if $r_z \ll r_o$.
We first give a brief description of the uncoupled OP solutions.
Next, we study the impact of the low order coupling energies in Eq.~(\ref{eq:Energy_all}) on these solutions.
We demonstrate that these energies can lead to a complete suppression of orientation selectivity.
%Uncoupled solutions
In the uncoupled case there are for $r_z>0$ two stable stationary solutions to the
amplitude equations Eq.~(\ref{eq:AmplEq_Coupled_thirdOrder}), namely OP stripes
\begin{equation}\label{eq:OPstripes_uncoupled_part3}
 z_{\operatorname{st}}=\mathcal{A} e^{\imath \left( \vec{k}_1 \cdot \vec{x} +\phi \right) }, \quad \mathcal{A}=\sqrt{r_z}\, ,
\end{equation}
and OP rhombic solutions
\begin{equation}\label{eq:OPrhombs_uncoupled_part3}
 z_{\operatorname{rh}}=\mathcal{A} \sum_{j=1}^2 \left( e^{\imath \vec{k}_j \cdot \vec{x}+\phi_j}+
e^{-\imath \vec{k}_j \cdot \vec{x}+\phi_{j^-}}\right) \, ,
\end{equation}
with $\phi_1+\phi_{1^-}=\phi_0$,  $\phi_2+\phi_{2^-}=\phi_0+\pi$, $\phi_0$ an arbitrary phase, and
$\mathcal{A}=\sqrt{r_z/5}\approx 0.447 \sqrt{r_z}$. 
In the uncoupled case the angle $\alpha=\arccos \vec{k_1} \cdot \vec{k_2}/k_c^2$ between the Fourier modes 
is arbitrary.
The stripe solutions are pinwheel free while the pinwheel density for the 
rhombic solutions varies as $\rho=4\sin \alpha$ and thus $0 \leq \rho \leq4$.
% Here : define PWC, rPWC, hPWC
For the rhombic solutions pinwheels are located on a regular lattice. We therefore
refer to these and other pinwheel rich solutions which are spatially
periodic as pinwheel crystals (PWC). In particular, we refer to pinwheel crystals with
as rhombic spatial layout as rPWC solutions and pinwheel crystals with a hexagonal layout
as hPWC solutions.
% Potential
Without inter-map coupling, the potential of the two solutions reads 
$V_{\operatorname{st}}=-\frac{1}{2} r_z^2 <V_{\operatorname{rh}}=-\frac{2}{5} r_z^2$, 
thus the stripe solutions
are always energetically preferred compared to rhombic solutions.\\
In the following we study three scenarios in which inter-map coupling can lead to
pinwheel stabilization. First, a deformation of the OP stripe solution can lead to the creation
of pinwheels in this solution. Second, inter-map coupling can energetically prefer the 
(deformed) OP  rhombic solutions compared to the stripe solutions. Finally, inter-map coupling
can lead to the stabilization of new PWC solutions.\\
For the low order
interaction terms the amplitude equations are given by 
$\partial_t A_i=-\delta V /\delta \overline{A}_i$, $\partial_t B_i=-\delta V /\delta \overline{B}_i$ with the potential
\begin{eqnarray}\label{eq:Potential_Cubic_part3}
V&=& V_A+V_B+\alpha \delta^2 \sum_j^3 \left(  |A_j|^2  + |A_{j^-}|^2 \right) \nonumber \\
&& +2\alpha \delta \left(A_1 \overline{A}_{2^-} B_3 + A_1 \overline{A}_{3^-} B_2 
 + A_{1^-} \overline{A}_{2} \overline{B}_3 + A_{1^-} 
\overline{A}_{3} \overline{B}_2 \right) \nonumber \\
&& +\sum_{i,j} g_{ij}^{(1)} |A_i|^2|B_j|^2 
 +\sum_{i\neq j} g_{ij}^{(2)}  A_i \overline{A}_j\overline{B}_i B_j 
+\sum_{i\neq j}g_{ij}^{(2)}  A_{i^-} \overline{A}_{j^-}B_i \overline{B}_j\nonumber \\
&&+\sum_{i,j} g_{ij}^{(3)} A_i\overline{A}_{j^-} \overline{B}_i \overline{B}_j +
\sum_{i,j} g_{ij}^{(3)} \overline{A}_i A_{j^-} B_i B_j \, ,
\end{eqnarray}
with the uncoupled contributions 
\begin{eqnarray}\label{eq:Potential_uncoupled}
 V_A&=&-r_z \sum_j^3 |A_j|^2 
+\frac{1}{2} \sum_{i,j}^3 g_{ij} |A_i|^2|A_j|^2
+\frac{1}{2} \sum_{i,j}^3 f_{ij} A_iA_{i^-}\overline{A}_j\overline{A}_{j^-} \nonumber \\
V_B&=&-r_o \sum_j^3 |B_j|^2 +\frac{1}{2} \sum_{i,j}^3 \widetilde{g}_{ij} |B_i|^2|B_j|^2 \, . 
\end{eqnarray}
The coupling coefficients read
$g_{ij}^{(1)}=2\alpha +2 \beta \cos ^2 (\alpha_{ij})$, 
$g_{ij}^{(2)}=2\alpha + \beta \left( 1+\cos ^2 (\alpha_{ij}) \right)$,
$g_{ij}^{(3)}=2\alpha + \beta \left( 1+\cos ^2 (\alpha_{ij}) \right)$,
$g_{ii}^{(3)}=\alpha + \beta $, where $\alpha_{ij}$ is the angle between the wavevector 
$\vec{k}_i$ and $\vec{k}_j$. 
%%%%%%%%%%%%%%%%%%%%%%%%%%%%%%%%%%%%%%%%%%%%%%%%%%%%%%%%%%%%%%%%%%%%%%%%%%%%%%%%%
\subsection{Product-type energy $U=\alpha o^2|z|^2$}
We first studied the impact of the low order product-type coupling energy.
Here, the constant $\delta(\gamma)$ enters explicitly in the amplitude 
equations, see Eq.~(\ref{eq:Potential_Cubic_part3}) and Eq.~(\ref{eq:delta}).
%%%%%%%%%%%
% a) OD stripes, OP stripes
% b) OD stripes, OP rhombs
% c) OD hex, OP stripes
% d) OD hex, OP rhombs (evtl distorted)
% e) OD hex, OP unif
\subsubsection{Stationary solutions and their stability}
In the case of OD stripes, see Methods, with $B_1=B,B_2=B_3=0$ we get the 
following amplitude equations
\begin{eqnarray}\label{eq:Amp_alp_ODstripes}
\partial_t \, A_1 &=& \left(r_z -\alpha \delta^2-2\alpha |B|^2 \right) A_1 -\alpha B^2 A_{1^-} 
+\text{nct}. \nonumber \\
\partial_t \, A_2 &=& \left(r_z -\alpha \delta^2-2\alpha |B|^2 \right) A_2 -2\alpha \delta A_{3^-} \overline{B} 
+\text{nct}. \nonumber \\
\partial_t \, A_3 &=& \left(r_z -\alpha \delta^2-2\alpha |B|^2 \right) A_3 -2\alpha \delta A_{2^-} \overline{B} 
+\text{nct}.
\end{eqnarray}
where nct. refers to non inter-map coupling terms 
$-\sum_j^3 g_{ij} |A_j|^2 A_i -\sum_{j\neq i}^3 f_{ij} A_j A_{j^-} \overline{A}_{i^-}$, resulting from the 
potential $V_A$, see Eq.~(\ref{eq:Potential_uncoupled}).
The equations for the modes $A_{i^-}$ are given by interchanging the modes
$A_i$ and $A_{i^-}$ as well as interchanging the modes $B_i$ and $\overline{B}_i$.
%a)
The OP stripe solution in case of inter-map coupling is given by
\begin{equation}
z=\mathcal{A}_1 e^{\imath (\vec{k}_1 \cdot \vec{x} +\phi_1)} +
\mathcal{A}_{1^-}e^{-\imath (\vec{k}_1 \cdot \vec{x}+\phi_{1^-})}\, ,
\end{equation}
with $\mathcal{A}_1=p^{3/2}/(2\sqrt{2}B^2\alpha)$, $\mathcal{A}_{1^-}=p/\sqrt{2}$, and \\ $p=r_z-2B^2\alpha 
-\alpha\delta^2-\sqrt{(r_z-\alpha\delta^2)(r_z-\alpha(4B^2+\delta^2))}$
and the phase relation \\ \mbox{$\phi_{1}-\phi_{1^-}=2\psi_1+\pi$}. 
In the uncoupled case ($\alpha=0$) they reduce to $\mathcal{A}_{1^-}=0$ and $\mathcal{A}_1=\sqrt{r_z}$.
With increasing inter-map coupling the amplitude $\mathcal{A}_{1^-}$ grows and the solutions 
are transformed, reducing the representation
of all but two preferred orientations.
The parameter dependence of this solution is shown in Fig.~\ref{fig:Amplitudes_alpha}\textbf{\sffamily{A}} 
for different values of the bias $\gamma$.
Both amplitudes become identical at $\alpha=r_z / (4B^2+\delta^2)$ with
\begin{equation}
\mathcal{A}_1=\mathcal{A}_{1^-}=\sqrt{\frac{r_z-\alpha B^2-\alpha\delta^2}{3}} \, .
\end{equation}
This pattern solution finally vanishes at
\begin{equation}\label{eq:alpha_crit_stripes}
\alpha_c=r_z/(B^2+\delta^2)=3 \, r_z/r_o \, . 
\end{equation}
%If we insert the dependence of $\delta$, \refeqn{eq:delta} and $B$, \refeqn{eq:stripes} on the 
%bias $\gamma$ we get 
%\begin{equation}
% \alpha_c=3r_z / r_o \, .
%\end{equation}
This existence border is thus independent of the OD bias $\gamma$.
Above this coupling strength only the trivial solution $A_j=0, \, \forall j$ is stable.\\
%b)
In addition to the OP stripe patterns there exist rhombic OP 
solutions, see Fig.~\ref{fig:Amplitudes_alpha}\textbf{\sffamily{B}}.
These rhombic solutions are pinwheel rich with a pinwheel density of $\rho=4\sin \pi/3\approx 3.46$
but are energetically not
preferred compared to the stripe solutions, see Fig.~\ref{fig:Amplitudes_alpha}\textbf{\sffamily{E}}.
The rhombic solutions in the uncoupled case 
$\mathcal{A}_1=\mathcal{A}_{1^-}=\mathcal{A}_2=\mathcal{A}_{2^-}$, $\mathcal{A}_3=\mathcal{A}_{3^-}=0$
are transformed by inter-map coupling.
The phase relations are given by 
\begin{eqnarray}\label{eq:Phases_alp_ODstripes_OPrhombs}
\phi_1+\phi_{1^-}&=\phi_0 ,    &   \phi_1-\phi_{1^-}=2\psi_1 +\pi \nonumber \\
\phi_2+\phi_{2^-}&=\phi_0 +\pi, &  \phi_3+\phi_{3^-}=\phi_0+\pi \nonumber \\
\psi_1+\phi_2+\phi_3&=\phi_0, & 
\end{eqnarray} 
where $\phi_0$ is an arbitrary phase.
Stationary amplitudes are given by
\begin{eqnarray}
 \mathcal{A}_1=\mathcal{A}_{1^-}&=&\sqrt{\frac{r_z+\alpha(\mathcal{B}^2-\delta^2)}{5}} \nonumber \\
\mathcal{A}_2=\mathcal{A}_{2^-}&=&\sqrt{\frac{3r_z-12\mathcal{B}^2\alpha-3\alpha\delta^2-1/3 q}{30}} \nonumber \\
\mathcal{A}_3=\mathcal{A}_{3^-}&=&\mathcal{A}_2\frac{3r_z-12\mathcal{B}^2\alpha-3\alpha\delta^2+1/3 q}{20\mathcal{B}\alpha \delta} \, ,
\end{eqnarray}
with $q=\sqrt{-3600\mathcal{B}^2\alpha^2\delta^2+(-9r_z+36\mathcal{B}^2\alpha+9\alpha\delta^2)^2}$.
With increasing inter-map coupling strength $\alpha$ the amplitudes $A_2=A_{2^-}$ are 
suppressed, see Fig.~(\ref{fig:Amplitudes_alpha})\textbf{\sffamily{B}}.
In addition, for nonzero bias $\gamma$, there is an increase of the amplitudes $A_3=A_{3^-}$.
The amplitudes $A_2$ and $A_3$ collapse at 
$\alpha/r_z=3/(12\mathcal{B}^2+20\mathcal{B}\delta+3\delta^2)$. A further increase of the inter-map coupling
strength leads to a suppression of these amplitudes and finally to the OP stripe pattern 
where $\mathcal{A}_2=\mathcal{A}_3=0$.\\
In the case the OD map is a constant, Eq.~(\ref{eq:ODconst}), the amplitude equations simplify to
\begin{equation}
 \partial_t \, A_i= \left(r_z-\alpha \delta^2\right) A_i -\sum_j g_{ij}|A_j|^2A_i -\sum_j f_{ij}
A_jA_{j^-}\overline{A}_{i^-} \, .
\end{equation}
Thus inter-map coupling in this case only renormalizes the bifurcation parameter and the energetic ground state 
is thus a stripe pattern with an inter-map coupling dependent reduction of the amplitudes
\begin{equation}
 \mathcal{A}_1=\sqrt{r_z-\alpha \delta^2}\, , \mathcal{A}_2=\mathcal{A}_3=0\, . 
\end{equation}
Therefore at $\alpha_c=r_z / \delta^2$ the stripe pattern ceases to exist and the only stable
solution is the trivial one i.e. $A_i=0$.
In addition, there is the rhombic solution with the stationary amplitudes
\begin{equation}
 \mathcal{A}_1=\mathcal{A}_{1^-}=\mathcal{A}_2=\mathcal{A}_{2⁻}=
\sqrt{\frac{r_z-\alpha \delta^2}{5}}\, , \mathcal{A}_3=0 \, .
\end{equation}
In the case of OD hexagons $B_i=\mathcal{B}e^{\imath \psi_i}$, Eq.~(\ref{eq:hex}), the amplitude 
equations read
\begin{eqnarray}\label{eq:Amp_alp_ODhex}
\partial_t \, A_i &=& \left(r_z-6\alpha \mathcal{B}^2 -\alpha \delta^2 \right) A_i
-\alpha \mathcal{B}^2 \left(A_{i^-}e^{2\imath \psi_i}+2A_{(i+1)^-} e^{\imath (\psi_i+\psi_{i+1})} +2A_{(i+2)^-} e^{\imath (\psi_i+\psi_{i+2})}\right) \nonumber \\
&& -2\alpha \mathcal{B}^2 \left( A_{i+1} e^{\imath (\psi_i-\psi_{i+1} )}+ A_{i+2} e^{\imath (\psi_i-\psi_{i+2} )}\right) \nonumber \\
&& -2\alpha\delta \mathcal{B} \left(A_{(i+1)^-}e^{-\imath \psi_{i+2}}+A_{(i+2)^-}e^{-\imath \psi_{i+1}}\right) 
+\text{nct}. \, ,
\end{eqnarray}
where the indices are cyclic i.e. $i+3=i$.
% c)
These amplitude equations have stripe-like solutions 
% d)
as well as solutions with a rhombic layout of the form  
$\mathcal{A}_1=\mathcal{A}_{1^-}=\mathcal{A}_2=\mathcal{A}_{2^-}$, $\mathcal{A}_3=\mathcal{A}_{3^-}$.
For both solutions the stationary phases depend on the inter-map coupling strength $\alpha$.
In this case stationary solutions of Eq.~(\ref{eq:Amp_alp_ODhex}) are calculated 
numerically using a Newton method and initial conditions close to these solutions.  
%while the phase relations are given by $\phi=\left( 0,\phi_0, 2\phi_0,\pi, \phi_0+\pi,2\phi_0+\pi\right)$.
% e)
In contrast to the case of OD stripes and OD constant solutions the amplitude 
equations (\ref{eq:Amp_alp_ODhex}) have
an additional type of PWC solution which have uniform amplitudes, $A_j=\mathcal{A}e^{\imath \phi_i}$.
The dynamics of their phases is given by 
\begin{eqnarray}
\partial_t \, \phi_i &=& 2\mathcal{A}^2 \sum_{j\neq i} \sin \left(\phi_i+\phi_{i^-}-\phi_j-\phi_{j^-} \right) \nonumber \\
&&-\mathcal{B}^2 \alpha \sum_{j\neq i} \left( 2\sin \left(\phi_i-\phi_j-\psi_i+\psi_j \right)
+2\sin \left(\phi_i-\phi_{j^-}-\psi_i-\psi_j \right) \right) \nonumber \\
&&-\mathcal{B}^2 \alpha \sin \left(\phi_i-\phi_{i^-}-2\psi_i\right) \nonumber \\  
&&-2\delta \alpha \mathcal{B} \left( \sin \left(\phi_i-\phi_{(i+1)^-}+\psi_{i+2}\right)
+\sin \left(\phi_i-\phi_{(i+2)^-}+\psi_{i+1}\right) \right)\, .
\end{eqnarray}
When solving the amplitude equations numerically we observe that the phase relations
vary with the inter-map coupling strength for non-uniform solutions. But for the uniform solution the
phase relations are independent of the inter-map coupling strength.
%Here, the stationary phases are independent of the inter-map coupling strength $\alpha$.
The phases of the uniform solution are determined up to a free phase $\varphi$ which 
results from the orientation
shift symmetry $z\rightarrow z \, e^{\imath \varphi}$ of Eq.~(\ref{eq:DynamicsSHcoupled_ref}).
We therefore choose $\phi_1=\psi_1$.
As an ansatz for the uniform solutions we use
\begin{eqnarray}\label{eq:UniformSol_part3}
\mathcal{A}_j &=& \mathcal{A}_{j^-}=\mathcal{A}, \quad  j=1,2,3  \nonumber \\
\phi_j &=& \psi_j+(j-1)2\pi/3+\Delta\, \delta_{j,2}  \nonumber \\
\phi_{j^-} &=& -\psi_j+(j-1)2\pi/3+\Delta\, \left( \delta_{j,1}+\delta_{j,3}\right)\, ,
\end{eqnarray}
where $\delta_{i,j}$ is the Kronecker delta and $\Delta$ a constant parameter.
Note, that $z(\mathbf{x})$ cannot become real since $\phi_j \neq -\phi_{j^-}$.
The equation for the uniform amplitudes is then given by
\begin{equation}
\partial_t \, \mathcal{A}= r_z \mathcal{A} -9\mathcal{A}^3-4\alpha\mathcal{B}^2\mathcal{A}
-\alpha \delta^2\mathcal{A}+\mathcal{A}\mathcal{B} \alpha \left(\mathcal{B}-2\delta \right)\cos \Delta\, ,
\end{equation}
while the phase dynamics reads 
\begin{equation}
 \partial_t \phi_j= -\mathcal{B}\alpha \left(\mathcal{B}-2\delta\right) \sin \Delta \, .
\end{equation}
The stationarity condition is fulfilled for an arbitrary $\delta$ only if $\Delta=0$ or $\Delta=\pi$.
The corresponding amplitudes are given by solving the stationarity condition for the real part and read
\begin{equation}\label{eq:StatAmp_unif_alp}
\mathcal{A}_{\Delta=0}=\sqrt{\frac{r_z-\alpha\left(3\mathcal{B}^2+2\mathcal{B}\delta+\delta^2\right)}{9}} \, , \quad
\mathcal{A}_{\Delta=\pi}=\sqrt{\frac{r_z-\alpha\left(5\mathcal{B}^2-2\mathcal{B}\delta+\delta^2\right)}{9}} \, .
\end{equation}
%%% Linear stability analysis
We calculate the stability properties of all solutions by linear stability analysis
considering perturbations of the amplitudes $\mathcal{A}_ j \rightarrow \mathcal{A} +a_j$, 
$\mathcal{A}_ {j^-} \rightarrow \mathcal{A} +a_{j^-}$ and
of the phases $\phi_j \rightarrow \phi_j + \varphi_j$, $\phi_{j^-} \rightarrow \phi_{j^-} + \varphi_{j^-}$. 
This leads to a perturbation matrix $M$.
Amplitude and phase perturbations in general do not decouple. We calculated the eigenvalues 
of the perturbation $M$ matrix numerically and checked the results by direct numerical simulation
of the amplitude equations.
In case of the uniform solutions Eq.~(\ref{eq:UniformSol_part3}) the perturbation matrix $M$
is explicitly stated in Text S3.\\
%%%%
The stability of the $\Delta=0 $and $\Delta=\pi$ uniform solutions 
depends on the coupling strength $\alpha$ and on the 
sign of $\left(\mathcal{B}-2\delta\right)$. As the solution of $B(\gamma)=2\delta(\gamma)$ is 
given by $\gamma=\gamma^*$ in the stability range of OD hexagons there is only 
one possible stable uniform solution, the $\Delta=\pi$ uniform solution.
This solution ceases to exist at 
$r_z<\alpha\left(5\mathcal{B}^2-2\mathcal{B}\delta+\delta^2\right)$.
This existence border is in fact independent of the bias $\gamma$ and given by 
\begin{equation}\label{eq:alpha_crit_hex}
 \alpha_{c}=3 r_z / r_o \, .
\end{equation} 
Thus the limit $r_z \rightarrow 0$ makes the uniform solution unstable for smaller and smaller
coupling strengths.
%%%%%%%%%%%%%%%%%%%
\subsubsection{Bifurcation diagram}
\begin{figure}[bt]
\begin{center}
\includegraphics[width=.9\linewidth]{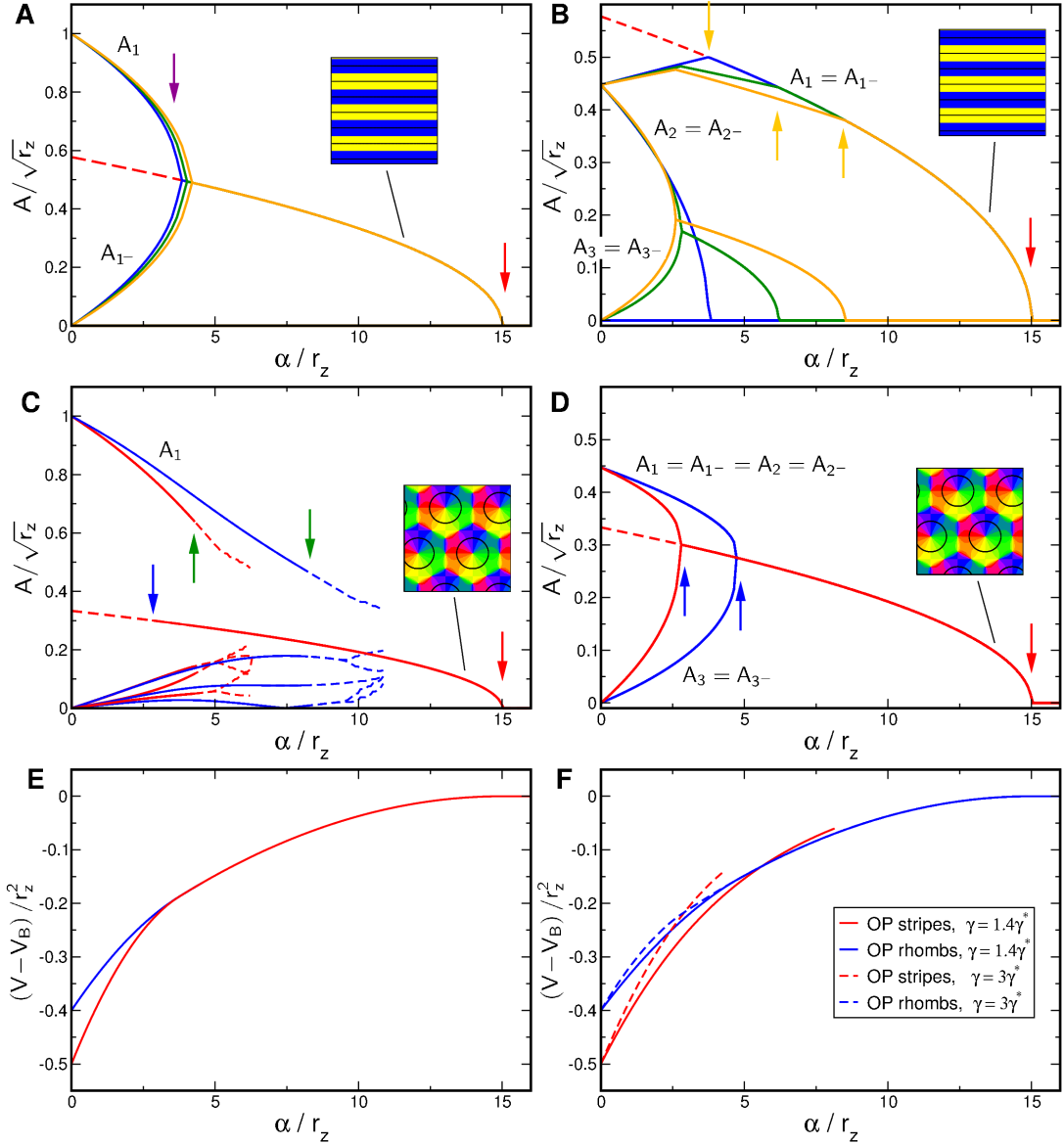} 
\end{center}
\caption{\textbf{Stationary amplitudes with coupling energy} $\mathbf{U=\alpha \, |z|^2 o^2}$, $r_o=0.2$.
Solid (dashed) lines: stable (unstable) solutions 
to Eq.~(\ref{eq:Amp_alp_ODstripes}) (OD stripes) and Eq.~(\ref{eq:Amp_alp_ODhex}) (OD hexagons).
\textbf{\sffamily{A,B}} OD stripes, \mbox{$\gamma=0$} (blue), $\gamma=\gamma^*$ (green), 
$\gamma=1.4 \gamma^*$ (orange). 
\textbf{\sffamily{C,D}} OD hexagons, $\gamma=1.4 \gamma^*$ (blue), $\gamma=3\gamma^*$ (red).
\textbf{\sffamily{A,C}} Transition from OP stripe solutions,
\textbf{\sffamily{B,D}} Transition from OP rhombic solutions.
\textbf{\sffamily{E}} Potential for OP stripes (red) and OP rhombs (blue) interacting with OD stripes, $\gamma=0$.
\textbf{\sffamily{F}} Potential for OP stripes and OP rhombs interacting with 
OD hexagons.
Arrows indicate corresponding lines in the phase diagram, Fig.~(\ref{fig:PD_alpha}).
}
\label{fig:Amplitudes_alpha}
\end{figure}

%OD stripes
The case of OP solutions when interacting with OD stripes is 
shown in Fig.~\ref{fig:Amplitudes_alpha}\textbf{\sffamily{A,B}}.
In case of OP stripes inter-map coupling suppresses the amplitude $A_1$ of the stripe pattern
while increasing the amplitude of the opposite mode $A_{1^-}$.
This transformation reduces the representation of all but two preferred orientations.
When both amplitudes collapse the resulting OP map is selective only to two orthogonal orientations namely 
$\vartheta=\phi_1$ and $\vartheta=\phi_1+\pi/2$.
We refer to these unrealistic solutions as \textit{orientation scotoma solutions}.
The phase relations ensure that OD borders that run parallel to the OP stripes
are located at the OP maxima and minima i.e. in the center of the orientation scotoma stripes.
With increasing inter-map coupling, this orientation scotoma pattern is suppressed until finally
all amplitudes are zero and only the homogeneous solution is stable.
In case of OP rhombs inter-map coupling makes the rhombic pattern more stripe-like by
reducing the amplitude $A_2=A_{2^-}$. The mode $A_3=A_{3^-}$ which is zero in the uncoupled case
increases and finally collapses with the mode $A_2$.
Increasing inter-map coupling more suppresses all but the two modes $A_1=A_{1^-}$, leading again
to the orientation scotoma stripe pattern.\\
%OD hex
The parameter dependence of OP solutions when interacting with 
OD hexagons is shown in Fig.~\ref{fig:Amplitudes_alpha}\textbf{\sffamily{C,D}}.
OP stripe solutions became above a critical inter-map coupling strength unstable against PWC solutions.
This critical coupling strength strongly depended on the OD bias.
OP rhombic solutions also became unstable against PWC but for a lower coupling strength than
the OP stripes. Thus there is at intermediate coupling strength a bistability between stripe-like
solutions and PWC solutions.
% Potential
The potential of the OP stripe and OP rhombic 
solutions is shown in Fig.~\ref{fig:Amplitudes_alpha}\textbf{\sffamily{E,F}}.
Stripes are energetically preferred in the uncoupled case as well as 
for small inter-map coupling strength for which they are stable.\\
% Summary (three scenarios of pinwheel stabilization)
To summarize, stripe solutions were deformed but no pinwheels were created for this solution.
The rhombic solutions were energetically not preferred for low inter-map coupling whereas for
intermediate inter-map coupling these solutions lose pinwheels and became stripe solutions.
Instead, additional pinwheel rich solutions with a crystal layout became stable for
intermediate inter-map coupling.
For large inter-map coupling orientation selectivity was completely suppressed.
\clearpage
%%%%
\subsubsection{Phase diagram}
\begin{figure}[bt]
\begin{center}
\includegraphics[width=.66\linewidth]{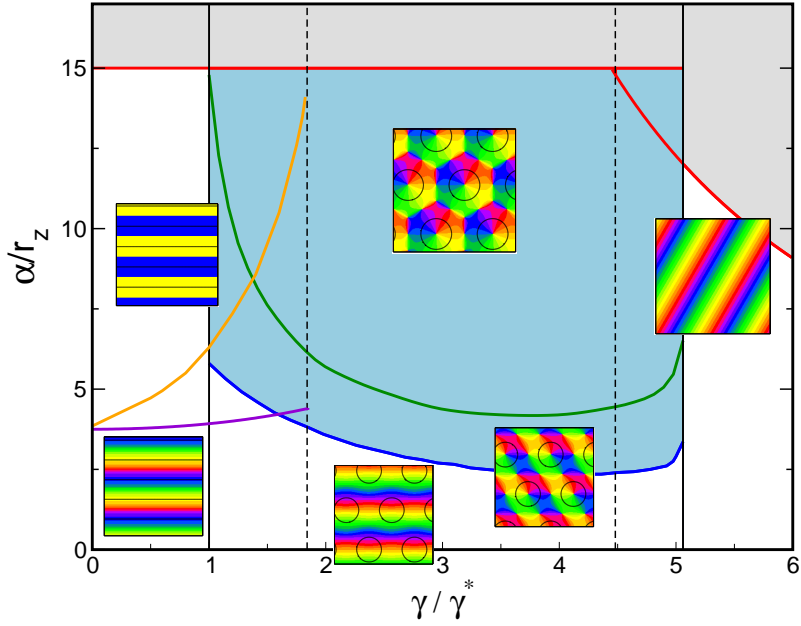} 
\end{center}
\caption{
\textbf{Phase diagram with the coupling energy} $\mathbf{U=\alpha \, o^2|z|^2}$, $r_o=0.2, r_z \ll r_o$.
Vertical lines: stability range of OD stripes, hexagons, and constant solution.
Magenta (orange) line: transition of stripes (rhombs) to the orientation scotoma solution.
Blue line: stability border for the $\Delta=\pi$ uniform solution (hPWC). 
Green line: stability line of stripe-like solutions.
Red line: pattern solutions ceases to exist, see Eq.~(\ref{eq:alpha_crit_stripes}) and Eq.~(\ref{eq:alpha_crit_hex}). 
Blue region: stability region of  hPWC, gray region: No pattern solution exists.
}
\label{fig:PD_alpha}
\end{figure}

The phase diagram as a function of the OD bias $\gamma$ and the inter-map coupling strength $\alpha$ 
for this coupling energy is shown in Fig.~\ref{fig:PD_alpha}.
%@ /home/reichl/UniformSol_AllCouplings/PWstab/Alp/
When rescaling the inter-map coupling strength as $\alpha/r_z$ the phase diagram is independent
of the bifurcation parameter of the OP map $r_z$.
Thus for fixed $r_o\gg r_z$ the phase diagram depends only on two control parameters $\gamma/\gamma^*$
and $\alpha/r_z$. 
% x-axis
The phase diagram contains the stability borders of the uncoupled OD solutions
$\gamma^*,\gamma_2^*,\gamma_3^*,\gamma_4^*$. They correspond to vertical lines,
as they are independent of the inter-map coupling in the limit $r_z \ll r_o$. 
At $\gamma=\gamma^*$ hexagons become stable. Stripe solutions become unstable at $\gamma=\gamma_2^*$.
At $\gamma=\gamma_3^*$ the homogeneous solution becomes stable while at
$\gamma=\gamma_4^*$ hexagons lose their stability. 
In the units $\gamma/\gamma^*$ the borders $\gamma_2^*, \gamma_3^*, \gamma_4^*$ 
vary slightly with $r_o$ , see Fig.~\ref{fig:constTerm}, and are drawn here for $r_o=0.2$.
% Colored lines
Colored lines correspond to the stability and existence borders of OP solutions.
%a) Stripes
In the region of stable OD stripes the OP stripes run parallel to the OD stripes.
With increasing inter-map coupling strength the orientation preference of all but two
orthogonal orientations is suppressed.
%b) Hexagon
In the region of stable OD hexagons stripe-like OP solutions dominate for low inter-map coupling strength.
Above a critical bias dependent coupling strength the $\Delta=\pi$ uniform solution becomes stable (blue line).
There is a region of bistability between stripe-like and uniform solutions until the stripe-like solutions lose 
their stability (orange line).
OP rhombic solutions lose their stability when the uniform solution becomes stable. Thus
there is no bistability between OP rhombs and OP uniform solutions. 
As in the case of OD stripes the uniform solution becomes unstable at $\alpha=r_z/(3\mathcal{B}^2)$.
Also in the case of OD hexagons the inter-map coupling leads to a transition towards the
trivial solution where there is no OP pattern at all.
In case of the OD constant solution the OP map is a stripe solution. 
Pinwheel rich solutions thus occur only in the region of stable OD hexagons. 
In the following we discuss the properties of these solutions. 
%%%%%%
\subsubsection{Interaction induced pinwheel crystals}
\begin{figure}[tb]
\begin{center}
\includegraphics[width=.8\linewidth]{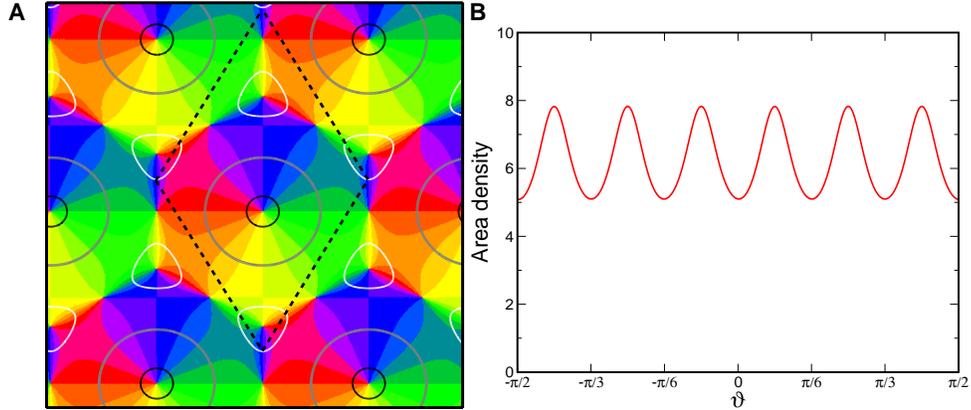}
\end{center}
\caption{\textbf{Ipsi-center pinwheel crystal}.
\textbf{\sffamily{A}} OP map, superimposed are the OD borders (gray),
90\% ipsilateral eye dominance (black), and 90\% contralateral eye dominance (white), 
$r_o=0.2, \gamma=3\gamma^*$.
Dashed lines mark the unit cell of the regular pattern.
\textbf{\sffamily{B}} Distribution of preferred orientations.}
\label{fig:Unif_d=pi_full}
\end{figure}

The uniform solution Eq.~(\ref{eq:UniformSol_part3}) with $\Delta=\pi$ is 
illustrated in Fig.~\ref{fig:Unif_d=pi_full}.
For all stationary solutions
the positions of the pinwheels are fixed by the OD map and there are no translational degrees of freedom. 
The unit cell (dashed line) contains 6 pinwheels which leads to a 
pinwheel density of $\rho=6\, \cos \pi/6 \approx 5.2$.
Two of them are located at OD maxima (contra center) while one is located at an OD minimum (ipsi center).
The remaining three pinwheels are located at OD saddle-points.
Therefore, all pinwheels are located where the gradient of the OD map is zero.
The pinwheel in the center of the OP hexagon is at the ipsilateral OD peak.
Because these pinwheels organize most of the map while the others 
essentially only match one OP hexagon to its neighbors
we refer to this pinwheel crystal as the \textit{Ipsi-center pinwheel crystal}.
The iso-orientation lines intersect the OD borders (gray) exactly with a right angle.
The intersection angles are, within the stability range of OD hexagons, independent of the bias $\gamma$.
The remarkable property of perfect intersection angles cannot be deduced directly from the coupling energy term.
The solution is symmetric under a combined rotation by $60^\circ$ and an orientation 
shift by $-60^\circ$. The symmetry of the pattern is reflected by the distribution of preferred 
orientations, see Fig.~\ref{fig:Unif_d=pi_full}\textbf{\sffamily{B}}.
Although the pattern is selective to all orientations the six 
orientations $\vartheta +n\frac{\pi}{6}$, $n=0,...,5$ are slightly overrepresented. \\
%%% Abschlusssatz:
To summarize, the low order product-type inter-map coupling leads in
case of OD hexagons to a transition from pinwheel free stripe solutions towards
pinwheel crystals. 
The design of the PWC is an example of an orientation hypercolumn dominated by one pinwheel.
With increasing inter-map coupling the PWC solution
is suppressed until only the homogeneous solution is stable. In case of
OD stripes or the constant solution the OP solutions are pinwheel free stripe pattern. 
%%%%%%
%%%%%%%%%%%%%%%%%%%%%%%%%%%%%%%%%%%%%%%%%%%%%%%%%%%%%%%%%%%%%%%%%%%%%%%%%%%%%%%%%%%%%%%%%%%%%%%%%%%%%%%%
%%%%%%%%%%%%%%%%%%%%%%%%%%%%%%%%%%%%%%%%%%%%%%%%%%%%%%%%%%%%%%%
\subsection{Gradient-type energy $U=\beta |\nabla z \dotp \nabla o|^2$}
When using a gradient-type inter-map coupling energy the interaction terms are independent of the 
OD shift $\delta$. In this case, the coupling strength can be rescaled as $\beta \mathcal{B}^2$ and is
therefore independent of the bias $\gamma$. The bias in this case only determines the
stability of OD stripes, hexagons or the constant solution.
%%%%%%%%
\subsubsection{Stationary solutions and their stability}
% a,b) OD stripes
A coupling to OD stripes is easy to analyze in the case of a gradient-type inter-map coupling.
The energetically preferred solutions are OP stripes with the direction perpendicular to the
OD stripes for which $U=0$. This configuration corresponds to the Hubel and Wiesel 
Ice-cube model \cite{HubelWiesel2}. In addition there are rPWC solutions
with the stationary amplitudes $\mathcal{A}_1=\mathcal{A}_{1^-}=\sqrt{(r_z+2\mathcal{B}^2\beta)/5}$,
$\mathcal{A}_2=\mathcal{A}_{2^-}=\sqrt{(r_z-\mathcal{B}^2\beta)/5}$,
$\mathcal{A}_3=\mathcal{A}_{3^-}=0$,
and the stationary phases as in Eq.~(\ref{eq:Phases_alp_ODstripes_OPrhombs}).
Increasing inter-map coupling strength $\beta$ leads to an increase of the 
amplitudes  $\mathcal{A}_1=\mathcal{A}_{1^-}$ while decreasing the amplitudes
$\mathcal{A}_2=\mathcal{A}_{2^-}$ thus making the rhombic solution more stripe-like.\\
%%%%
In the case the OD map is a constant, Eq.~(\ref{eq:ODconst}), the gradient-type inter-map coupling 
leaves the OP dynamics unaffected. 
The stationary states are therefore OP stripes with an arbitrary direction and 
rPWC solutions as in the uncoupled case.\\
In the case of OD hexagons the amplitude equations read
\begin{eqnarray}\label{eq:Amp_bet_ODhex}
 \partial_t \, A_i &=& \left(r_z-3\beta \mathcal{B}^2 \right)A_i
+\frac{5}{4}\beta \mathcal{B}^2 \left(A_{i+1}e^{\imath (\psi_i-\psi_{i+1})}+
A_{i+2}e^{\imath (\psi_i-\psi_{i+2})}\right) \nonumber \\
&& -\beta \mathcal{B}^2 \left( A_{i^-}e^{2\imath \psi_i}+\frac{5}{4}A_{(i+1)^-}
e^{\imath(\psi_i+\psi_{i+1})}+\frac{5}{4}A_{(i+2)^-}
e^{\imath(\psi_i+\psi_{i+2})} \right)+\text{nct.}\, .
\end{eqnarray}
Using $A_i=\mathcal{A}_i e^{\imath \phi_i}$ we obtain the phase equations
\begin{eqnarray}
\mathcal{A}_i \partial_t \, \phi_i &=& 
 \sum_{j\neq i} \mathcal{A}_j \mathcal{A}_{j^-} \mathcal{A}_{i^-}  \sin \left(\phi_i+\phi_{i^-}-\phi_j-\phi_{j^-} \right) \nonumber \\
&&-\mathcal{B}^2 \beta \sum_{j\neq i} \left( \frac{5}{4}\mathcal{A}_j \sin \left(\phi_i-\phi_j-\psi_i+\psi_j \right)
+\frac{5}{4}\mathcal{A}_{j^-} \sin \left(\phi_i-\phi_{j^-}-\psi_i-\psi_j \right) \right) \nonumber \\
&&-\mathcal{B}^2 \beta \mathcal{A}_{i^-}\sin \left(\phi_i-\phi_{i^-}-2\psi_i\right)\, .
\end{eqnarray}
% c,d,e)
These amplitude equations have stripe-like and rhombic solutions with inter-map coupling
dependent phase relations. We therefore calculate their stationary phases and amplitudes numerically
using a Newton method and initial conditions close to these solutions.
Besides stripe-like and rhombic solutions these amplitude equations have uniform solutions.
Again we find that the ansatz Eq.~(\ref{eq:UniformSol_part3}) can satisfy the stationarity condition.
The phase dynamics in this case reads
\begin{equation}
 \partial_t \, \phi_i =-\frac{1}{4}\mathcal{B}^2 \beta \sin \Delta\, .
\end{equation}
As in the case of the product-type inter-map coupling energy
stationary solutions are $\Delta=0$ and $\Delta=\pi$
with the stationary amplitudes
\begin{equation}\label{eq:Unif_d_beta}
 \mathcal{A}_{\Delta=0}=\sqrt{\frac{r_z-3/2 \mathcal{B}^2\beta}{9}}\, , \quad
 \mathcal{A}_{\Delta=\pi}=\sqrt{\frac{r_z-2 \mathcal{B}^2\beta}{9}} \, .
\end{equation}
We studied the stability properties of both stationary solutions by linear stability analysis where
amplitude and phase perturbations in general do not decouple.
The stability matrix of the uniform solutions is given in Text S3. The eigenvalues
are calculated numerically.
It turned out that the $\Delta=\pi$ solution is unstable for $\beta>0$ while 
the $\Delta=0$ solution becomes stable for $\beta \approx 0.05 r_z/\mathcal{B}^2$.
The $\Delta=0$ solution loses its stability above 
\begin{equation}
\beta_c \, \mathcal{B}^2=2 r_z /3 \, .
\end{equation}
From thereon only the homogeneous solution $A_j=0$ is stable.
%%%
\subsubsection{Bifurcation diagram}
\begin{figure}[bt]
\begin{center}
\includegraphics[width=.5\linewidth]{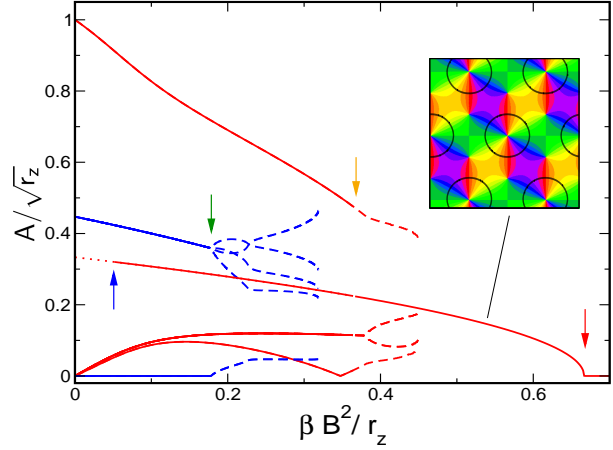} 
\end{center}
\caption{\textbf{Stationary amplitudes with} $\mathbf{U=\beta |\nabla z \dotp \nabla o|^2}$ and OD hexagons.
Solid (dashed) lines: Stable (unstable) solutions to Eq.~(\ref{eq:Amp_bet_ODhex}).
%Dashed line: $\Delta=0$ uniform solution, Eq.~(\ref{eq:Unif_d_beta}).
Transition from OP stripes towards the uniform solution (red),
transition from OP rhombs towards the uniform solution (blue).
Arrows indicate corresponding lines in the phase diagram, Fig.~(\ref{fig:PD_beta}).
}
\label{fig:Amplitudes_beta}
\end{figure}

The course of the stationary amplitudes when interacting with OD hexagons 
is shown in Fig.~\ref{fig:Amplitudes_beta}.
% OP rhombs
The OP rhombic solution is almost unchanged by inter-map coupling but 
above a critical coupling strength the rhombs decay into a stripe-like solution.
% OP stripes
The amplitude of the OP stripe solution is suppressed by inter-map coupling
and finally becomes unstable against the $\Delta=0$ uniform solution.
Thus for large inter-map coupling only the uniform solution is stable.
% d=0 uniform
A further increase in the inter-map coupling suppresses the amplitude of this uniform solution
until finally only the homogeneous solution is stable.
%%%%
\subsubsection{Phase diagram}
\begin{figure}[bt]
\begin{center}
\includegraphics[width=.66\linewidth]{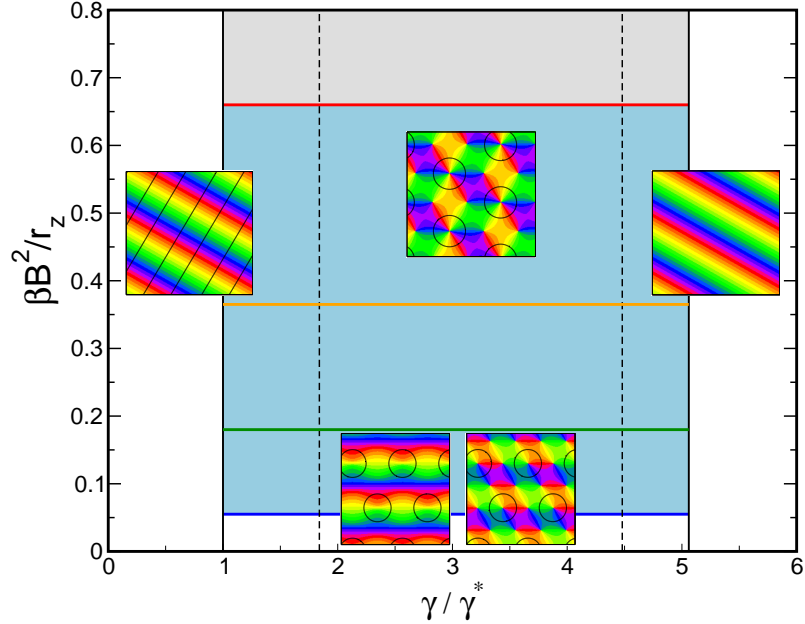} 
\end{center}
\caption{
\textbf{Phase diagram with the coupling energy} $\mathbf{U=\beta |\nabla z \dotp \nabla o|^2}$, $r_z \ll r_o$.
Vertical black lines: stability range of OD stripes, hexagons, and constant solution.
Blue line: stability border for the $\Delta=0$ uniform solution.
Green line: rhombic solutions become unstable.
Orange line: stripe-like solutions become unstable.
Red line: pattern solutions cease to exist, see Eq.~(\ref{eq:Unif_d_beta}).
Gray region: No pattern solution exists.
}
\label{fig:PD_beta}
\end{figure}

The phase diagram of this coupling energy is shown in Fig.~\ref{fig:PD_beta}.
%x-axis
% y-axis
We rescaled the inter-map coupling strength as $\beta \mathcal{B}^2/r_z$, where
$\mathcal{B}$ is the stationary amplitude of the OD hexagons.
The stability borders are then independent of the OD bias in the OD solutions and further independent of the
bifurcation parameter $r_z$. 
This simplifies the analysis since
the OP solutions and their stability depend on $\gamma$ only indirect via the amplitudes $B$.
In case of OD stripes or OD constant solution there is no pinwheel crystallization. Instead
the OP solutions are pinwheel free stripes.
In case of OD hexagons hPWCs become stable above
$\beta \approx 0.05 r_z/\mathcal{B}^2$ (blue line).
Rhombic OP patterns become unstable at $\beta \mathcal{B}^4/r_z \approx 0.17$ and decay 
into a stripe-like solution (green line). At $\beta \mathcal{B}^4/r_z \approx 0.36$ these 
stripe-like solutions become unstable (orange line). 
Thus above $\beta \mathcal{B}^4/r_z \approx 0.36$ the hPWC is the 
only stable solution. At $\beta \mathcal{B}^2/r_z=2/3$ the pattern solution ceases to exist (red line).  
%%%%
\subsubsection{Interaction induced pinwheel crystals}\label{subsec:Braitenberg_part3}
\begin{figure}[bt]
\begin{center}
\includegraphics[width=0.8\linewidth]{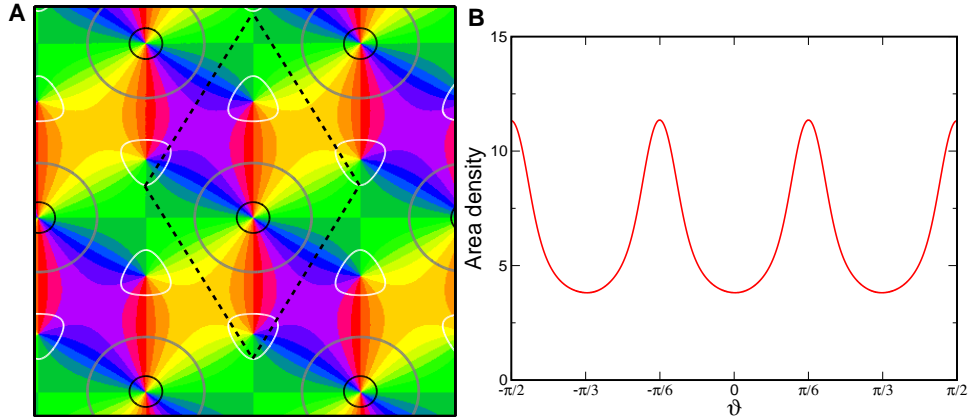} 
\end{center}
\caption{\textbf{The Braitenberg pinwheel crystal}, $\Delta=0$ uniform solution of Eq.~(\ref{eq:UniformSol_part3}).
\textbf{\sffamily{A}} OP map,
superimposed are the OD borders (gray),
90\% ipsilateral eye dominance (black), and 90\% contralateral eye dominance (white), 
$r_o=0.2, \gamma=3\gamma^*$.
Dashed lines mark the unit cell of the regular pattern.
\textbf{\sffamily{B}} Distribution of preferred orientations.
}
\label{fig:Unif_d=0_full}
\end{figure}

The uniform solution Eq.~(\ref{eq:UniformSol_part3}), $\Delta=0$ is illustrated in Fig.~\ref{fig:Unif_d=0_full}.
This PWC contains only three pinwheels per unit cell leading to a pinwheel density
of $\rho=3\cos \pi/6 \approx 2.6$. Two of the three pinwheels are located at maxima of the
OD map (contra peak) while the remaining pinwheel is located at the minimum (ipsi peak) of the OD map.
A remarkable property of this solution is that 
the pinwheel located at the OD minimum, carries a topological charge of 1 such that each 
orientation is represented twice around this pinwheel.
Pinwheels of this kind have not yet been observed in experimentally recorded OP maps. 
This kind of uniform solution corresponds to the structural pinwheel model by Braitenberg \cite{Braitenberg}. 
We therefore refer to this solution as the \textit{Braitenberg pinwheel crystal}.\\
The iso-orientation lines are again perfectly perpendicular to OD borders and this is independent of the 
bias $\gamma$.
The solution is symmetric under a combined rotation by $120^\circ$ and an orientation shift by $-2\pi/3$. 
Further it is symmetric under a rotation by $180^\circ$.
The pattern is selective to all orientations but the distribution of represented orientations is not uniform. 
The three orientations $\vartheta +n\frac{\pi}{3}$, $n=0,...,2$ are 
overrepresented, see Fig.~\ref{fig:Unif_d=0_full}\textbf{\sffamily{B}}. \\
Overall this OP map is dominated by uniform regions around hyperbolic points. In contrast to the ipsi center 
PWC all pinwheels in this OP map organize a roughly similar fraction of the cortical surface.
%%%%%%%%%%%%%%%%%%%%%%%%%%%%%%%%%%%%%%%%%%%%%%%%%%%%%%%%%%%%%%%%%%%%%%%%%%%%%%%%%%%%%%%%

%%%%%%%%%%%%%%%%%%%%%%%%%%%%%%%%%%%%%%%%%%%%%%%%%%%%%%%%%%%%%%%%%%%%%

%%% Summary (all 4  energies)%%%%%%%%%%%%%%%%%%%%%%%%%%%%%%%%%%%%%%%%
%%%%%%%%%%%%%%%%%%%%%%%%%%%%%%%%%%%%%%%%%%%%%%%%%%%%%%%%%%%%%%%%%%%%%%%%%%%
%%%%%%%%%%%%%%%%%%%%%%%%%%%%%%%%%%%%%%%%%%%%%%%%%%%%%%%%%%%%%%%%%%%%%%%%%%%
%%%%%%%
\section*{Discussion}
%%%%%%%%%%%%%%%%%%%%%%%%
%% Summary: our approach, our results
%%%%%%%%%%%%%%%%%%%%%%%%
%%%%%%%%%
\textbf{Summary of results}\\
In this study we presented a symmetry-based analysis of models formalizing that 
visual cortical architecture is shaped by the coordinated
optimization of different functional maps. 
In particular, we focused on the question of whether and how different 
optimization principles specifically impact on the spatial layout of
functional columns in the primary visual cortex.
We identified different representative candidate optimization principles.
We developed a dynamical systems approach for analyzing the simultaneous optimization 
of interacting maps and examined how their layout is 
influenced by coordinated optimization.
In particular, we found that inter-map coupling can stabilize pinwheel-rich layouts even
if pinwheels are intrinsically unstable in the weak coupling limit.
We calculated and analyzed the stability properties of solutions forming 
spatially regular layouts with pinwheels arranged in a crystalline array.
We analyzed the structure of these pinwheel crystals in terms of their stability properties, spatial layout, and 
geometric inter-map relationships. 
For all models, we calculated phase diagrams showing the stability of the pinwheel crystals depending on the 
OD bias and the inter-map coupling strength.
Although differing in detail and exhibiting distinct pinwheel crystal phases for strong coupling,
the phase diagrams exhibited many commonalities in their structure.
These include the general fact that the hexagonal PWC phase is preceded by a phase of
rhombic PWCs and that the range of OD biases over which pinwheel crystallization
occurs is confined to the stability region of OD patch solutions.\\
%%%%%%%%%%%%%%%%%%%
% Relation to previous work
%%%%%%%%%%%%%%%%%%%
\textbf{Comparison to previous work}\\
Our analytical calculations of attractor and ground states 
close a fundamental gap in the theory of visual cortical architecture and its development.
They rigorously establish that models of interacting OP and OD maps in principle offer 
a solution to the problem of pinwheel stability \cite{Wolf3,Wolf2}.
This problem and other aspects of the influence of OD segregation on OP maps have previously been studied 
in a series of models such as elastic 
net models \cite{Wolf3,Hoffsummer,Goodhill,Goodhill2,Sur5,Schulten},
self-organizing map models \cite{Obermayer,Sur2,Swindale1,Swindale5,Bednar}, 
spin-like Hamiltonian models \cite{Cho1,Cho2}, spectral filter models \cite{Grossberg}, correlation based 
models \cite{Tanaka,Miller}, and evolving field models \cite{Pierre}.
% describe in more detail
Several of these simulation studies found a higher number of pinwheels per hypercolumn 
if the OP map is influenced by strong OD segregation compared to the OP layout in isolation 
or the influence of weak OD segregation \cite{Wolf3,Pierre,Cho1}.
In such models, large gradients of OP and OD avoid each other \cite{Swindale1,Obermayer}. 
As a result, pinwheel centers tend to be located at centers of 
OD columns as seen in experiments \cite{Crair2,Bonhoeffer,Loewel,Matsuda,Tanaka}. 
By this mechanism, pinwheels are spatially trapped and pinwheel annihilation can be reduced \cite{Wolf3}.
Moreover, many models appear capable of reproducing realistic 
geometric inter-map relationships such as perpendicular intersection angles between OD borders and
iso-orientation lines \cite{Tanaka,Grossberg,Hoffsummer}.
Tanaka et al. reported from numerical simulations that the relative positioning of orientation pinwheels 
and OD columns was dependent on model parameters \cite{Tanaka}.
Informative as they were, almost all of 
these previous studies entirely relied on simulation 
methodologies that do not easily permit to assess the progress and convergence of solutions.  
Whether the reported patterns were attractors or just snapshots of transient states 
and whether the solutions would further develop towards pinwheel-free solutions 
or other states thus remained unclear.
Moreover, in almost all previous models, a continuous variation of the inter-map
coupling strength was not possible which makes it hard to disentangle the 
contribution of  inter-map interactions from intrinsic mechanisms.
The only prior simulation study of a coordinated optimization model that tracked the number
of pinwheels during the optimization process did not provide evidence that pinwheel
annihilation could be stopped but only reported a modest reduction in annihilation efficiency \cite{Wolf3}. 
From this perspective, the prior evidence for coordination induced pinwheel stabilization
appears relatively limited. 
Our analytical results leave no room to doubt that map interactions can stabilize an intrinsically
unstable pinwheel dynamics. They also reveale that interaction of orientation preference with a
stripe pattern of OD is per se not capable of stabilizing pinwheels.\\ 
%%%%%%%%%%%%%%%%%%%%%%%%%%%
% Theoretical structure of OP - OD models
%%%%%%%%%%%%%%%%%%%%%%%%%%%
\noindent
\textbf{The mathematical structure of interaction models}\\
Independent of its predictions, our study clarifies the general mathematical structure 
of interaction dominated optimization models.
To the best of our knowledge our study for the first time describes an analytical approach 
for examining the solutions of  
coordinated optimization models for OP and OD maps. 
Our symmetry-based phenomenological analysis of conceivable coupling terms provides a general 
classification and parametrization of biologically plausible coupling terms.
To achieve this we mapped the optimization problem to a 
dynamical systems problem which allows for a perturbation expansion of 
fixed points, local minima, and optima.
Using weakly nonlinear analysis, we derived amplitude equations as an 
approximate description near the symmetry breaking transition.
We identified a limit in which inter-map coupling becomes effectively unidirectional 
enabling the use of the uncoupled OD patterns.
We studied fixed points and calculated their stability properties for different types
of inter-map coupling energies.
This analysis revealed a fundamental difference between high and low order coupling energies.
For the low order versions of these energies, a strong inter-map coupling
typically leads to OP map suppression, causing the orientation selectivity of
all neurons to vanish.
In contrast, the higher order variants of the coupling energies do generally not cause map suppression
but only influence pattern selection, see Text S1.
We did not consider an interaction with the retinotopic map.
Experimental results on geometric relationships between the retinotopic map and the OP map
are ambiguous.
In case of ferret visual cortex high gradient regions of both maps avoid each other \cite{Sur5}.
In case of cat, however, high gradient regions overlap \cite{Das2}.
%Compared to other feature maps, the retinotopic map seems to be a special case with respect to 
%its geometric relationships. 
%It has been observed experimentally that the correlations between OP and retinotopic map are such 
%that high gradient regions do not avoid each other \cite{Das2}. 
Such positive correlations cannot be easily treated with dimension reduction models, see \cite{Swindale8}.
It is noteworthy that our phenomenological analysis identified coupling terms that could induce an attraction 
of high gradient regions.
Such terms contain the gradient of only one field and can thus be considered as a mixture of
the gradient and the product-type energy.\\
%
%%%%%%%%%%%%%%%%%%%%%%%%%%%%%%%%%%%%%%%%%%%%%%
%%%%%%%%%%%%%%%%%%%%%%%%%%%%%%%%%%%%%%%%%%%%%%%%%%%%%
% Patchy layout important
%%%%%%%%%%%%%%%%%%%%%%%%%%%%%%%%%%%%%%%%%%%%%%%%%%%%%
\textbf{Conditions for pinwheel stabilization}\\
Our results indicate that a patchy layout of a second visual map interacting with the OP map
is important for the effectiveness of pinwheel stabilization by inter-map coupling.
Such a patchy layout can be easily induced by an asymmetry in the representation of the corresponding stimulus
feature such as eye dominance or spatial frequency preference.
In spatial frequency maps, for instance, low spatial frequency patches tend to form islands 
in a sea of high spatial frequency preference \cite{Bonhoeffer}.
In our model, the patchy layout results from the overall dominance of one eye.
In this case, OD domains form a system of hexagonal patches rather than stripes enabling
the capture and stabilization of pinwheels by inter-map coupling.
The results from all previous models did not support the view that OD 
stripes are capable of stabilizing pinwheels \cite{Cho1,Pierre,Wolf3}. 
Our analysis shows that OD stripes are indeed not able to stabilize pinwheels, 
a result that appears to be independent of the specific type of map interaction. 
%and further independent of 
%a detuning of typical wavelength.
In line with this, several other theoretical studies, using 
numerical simulations \cite{Cho1,Pierre,Wolf3}, indicated that 
more banded OD patterns lead to less pinwheel rich OP maps.
For instance, in simulations using an elastic net model, the average pinwheel density 
of OP maps interacting with a patchy OD layout was reported substantially higher (about 2.5 pinwheels per hypercolumn) 
than for OP maps interacting with a more stripe-like 
OD layout (about 2 pinwheels per hypercolumn) \cite{Wolf3}.\\      
%%%%
\textbf{Experimental evidence for pinwheel stabilization by inter-map coupling}\\
Several lines of biological evidence appear to support the picture of interaction induced pinwheel
stabilization.  
Supporting the notion that pinwheels might be stabilized by the interaction with patchy OD columns,
visual cortex is indeed dominated by one eye in early postnatal development and has a pronounced 
patchy layout of OD domains \cite{Crair,Stryker3,Horton}.
Further support for the potential relevance of this picture comes from experiments in which the OD map
was artificially removed resulting apparently in a significantly smoother OP map \cite{Sur2}.
In this context it is noteworthy that macaque visual cortex appears to exhibit all three fundamental
solutions of our model for OD maps: stripes, hexagons, and a monocular solution, which are
stable depending on the OD bias. 
%All three patterns are found to resemble patterns observed in physiological OD maps.
In the visual cortex of macaque monkeys, all three types of patterns are 
found near the transition to the monocular segment, see \cite{Horton} and Fig.~(\ref{fig:constSolutions}). 
Here, OD domains form bands in the binocular region and 
a system of ipsilateral eye patches at the transition zone to the monocular region where
the contralateral eye gradually becomes more dominant.
If pinwheel stability depends on a geometric coupling to the system of OD columns one predicts
systematic differences in pinwheel density between these three 
zones of macaque primary visual cortex.
Because OD columns in the binocular region of macaque visual cortex are predominantly arranged
in systems of OD stripes our analysis also indicates that pinwheels in these regions are either
stabilized by other patchy columnar systems or intrinsically stable.\\ 
%%%%%%%%%%%%%%%%%%%%%%%%%%%%%%%%
% Structure of the maps with PWC: IS, PW positions
%%%%%%%%%%%%%%%%%%%%%%%%%%%%%%%%
\textbf{The geometry of interaction induced pinwheel crystals}\\
One important general observation from our results is that
map organization was often not inferable by simple
qualitative considerations on the energy functional.
The organization of interaction induced hexagonal pinwheel crystals
reveals that the relation between coupling energy and resulting map structure 
is quite complex and often counter intuitive.
We analyzed the stationary patterns with respect to intersection angles and pinwheel positions.
%IS: 
In all models, intersection angles of iso-orientation lines and OD borders
have a tendency towards perpendicular angles whether the energy term mathematically depends
on this angle, as for the gradient-type energies, or not, as for the product-type energies.
Intersection angle statistics thus are not a very sensitive indicator of the
type of interaction optimized.
Mathematically, these phenomena result from the complex interplay between the single map energies and
the interaction energies.
%PWatOD
In case of the low order gradient-type inter-map coupling energy all pinwheels are
located at OD extrema, as expected from the used coupling energy.
For other analyzed coupling energies, however, the remaining pinwheels are
located either at OD saddle-points (low order product-type energy)
or near OD borders (higher order gradient-type energy), in contrast
to the expection that OD extrema should be energetically preferred.
%Half of the pinwheels in the analyzed hPWC solutions are located at OD extrema, as expected from
%the used coupling energies.
%However, the higher order gradient-type interaction term positioned half of the
%pinwheels close to OD borders although examination of the interaction term per se
%suggests that OD extrema should be energetically preferred.
%Moreover, for some PWCs some pinwheels are located at OD saddle-points.
Remarkably, such correlations, which are expected from the gradient-type coupling energies, occur
also in the case of the product-type energies.
Remarkably, in case of product type energies pinwheels are located at OD saddle-points.
which is not expected per se and presumably result from
the periodic layout of OP and OD maps.
Correlations between pinwheels and OD saddle-points have not yet 
been studied quantitatively in experiments and may thus provide valuable information on the
principles shaping cortical functional architecture.\\
%%%%%%%%%%%%%%%%%%%%%%%%%%%%%
% Questions of optimization approaches in general
%%%%%%%%%%%%%%%%%%%%%%%%%%%%%
\textbf{How informative is map structure?}\\
Our results demonstrate that, 
although distinct types of coupling energies can leave distinguishing signatures in
the structure of maps shaped by interaction (as the OP map in our example), 
drawing precise conclusions about the coordinated optimization principle from observed map
structures is not possible for the analyzed models.  
% 1 Introduction (1): Previous attempts 
In the past numerous studies have attempted to identify signatures of coordinated optimization
in the layout of visual cortical maps and to infer the validity of specific optimization models from
aspects of their coordinated geometry 
\cite{Wolf3,Hoffsummer,Goodhill,Goodhill2,Sur5,Schulten,Obermayer,Sur2,
Swindale1,Swindale5,Bednar,Cho1,Cho2}.
It was, however, never clarified theoretically in which respect and to which degree map
layout and geometrical factors of inter-map relations are informative with respect to an 
underlying optimization principle.
Because our analysis provides complete information of the detailed relation between map geometry
and optimization principle for the different models our results enable to critically assess whether 
different choices of energy functionals specifically impact on the predicted 
map structure and conversely what can be learned about the underlying optimization
principle from observations of map structures.\\
% 2 Introduction (2): What we did
We examined the impact of different interaction energies on the structure
of local minima and ground states of models for the coordinated optimization of a complex and a real
scalar feature map such as OP and OD maps.
The models were constructed such that in the absence of interactions, the maps reorganized
into simple stripe or blob pattern. In particular, the complex scalar map without interactions would form 
a periodic stripe pattern without any phase singularity.
In all models, increasing the strength of interactions could eventually stabilize
qualitatively different more complex and biologically more realistic patterns containing 
pinwheels that can become the energetic
ground states for strong enough inter-map interactions.
The way in which this happens provides fundamental insights into the relationships between
map structure and energy functionals in optimization models for visual cortical functional architecture.\\ 
%  3 Different solutions-> different energies
Our results demonstrate that the structure of maps shaped by inter-map interactions is in principle informative
about the type of coupling energy.
The organization of the complex scalar map that optimizes the 
joined energy functional was in general different 
for all different types of coupling terms examined.
We identified a class of hPWC solutions which become stable for large inter-map coupling.
This class depends on a single parameter which is specific to the used inter-map
coupling energy.
Furthermore, as shown in Text S1, pinwheel positions in rPWCs, tracked while increasing 
inter-map coupling strength, were different
for different coupling terms examined and thus could in principle serve as a trace of the 
underlying optimization principle.
This demonstrates that, although pinwheel stabilization is not restricted to a particular choice 
of the interaction term, each analyzed phase diagram is specific to the used coupling energy.
In particular, in the strong coupling regime substantial information can be obtained from a detailed
inspection of solutions.\\
% 4 Phase diagrams
In the case of the product-type coupling energies, the resulting phase diagrams
are relatively complex as stationary solutions and stability borders depend
on the magnitude of the OD bias.
Here, even quantitative values of model parameters can in principle be constrained by analysis
of the map layout.
In contrast, for the gradient-type coupling energies, the bias dependence can be absorbed into
the coupling strength and only selects the stationary OD pattern. 
This leads to relatively simple phase diagrams. For these models map layout is thus uninformative
of quantitative model parameters.
We identified several biologically very implausible OP patterns.
In the case of the product-type energies, we found orientation scotoma solutions which are selective to
only two preferred orientations.
In the case of the low order gradient-type energy, we found OP patterns containing pinwheels
with a topological charge of 1 which have not yet been observed in experiments.
If the relevant terms in the coupling energy could be determined by other means, the parameter
regions in which these patterns occur could be used to constrain model parameters by theoretical bounds.\\
%6 Information rather qualitative than quantitative
The information provided by map structure overall appears qualitative rather than quantitative.
In both low order inter-map coupling 
energies (and the gradient-type higher order coupling energy, see Text S1), hPWC patterns 
resulting from strong interactions were fixed,
not exhibiting any substantial dependence on the precise choice of interaction coefficient.
In principle, the spatial organization of stimulus preferences in a map is an infinite dimensional
object that could sensitively depend in distinct ways to a large number of model parameters.
It is thus not a trivial property that this structure often gives essentially no information
about the value of coupling constants in our models.
The situation, however, is reversed when considering the structure of rPWCs.
These solutions exist and are stable although energetically not favored in the 
absence of inter-map interactions.
Some of their pinwheel positions continuously depend on the strength of inter-map interactions.
These solutions and their parameter dependence nevertheless are also largely uninformative about the
nature of the interaction energy.
This results from the fact that rPWCs are fundamentally uncoupled system solutions that are only
modified by the inter-map interaction.
As pointed out before, preferentially orthogonal intersection angles between iso-orientation lines and
OD borders appear to be a general feature of coordinated optimization models in the strong
coupling regime. Although the detailed form of the intersection angle histogram is
solution and thus model specific, our analysis does not corroborate attempts to use this
feature to support specific optimization principles, see also \cite{Swindale4,Swindale9,Goodhill4}.
%7 Other coupling energies
The stabilization of pinwheel crystals for strong inter-map coupling appears to be universal and
provides per se no specific information about the underlying optimization principle.
In fact, the general structure of the amplitude equations is universal and only the coupling coefficients
change when changing the coupling energy.
It is thus expected that also for other coupling energies, respecting the proposed set of symmetries, 
PWC solutions can become stable for large enough inter-map coupling.\\
%%%%%%%%%%%%%%%%%%%%%%%%%%%%%%%%%%%%%%%%%%%%%%%%
%%%%%%%%%%%%%%%%%%%
% Higher feature dimensions
%%%%%%%%%%%%%%%%%%%
%\textbf{Additional feature dimensions}\\
%%%%%%%%%%%%%%%%%%%%%%%%%%%%%%%%%%%%%%%%%%%%%%%%
%%%%%%%%%%%%%%%%%%%%%%%%%%%%%%%%%%%%%%%%%%%%%%%%%%%%%%
\textbf{Conclusions}\\
Our analysis conclusively demonstrates that OD segregation can stabilize pinwheels and
induce pinwheel-rich optima in models for the coordinated optimization of OP and OD maps
when pinwheels are intrinsically unstable in the uncoupled dynamics of the OP map.
This allows to systematically assess the possibility that inter-map coupling might be the 
mechanism of pinwheel stabilization in the visual cortex.
The analytical approach developed here is independent of details of specific optimization principles 
and thus allowed 
to systematically analyze how different optimization principles impact on map layout.
Moreover, our analysis clarifies to which extend the
observation of the layout in physiological maps can provide information about optimization principles 
shaping visual cortical organization.\\
% fehlt: However: common design
The common design observed in experimental OP maps \cite{Kaschube6} is, however, not 
reproduced by the optima of
the analyzed optimization principles. 
Whether this is a consequence of the applied weakly nonlinear analysis or of the low number of
optimized feature maps or should be considered a generic feature of coordinated optimization models
will be examined in part (II) of this study \cite{Reichl6}.
In part (II) we complement our analytical studies by numerical simulations of the
full field dynamics. Such simulations allow to study the rearrangement of maps during the optimization
process, to study the timescales on which optimization is expected to take place, and
to lift many of the mathematical assumptions employed by the above analysis.
In particular, we concentrate on the higher order inter-map coupling energies for which
the derived amplitude equations involved several simplifying conditions, see Text S1.
%%%%%%%
% You may title this section "Methods" or "Models". 
% "Models" is not a valid title for PLoS ONE authors. However, PLoS ONE
% authors may use "Analysis" 
\section*{Methods}\label{sec:MM}
%%%%%%%%%%%%%%%%%%%%%%
\subsection*{Intersection angles}\label{sec:Intersect}
We studied the intersection angles between iso-orientation lines and OD borders.
The intersection angle of an OD border with an iso-orientation contour $\alpha(\mathbf{x})$ is
given by 
\begin{equation}\label{eq:IntersectionAngle}
\alpha(\mathbf{x})=\cos^{-1} \left( \frac{\nabla o(\mathbf{x}) \cdot\nabla \vartheta(\mathbf{x})}
{|\nabla o(\mathbf{x})| |\nabla \vartheta(\mathbf{x})|} \right), 
\end{equation}
where $\mathbf{x}$ denotes the position of the OD zero-contour lines. 
A continuous expression for the OP gradient is given by $\nabla \vartheta =\operatorname{Im} \nabla z /z$. 
We calculated the frequency of intersection angles in the range $[0, \pi/2]$.
In this way those parts of the maps are emphasized from which the most significant information about
the intersection angles can be obtained \cite{Loewel}.
These are the regions where the OP gradient is high and thus every intersection angle receives
a statistical weight according to $|\nabla  \vartheta|$. For an alternative method see \cite{Blasdel}.
%%%%%%%%%
\subsection*{The transition from OD stripes to OD blobs}\label{sec:UncupledDyn_part3}
We studied how the emerging OD map depends on the overall eye dominance.
To this end we mapped the uncoupled OD dynamics to a Swift-Hohenberg equation containing 
a quadratic interaction term instead of a constant bias. 
This allowed for the use of weakly nonlinear analysis to derive amplitude
equations as an approximate description of the shifted OD dynamics near the bifurcation point.
We identified the stationary solutions and studied their stability properties.
Finally, we derived expressions for the fraction of contralateral eye dominance for the stable solutions.
%%%%%%%%%%%%%%%%%%%%%%%%%%%%%%%%%%%%%%%%%%%%%%%%%%%%%%%%%%%%%%%%%%%%%%%%%
\subsubsection*{Mapping to a dynamics with a quadratic term}\label{sec:Mapping_part3}
Here we describe how to map the Swift-Hohenberg equation
\begin{equation}\label{eq:Uncoupled_O_dynamics}
\partial_t \, o(\mathbf{x},t) = \hat{L}\, o(\mathbf{x},t) -o(\mathbf{x},t)^3+\gamma \, ,
\end{equation}
to one with a quadratic interaction term. To eliminate the constant term
we shift the field by a constant amount $o(\mathbf{x},t)=\widetilde{o}(\mathbf{x},t)+\delta$.
This changes the linear and nonlinear terms as
\begin{eqnarray}
\hat{L}\,o &\rightarrow& \hat{L} \, \widetilde{o}-\left(k_{c}^4-r_o\right)\delta \nonumber \\
o^3 &\rightarrow& -\widetilde{o}^{\,3}+3\delta\widetilde{o}^{\,2}+3\delta^2\widetilde{o}+\delta^3 \, .
\end{eqnarray}
We define the new parameters $\widetilde{r}_o=r_o-3\delta^2$ and $\widetilde{\gamma}=-3\delta$.
This leads to the new dynamics
\begin{equation}
 \partial_t \, \widetilde{o}=\widetilde{r}_o \,  \widetilde{o}-\left(k_{c}^2+\Delta\right)^2\tilde{o}
+\tilde{\gamma}\widetilde{o}^{\,2}-\widetilde{o}^{\,3}-\delta^3-\left(k_{c,o}^4-r_o\right)\delta+\gamma \, .
\end{equation}
The condition that the constant part is zero is thus given by
\begin{equation}\label{eq:Mapping_deltaeq_part3}
 -\delta^3-\left(k_{c}^4-r_o\right)\delta+\gamma=0\, .
\end{equation}
%For simplicity we set $k_{c}=1$.
For $r_o<1$ the real solution to Eq.~(\ref{eq:Mapping_deltaeq_part3}) is given by 
\begin{equation}\label{eq:delta}
\delta=\frac{2^{1/3}\left(k_c-r_o \right)}{\beta}-\frac{\beta}{3\, 2^{1/3}} \, ,
\end{equation}
with $\beta=\left(-27\gamma+\sqrt{108\left(r_o-k_c\right)^3+729\gamma^2}\right)^{1/3}$.
For small $\gamma$ this formula is approximated as
\begin{equation}
\delta \approx \gamma \frac{1}{k_c^4-r_o}-\gamma^3 \frac{1}{(k_c^4-r_o)^4}+3\gamma^5\frac{1}
{(k_c^4-r_o)^7}+\dots 
\end{equation}
The uncoupled OD dynamics we consider in the following is therefore given by
\begin{equation}\label{eq:OD_uncoupled}
 \partial_t \, \widetilde{o}=\widetilde{r}_o \widetilde{o}-\left(k_{c}^2+\Delta\right)^2\widetilde{o}
+\widetilde{\gamma}\widetilde{o}^{\,2}-\widetilde{o}^{\,3}\, .
\end{equation}
This equation has been extensively studied in pattern formation literature \cite{Soward}. 
%%%%%%%%%%%%%%%%%%%%%%%%%%%%%%%%%%%%%%%%%%%%%%%%%%%%%%%%%%%%%%%%%%%%%%%%%%%%%%%%%%%%%%%%%%%%%%%%%%%%%%%%
\subsubsection*{Amplitude equations for OD patterns}\label{sec:AmplOD}
We studied Eq.~(\ref{eq:OD_uncoupled}) using weakly nonlinear analysis. This method
leads to amplitude equation as an approximate description of the full field 
dynamics Eq.~(\ref{eq:OD_uncoupled}) near the bifurcation point $\widetilde{r}_o=0$.
We summarize the derivation of the amplitude equations for the OD dynamics which is of the form
\begin{equation}\label{eq:ODuncoupled}
\partial_t \, o(\mathbf{x},t) = \hat{L} \, o(\mathbf{x},t)+N_2[o,o] -N_3[o,o,o]\, ,
\end{equation}
with the linear operator $\hat{L}=r_o-\left(k_{c,o}^2+\Delta \right)^2$. 
In this section we use for simplicity the variables $\left(o,r_o,\gamma\right)$
instead of $\left(\tilde{o},\tilde{r}_o,\tilde \gamma\right)$.
The derivation is performed for general quadratic and cubic nonlinearities but are specified later
according to Eq.~(\ref{eq:DynamicsSHcoupled_ref}) as $N_3[o,o,o]=o^3$ and $N_2[o,o]=\gamma \, o^2$.
For the calculations in the following, it is useful to separate $r_o$ from the linear operator
\begin{equation}
\hat{L}=r_o+\hat{L}^0 \, , 
\end{equation}
therefore the largest eigenvalue of $\hat{L}^0$ is zero.
The amplitude of the field $o(\mathbf{x},t)$ is assumed to be small near the onset $r_o=0$ and thus the nonlinearities are small.
We therefore expand both the field $o(\mathbf{x},t)$ and the control parameter $r_o$ in powers 
of a small expansion parameter $\mu$ as
\begin{equation}\label{eq:Expand_o}
o(\mathbf{x},t)=\mu \, o_1(\mathbf{x},t)+\mu^2 o_2(\mathbf{x},t)+\mu^3 o_3(\mathbf{x},t)+\dots \, ,
\end{equation}
and
\begin{equation}\label{eq:Expand_mu}
r_o=\mu r_1 + \mu^2 r_2 + \mu^3 r_3 + \dots 
\end{equation}
%Due to the quadratic interaction term there is no inversion symmetry $o(\mathbf{x},t)\rightarrow -o(\mathbf{x},t)$. 
%Therefore the field $o(\mathbf{x},t)$ and the control parameter $r_o$ are expanded in powers of $\mu=r_o$ as
The dynamics at the critical point $r_o = 0$ becomes arbitrarily slow since
the intrinsic timescale $\tau=r_o^{-1}$ diverges at the critical point. To compensate
we introduce a rescaled time scale $T$ as
\begin{equation}
T=r_o \, t, \quad \partial_t = r_o \, \partial_T \, .
\end{equation}
In order for all terms in Eq.~(\ref{eq:ODuncoupled}) to be of the same order the quadratic interaction
term $N_2$ must be small. We therefore rescale $N_2$ as $\sqrt{r_o} N_2$.
This preserves the nature of the bifurcation compared to the case $N_2=0$.\\
We insert the expansion Eq.~(\ref{eq:Expand_o}) and Eq.~(\ref{eq:Expand_mu}) in 
the dynamics Eq.~(\ref{eq:ODuncoupled}) and get
\begin{eqnarray}\label{eq:Expansion_part3}
0 & = &  \mu  \,  \hat{L}^0 o_1 \nonumber \\
   &  +   & \, \mu^2 \left(-\hat{L}^0 o_2 -r_1 \partial_T o_1+r_1 o_1+\sqrt{\mu r_1+\mu^2 r_2 +\dots}\, N_2[o_1,o_1]\right) \nonumber \\
   &  +   & \, \mu^3   \left(-\hat{L}^0 o_3+r_1 \left(o_2-\partial_T o_2 \right) +r_2 \left(o_1-\partial_T o_1 \right)-N_3[o_1,o_1,o_1] \right)\nonumber \\
 &     \vdots &
\end{eqnarray}
We sort and collect all terms in order of their power in $\mu$.
The equation can be fulfilled for $\mu>0$ only if each of these terms is zero. 
We therefore solve the equation order by order.
In the leading order we get the homogeneous equation 
\begin{equation}\label{eq:LeadingOrder_o}
\hat{L}^0 o_1=0 \, . 
\end{equation}
Thus $o_1$ is an element of the kernel of $\hat{L}^0$. The kernel contains linear
combinations of modes with wavevector $\vec{k}_j$ on the critical circle $|\vec{k}_j|=k_{c,o}$.
At this level any of such wavevectors is possible. We choose
\begin{equation} \label{eq:kernelOD}
o_1=\sum_{j}^{n} B_j(T)e^{i \vec{k}_j \cdot \vec{x}}
+\sum_{j}^{n} \overline{B}_{j}(T)e^{-i \vec{k}_j \cdot \vec{x}} \, ,
\end{equation} 
where the wavevectors are chosen to be equally spaced 
$\vec{k}_j=k_{c,o} \left(\cos (j \pi/n), \sin (j \pi/n) \right)$ and the complex amplitudes 
$B_j=\mathcal{B}_j e^{\imath \psi_j}$.
%We do not introduce additional space variables $B(X,Y,T)$, needed when considering
%large-scale pattern modulations.\\
The homogeneous equation leaves the amplitudes $B_j$ undetermined. These amplitudes
are fixed by the higher order equations. 
Besides the leading order homogeneous equation we get inhomogeneous equations of the form
\begin{equation}\label{eq:Inhom_o_part3}
\hat{L}^0 o_m = F_m  
\end{equation}
To solve this inhomogeneous equation we first apply a solvability condition.
We thus apply the \textit{Fredholm Alternative theorem} to Eq.~(\ref{eq:Inhom_o_part3}).
Since the operator $\hat{L}^0$ is self-adjoint $\hat{L}^0=\hat{L}^{0 \, \dagger}$, the equation is 
solvable if and only if $F_m$ is orthogonal
to the kernel of $\hat{L}^0$ i.e.
\begin{equation}\label{eq:SolveCond_o}
\langle F_m, \tilde{o} \rangle=0, \quad \forall \,  \hat{L}^0 \tilde{o}=0 
\end{equation}
The orthogonality to the kernel can be expressed by a projection operator $\hat{P}_c$
onto the kernel and the condition $\langle F,\tilde{o} \rangle=0$ can
be rewritten as $\hat{P}_c F=0$.\\
At second order we get
\begin{equation}
\hat{L}^0 o_2=r_1 \left(o_1 -\partial_T o_1 \right) \, .
\end{equation}
Applying the solvability condition Eq.~(\ref{eq:SolveCond_o}) we see that this 
equation can be fulfilled only for $r_1=0$.
At third order we get
\begin{equation}
\hat{L}^0 o_3=r_2 \left( o_1 -\partial_T o_1\right) + N_2[o_1,o_1] -N_3[o_1,o_1,o_1] \, .
\end{equation}
The parameter $r_2$ sets the scale in which $o_1$ is measured and we can set $r_2=1$.
We apply the solvability condition and get 
\begin{equation}
\partial_T \, o_1 = o_1 + \hat{P}_c N_2[o_1,o_1]-\hat{P}_c N_3[o_1,o_1,o_1] \, .
\end{equation}
We insert our ansatz Eq.~(\ref{eq:kernelOD})
which leads to the amplitude equations at third order
\begin{equation}
 \partial_T B_i=B_i+\hat{P}_i  \sum_{j,k} B_j B_k e^{-\imath \vec{k}_i \vec{x}} N_2[e^{\imath 
\vec{k}_j \vec{x}},e^{\imath \vec{k}_k \vec{x}}] 
-\hat{P}_i \sum_{j,k} B_j B_k  B_l e^{-\imath \vec{k}_i \vec{x}} N_3[e^{\imath 
\vec{k}_j \vec{x}},e^{\imath \vec{k}_k \vec{x}},e^{\imath \vec{k}_l \vec{x}} ] \, ,
\end{equation}
where $\hat{P}_i$ is the projection operator onto the subspace 
$\lbrace e^{\imath \vec{k}_i \vec{x}} \rbrace$
of the kernel. 
$\hat{P}_i$ picks out all combinations of the modes which have their
wavevector equal to $\vec{k}_i$.
In our case the three active modes form a so called triad resonance
$\vec{k}_1+\vec{k}_2+\vec{k}_3=0$. The quadratic coupling terms which
are resonant to the mode $B_1$ are therefore given by
\begin{equation}
 \overline{B}_2 \overline{B}_3  e^{-\imath \vec{k}_1 \vec{x}}\left(
N_2[e^{-\imath \vec{k}_2 \vec{x}},e^{-\imath \vec{k}_3 \vec{x}}]+
N_2[e^{-\imath \vec{k}_3 \vec{x}},e^{-\imath \vec{k}_2 \vec{x}}] \right)\, .
\end{equation}
Resonant contributions from the cubic nonlinearity result from terms 
of the form $|B_j|^2B_i$. Their coupling coefficients are given by
\begin{eqnarray}\label{eq:CouplCoeff_tildegij}
\widetilde{g}_{ij}&=&N_3[e^{\imath \vec{k}_i \vec{x}},e^{\imath \vec{k}_j \vec{x}},e^{-\imath \vec{k}_j \vec{x}}]+
N_3[e^{\imath \vec{k}_i \vec{x}},e^{-\imath \vec{k}_j \vec{x}},e^{\imath \vec{k}_j \vec{x}}]+
N_3[e^{\imath \vec{k}_j \vec{x}},e^{\imath \vec{k}_i \vec{x}},e^{-\imath \vec{k}_j \vec{x}}]+\nonumber \\
&&N_3[e^{-\imath \vec{k}_j \vec{x}},e^{\imath \vec{k}_i \vec{x}},e^{\imath \vec{k}_j \vec{x}}]+
N_3[e^{\imath \vec{k}_j \vec{x}},e^{-\imath \vec{k}_j \vec{x}},e^{\imath \vec{k}_i \vec{x}}]+
N_3[e^{-\imath \vec{k}_j \vec{x}},e^{\imath \vec{k}_j \vec{x}},e^{\imath \vec{k}_i \vec{x}}]\, ,
\end{eqnarray}
and
\begin{equation}
 \widetilde{g}_{ii}=N_3[e^{\imath \vec{k}_i \vec{x}},e^{\imath \vec{k}_i \vec{x}},e^{-\imath \vec{k}_i \vec{x}}]+
N_3[e^{\imath \vec{k}_i \vec{x}},e^{-\imath \vec{k}_i \vec{x}},e^{\imath \vec{k}_i \vec{x}}]+
N_3[e^{-\imath \vec{k}_i \vec{x}},e^{\imath \vec{k}_i \vec{x}},e^{\imath \vec{k}_i \vec{x}}]\, .
\end{equation}
When specifying the nonlinearities Eq.~(\ref{eq:DynamicsSHcoupled_ref}) the coupling coefficients 
are given by $\widetilde{g}_{ij}=6, \widetilde{g}_{ii}=3$.
Finally, the amplitude equations (here in the shifted variables ($\tilde{r}_o,\tilde{\gamma}$) 
are given by
\begin{equation}\label{eq:Ampl_OD_uncoupled}
\partial_t B_1 = \widetilde{r}_o B_1-3|B_1|^2B_1-6\left(|B_2|^2+|B_3|^2\right)B_1
+2\widetilde{\gamma} \overline{B}_2\overline{B}_3 \, ,
\end{equation}
where we scaled back to the original time variable $t$.
Equations for $B_2$ and $B_3$ are given by cyclic permutation of the indices.
%%%%%%%%%%%%%%%%%%%%%%%%%%%%%%%%%%%%%%%%%%%%%%%%%%%%%%%%%%%%%%%%%%%%%%%%%
%%%%%%%%%%%%
\subsubsection*{Stationary solutions}
\begin{figure}[t]
\begin{center}
\includegraphics[width=\linewidth]{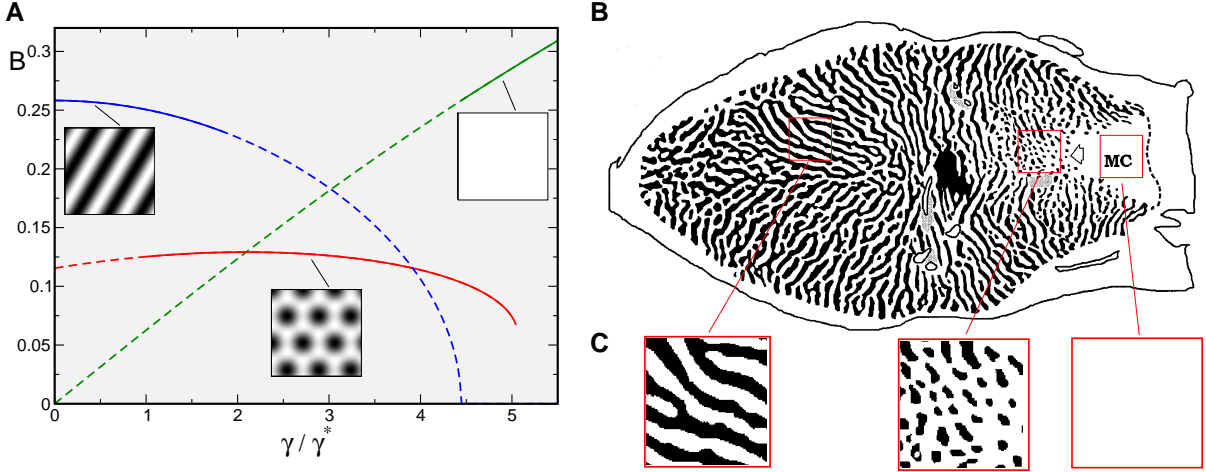}
\end{center}
\caption{
\textbf{OD patterns.}
\textbf{\sffamily{A}} Stationary amplitudes of the OD dynamics.
The course of $\mathcal{B}_{st}(\gamma)$ Eq.~(\ref{eq:stripes}) (blue), 
$\mathcal{B}_{hex}(\gamma)$ Eq.~(\ref{eq:hex}) (red), and of $\delta(\gamma)$ Eq.~(\ref{eq:delta}) (green) 
for $r_o=0.2$. 
The solutions are plotted in solid lines within their stability ranges.
\textbf{\sffamily{B}} OD map of macaque monkey. Adapted from \cite{Horton}.
\textbf{\sffamily{C}} Details of \textbf{\sffamily{B}} with stripe-like, patchy, and homogeneous layout.
}
\label{fig:constSolutions}
\end{figure}

The amplitude equations (\ref{eq:Ampl_OD_uncoupled}) have three types of 
stationary solutions, namely
OD stripes 
\begin{equation}\label{eq:stripes}
o_{st}(\mathbf{x})=2\mathcal{B}_{st}\cos \left(x+\psi \right) +\delta, 
\end{equation}
with $\mathcal{B}_{st}=\sqrt{\tilde{r}/3}$,
hexagons
\begin{equation}\label{eq:hex}
o_{hex}(\mathbf{x})=\mathcal{B}_{hex}\sum_{j=1}^3 e^{\imath \psi_j}e^{\imath \vec{k}_j \cdot \vec{x}}+c.c. +\delta, 
\end{equation}
with the resonance condition $\sum_j^3 \vec{k}_j=0$
and $\mathcal{B}_{hex}=-\tilde{\gamma}/15+\sqrt{\left(\tilde{\gamma}/15\right)^2+\tilde{r}/15}$.
Finally, there is a homogeneous solution with spatially constant eye dominance 
\begin{equation}\label{eq:ODconst}
o_c(\mathbf{x})=\delta.
\end{equation}
The spatial average of all solutions is $\langle o(\mathbf{x}) \rangle =\delta$.
The course of $\mathcal{B}_{st}$, $\mathcal{B}_{hex}$, and of $\delta(\gamma)$ is 
shown in Fig.~\ref{fig:constSolutions}.%\textbf{\sffamily{A}}.
%\begin{figure}[t]
%\begin{center}
%\includegraphics[width=0.45\linewidth]{pics/ConstSolutions/ConstSolutions_all2}
%\end{center}
%\caption{
%%\textbf{OD patterns.}
%\textbf{ Stationary amplitudes of the OD dynamics.}
%The course of $\mathcal{B}_{st}(\gamma)$ Eq.~(\ref{eq:stripes}) (blue), 
%$\mathcal{B}_{hex}(\gamma)$ Eq.~(\ref{eq:hex}) (red), and of $\delta(\gamma)$ Eq.~(\ref{eq:delta}) (green) 
%for $r_o=0.2$. 
%The solutions are plotted in solid lines within their stability ranges.
%}
%\label{fig:constSolutions}
%\end{figure}
%%%%%%%%%%%%%%%%%%%%%%%%%%%%%%%%%%%%%%%%%%%%%%%%%%%%%%%%%%
\subsubsection*{Linear stability analysis for OD patterns}
\begin{figure}[t]
\includegraphics[width=\linewidth]{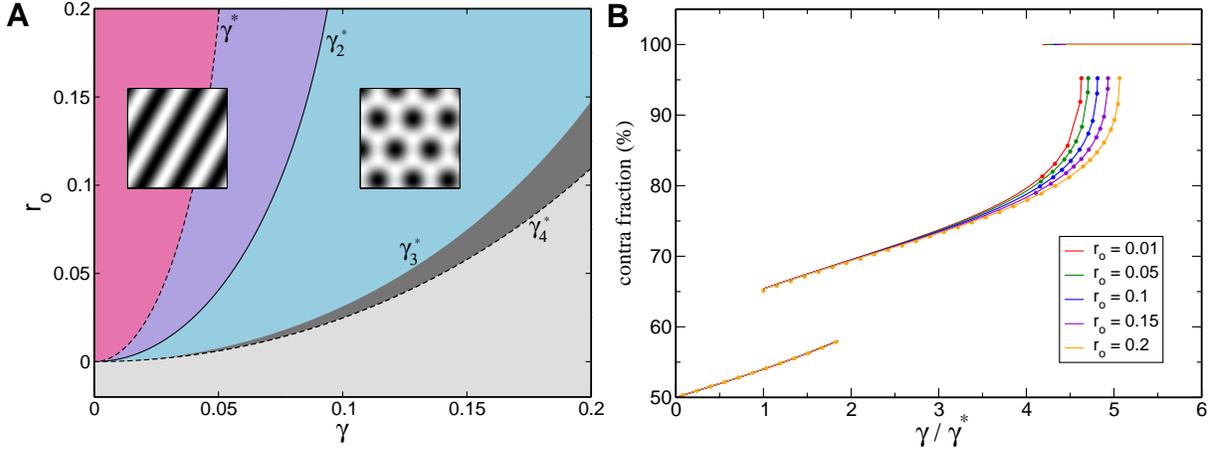}
\caption{\textbf{The uncoupled OD dynamics}.
\textbf{\sffamily{A}} Phase diagram of the OD model Eq.~(\ref{eq:Uncoupled_O_dynamics}). Dashed lines:
stability border of hexagon solutions, 
%\mbox{$\gamma^*=\left(51-50\, r_o\right)\sqrt{r_o}/51^{3/2}, \gamma^*_4=\left(12-7\, r_o\right)\sqrt{r_o}\sqrt{5}/(24\sqrt{3}).$  }
solid line: stability border of stripe solution, % $\gamma^*_2=\left(15-14\,r_o\right)\sqrt{r_o}/15^{3/2}$,
gray regions: stability region of constant solution %$\gamma^*_3=\left(3-2\, r_o\right)\sqrt{r_o}/3^{3/2} $.
\textbf{\sffamily{B}} Percentage of neurons dominated by the
contralateral eye plotted for the three stationary solutions. 
Circles: numerically obtained values, solid lines: $C_{st}$ and $C_{hex}$. 
}
\label{fig:constTerm}
\end{figure}

We decomposed the amplitude equations (\ref{eq:Ampl_OD_uncoupled}) into the
real and imaginary parts. From the imaginary part we get the phase
equation
\begin{equation}
\partial_t \psi_1 = -2\widetilde{\gamma} \sin \left(\psi_1+\psi_2+\psi_3 \right) \, ,
\end{equation}
and equations for $\psi_2,\psi_3$ by cyclic permutation of the indices.
The stationary phases are given by $\psi_1+\psi_2+\psi_3=\{ 0,\pi \}$.
The phase equation can be derived from the potential 
$V[\psi]=-2\widetilde{\gamma} \cos (\psi_1+\psi_2+\psi_3)$.
We see that the solution $\psi_1+\psi_2+\psi_3=0$ is stable for $\widetilde{\gamma}>0 (\gamma<0)$
and the solution $\psi_1+\psi_2+\psi_3=\pi$ is stable for $\widetilde{\gamma}<0 (\gamma>0)$.\\
We calculate the stability borders of the OD stripe, hexagon, and constant solution 
in the uncoupled case. This treatment follows \cite{Soward}. 
In case of stripes the three modes of the amplitude equations are perturbed as 
\begin{equation}
B_1  \rightarrow  B_{st}+b_1\, , \quad 
B_2  \rightarrow  b_2 \, , \quad 
B_3  \rightarrow  b_3, 
\end{equation}
assuming small perturbations $b_1,b_2$, and $b_3$.
This leads to the linear equations $\partial_t \vec{b}=M \vec{b}$ with the stability matrix
\begin{equation}
 M=\left(
  \begin{array}{ c c c}
     \widetilde{r}-9B_{st}^2 & 0 & 0\\
     0 & \widetilde{r}-6B_{st}^2 & 2\widetilde{\gamma}B_{st}\\
    0 & 2\widetilde{\gamma} B_{st} & \widetilde{r}-6B_{st}^2
  \end{array} \right)\, .
\end{equation}
The corresponding eigenvalues are given by
\begin{equation}
\lambda=\left(-2\widetilde{r},-\widetilde{r}-2\sqrt{\widetilde{r}/3}\widetilde{\gamma}, 
-\widetilde{r}+2\sqrt{\widetilde{r}/3}\widetilde{\gamma} \right) \, .
\end{equation}
This leads to the two borders for the stripe stability
\begin{equation}
\widetilde{r}=0, \quad \widetilde{r}=\frac{4}{3}\widetilde{\gamma}^2 \, . 
\end{equation}
In terms of the original variables $r_o,\gamma$ the borders are
given by ($0<r_o<1$)
\begin{equation}
\gamma_3^*=\frac{\left(3-2r_o\right)\sqrt{r_o}}{3^{3/2}}  ,\quad \gamma_2^*=\frac{\left(15-14r_o\right)\sqrt{r_o}}{15^{3/2}} \, .
\end{equation}
To derive the stability borders for the hexagon solution $o_{hex}(\mathbf{x})$ we perturb the amplitudes as
\begin{equation}
B_1  \rightarrow  B_{hex}+b_1 \, , \quad
B_2  \rightarrow  B_{hex}+b_2 \, , \quad
B_3  \rightarrow  B_{hex}+b_3 \, .
\end{equation}
The stability matrix is then given by
\begin{equation}
 M=\left(
  \begin{array}{ c c c}
     -21B_{hex}+\widetilde{r} & -12B_{hex}^2-2B_{hex}\widetilde{\gamma} & -12B_{hex}^2-2B_{hex}
\widetilde{\gamma}\\
    -12B_{hex}^2-2B_{hex}\widetilde{\gamma}  & -21B_{hex}+\widetilde{r} & -12B_{hex}^2-2B_{hex}
\widetilde{\gamma} \\
    -12B_{hex}^2-2B_{hex}\widetilde{\gamma} & -12B_{hex}^2-2B_{hex}\widetilde{\gamma} & -21B_{hex}
+\widetilde{r}
  \end{array} \right)\, ,
\end{equation}
and the corresponding eigenvalues are given by
\begin{equation}
\lambda=\left(-45B_h^2+\widetilde{r}-4B_h\widetilde{\gamma},-9B_h^2+\widetilde{r}+2B_h\widetilde{\gamma},
-9B_h^2+\widetilde{r}+2B_h\widetilde{\gamma}\right) \, .
\end{equation}
The stability borders for the hexagon solution are given by
\begin{equation}
\widetilde{r}=-\frac{1}{15}\widetilde{\gamma}^2  , \quad \widetilde{r}=\frac{16}{3}\widetilde{\gamma}^2 \, . 
\end{equation}
In terms of the original variables we finally get
\begin{equation}
\gamma_4^*=\frac{\left(12-7r_o\right)\sqrt{r_o}\sqrt{5}}{24\sqrt{3}}  , \quad \gamma^*=\frac{\left(51-50r_o\right)\sqrt{r_o}}{51^{3/2}} \, .
\end{equation}
The phase diagram of this model is depicted in Fig.~\ref{fig:constTerm}\textbf{\sffamily{A}}.
It shows the stability borders $\gamma^*,\gamma_2^*,\gamma_3^*$, and $\gamma_4^*$ for the three 
solutions obtained by linear stability analysis.
Without a bias term the OD map is either constant, for $r_o<0$, or has a stripe layout, for $r_o>0$.
For positive $r_o$ and increasing bias term there are two transition regions, first a transition
region from stripes to hexagons and second a transition region from hexagons to the constant solution.\\
The spatial layout of the OD hexagons consists of
hexagonal arrays of ipsilateral eye dominance blobs in a sea of contralateral eye 
dominance, see Fig.~\ref{fig:constTerm}\textbf{\sffamily{A}}.
%%%%%%%%%%%%%%%%%%%%%%%
\subsubsection*{Contralateral eye fraction}\label{sec:Contrafraction_part3}
To compare the obtained solutions with physiological OD maps we 
quantified the fraction of neurons selective to the contralateral eye inputs.
For stripe and hexagon solutions 
we thus calculated the fraction of contralateral eye dominated 
territory $C_{st}$ and $C_{hex}$. In case of stripes this is a purely one-dimensional problem. The zeros of the field
are given by
\begin{equation}
o_{st}(x)=2\mathcal{B}_{st}\cos \left(x+\psi \right) +\delta =0 \, ,
\end{equation}
with the solution
\begin{equation}
x=\arccos \left(\frac{-\delta}{2\mathcal{B}_{st}}\right)\, .
\end{equation}
As the field has a periodicity of $\pi$ the area fraction is given by
\begin{equation}
C_{st}=\arccos \left(\frac{-\delta}{2\mathcal{B}_{st}}\right) /\pi \, .
\end{equation}
In case of hexagons we observe that the territory of negative $o(\mathbf{x})$ values is
approximately a circular area. We obtain the fraction of negative $o(\mathbf{x})$ values
by relating this area to the area of the whole hexagonal lattice. 
In case of hexagons the field is given by
\begin{equation}
o_{hex}(\mathbf{x})=2\mathcal{B}_{hex}\sum_j \cos \left(\vec{k}_j \vec{x}+\psi_j \right)+\delta \, .
\end{equation}
As an approximation we project the field onto the $x$-axis and choose for simplicity $\psi_j=0, \, \forall j$. 
The field has its maximum at the origin $o_{hex}(0,0)=6+\delta$.
The projection leads to 
\begin{equation}\label{eq:zerosofhex}
f(x)=2\mathcal{B}_{hex}\left(\cos x +2\cos (1/2) x \right)+\delta \, .
\end{equation}
The zeros $f(x_1)=0$ are located at
\begin{equation}
x_1= 2 \arccos \left(\frac{1}{2} \left(-1+\sqrt{3+\frac{\delta}{\mathcal{B}_{hex}}}\right)\right)\, .
\end{equation}
The circular area of positive $o_{hex}(\mathbf{x})$ values is now given by $A_c=\pi x_1^2$.
The periodicity of the hexagonal pattern is given by $f(x_2)_{\delta=0}=\min (f)_{\delta=0}=-3$. This
leads to $x_2=4\pi/3$. The area of the hexagon is therefore given by
$A_{hex}=3 x_2^2\sqrt{3}/2$. The contra fraction is finally given by
\begin{equation}\label{eq:Chex}
1-C_{hex}\approx \frac{A_c}{A_{hex}}=\frac{\sqrt{3}}{2\pi} \arccos \left(\frac{1}{2} \left(-1+\sqrt{3+\frac{\delta}{\mathcal{B}_{hex}}}\right)\right)^2 \, .  
\end{equation}
The course of the fractions $C_{st}$ and $C_{hex}$ is shown in Fig.~\ref{fig:constTerm}\textbf{\sffamily{B}}.
At the border $\gamma=\gamma^*$, where hexagons become stable $C_{hex}\approx 65.4\%$.
At the border $\gamma=\gamma_4^*$, where hexagons loose stability $C_{hex}\approx 95.2\%$.
Both quantities are independent of $r_o$.
We confirmed our results by direct numerical calculation of the fraction of positive $o_{hex}(\mathbf{x})$ pixel 
values.
Deviations from the result Eq.~(\ref{eq:Chex}) are small. For $\gamma/\gamma^* \approx 1$ the zeros
of Eq.~(\ref{eq:zerosofhex}) are not that well approximated with a circular shape and the projection described
above leads to the small deviations which decrease with increasing bias $\gamma$.
%%%%%%%%%%%%%%%%%%%%%%%%%%%%%%%%%%%%%%%%%%%%%%%%%%%%%%%%%%%%%%%%%%%%%%%
%%%%%%%%%%%%%%%%%%%%%%%%%%%%%%%%%%%%%%%%%%%%%%%%%%%%%%%%%%%%%%%%%%%%%%%
%%%%%%%%%%%%%%%%%%%%%%%%%%%%%%%%
%%%%
%%%%%%%%%%%%%%%%%%%%%%%%%%%%%%%%%%%%%%%%%%%%%%%%%%%%%%
\newpage
\begin{flushleft}
{\Large
\textbf{Text S1: Supporting Information for\\
Coordinated optimization of visual cortical maps\\
(I) Symmetry-based analysis}
}
\end{flushleft}
%\section*{Text S12}
\section{Introduction}
In case of the low order inter-map coupling energies strong inter-map coupling
leads to a suppression of OP selectivity.
This suppressive effect can be avoided by restricting coupling strengths.
One aim of this article is to test different optimization principles and potentially rule out
some optimization principles.
When comparing our results from different optimization principles to biological 
data such parameter tuning reduces the practicability. 
In this supporting information we complement our study using the high order
inter-map coupling energies.
We show that in this case a suppression of OP selectivity cannot occur.
We derive coupled amplitude equations which, however, involve several
mathematical assumptions.
A systematic treatment as it is shown in the main article would imply that low order and
higher order inter-map coupling energies are in general non-zero.
Low order energy terms would enter at third order in the expansion and higher
order corrections could potentially alter the stability properties.
In addition, higher order inter-map coupling energies can affect the stability of patterns.
In the following, we assume all low order inter-map coupling energies to be zero
and that contributions entering the amplitude equations at higher orders can be neglected.
The obtained results are confirmed numerically in part (II) of this study.
%%%%%%%%%%%%%%%%%%%%%%%%%%%%%%%%%%%
\section{Coupled amplitude equations: Higher order terms}\label{sec:CoupledAmplitudeEquations_part3}
We studied the coupled Swift-Hohenberg equations
\begin{eqnarray}\label{eq:fieldEq_forAmplDerivS1}
 \partial_t \, z(\mathbf{x},t) &=& r_z z(\mathbf{x},t) -\hat{L}_z^0\, z(\mathbf{x},t) -N_{3,u}[z,z,\overline{z}] -N_{7,c}[z,z,\overline{z},o,o,o,o] \nonumber \\
 \partial_t \, o(\mathbf{x},t) &=& r_o o(\mathbf{x},t) -\hat{L}_o^0 \, o(\mathbf{x},t) +N_{2,u}[o,o]-N_{3,u}[o,o,o] -\tilde{N}_{7,c}[o,o,o,z,z,\overline{z},\overline{z}]\, ,
\end{eqnarray}
with the higher order inter-map coupling energies
\begin{equation}\label{eq:Energy_HO}
U=\tau \, o^4 |z|^4 +\epsilon\, |\nabla z \dotp \nabla o|^4 \, ,
\end{equation}
using weakly nonlinear analysis.
We study Eq.~(\ref{eq:fieldEq_forAmplDerivS1}) close to the pattern forming bifurcation where
$r_z$ and $r_o$ are small. 
We therefore expand both control parameters in powers of the small expansion parameter $\mu$
\begin{eqnarray}
 r_z &=&\mu r_{z1}+\mu^2 r_{z2} +\mu^3 r_{z3}+\dots \nonumber \\
 r_o &=& \mu r_{o1} +\mu^2 r_{o2} +\mu^3 r_{o3}+\dots  \, .
\end{eqnarray}
Close to the bifurcation the fileds are small and thus nonlinearities are weak.
We therefore expand both fields as
\begin{align}
o(\mathbf{x},t)&=\mu o_1(\mathbf{x},t)+\mu^2 o_2(\mathbf{x},t)+\mu^3 o_3(\mathbf{x},t)+\dots \nonumber \\
z(\mathbf{x},t)&=\mu z_1(\mathbf{x},t)+\mu^2 z_2(\mathbf{x},t)+\mu^3 z_3(\mathbf{x},t)+\dots
\end{align}
We further introduced a common slow timescale $T=r_z t$ and insert the expansions 
in Eq.~(\ref{eq:fieldEq_forAmplDerivS1}) and get
\begin{eqnarray}\label{eq:Expansion_z_coupled}
 0&=&\mu \hat{L}^0 z_1 \nonumber \\
&&+\mu^2 \left(-\hat{L}^0z_2+r_{z1} z_1-r_{z1} \partial_T z_1 \right)\nonumber\\
&&+\mu^3 \left(-r_{z2} \partial_Tz_1+r_{z2} z_1+r_{z1} z_2-r_{z1} \partial_T z_2-\hat{L}^0z_3-N_{3,u}[z_1,z_1,\overline{z}_1]\right) \nonumber \\
&&\vdots \nonumber \\
&&+\mu^7 \left( -\hat{L}^0 z_7 +r_{z2} z_5 +r_{z4}z3+r_{z6}z_1 +\dots +N_{3,u}[z_5,z_1,\overline{z}_1] \right) \nonumber\\
&&+\mu^7 \left(- N_{7,c}[z_1,z_1,\overline{z}_1,o_1,o_1,o_1,o_1]\right) \nonumber \\
&& \vdots
\end{eqnarray}
%\clearpage{} 
and
\begin{eqnarray}\label{eq:Expansion_o_coupled}
 0&=&\mu  \hat{L}^0 o_1 \nonumber \\
&&+\mu^2\left(-\hat{L}^0 o_2+r_{o1} o_1- r_{z1} \partial_T o_1 +\sqrt{\mu r_{o1}+\mu^2 r_{o2}+\dots}\tilde{N}_{2,u}[o_1,o_1] \right)\nonumber\\
&&+\mu^3\left(- r_{z2} \partial_T o_1+r_{o2} o_1+r_{o1} o_2- r_{z1} \partial_T o_2-\hat{L}^0o_3-\tilde{N}_{3,u}[o_1,o_1,o_1]\right) \nonumber \\
&&\vdots \nonumber \\
&&+\mu^7 \left(-\hat{L}^0 o_7+r_{o2} o_5+r_{o4} o_3+r_{o6} o_1+\dots -\tilde{N}_{3,u}[o_5,o_1,o_1]-\tilde{N}_{2,u}[o_1,o_5]-\dots \right)\nonumber \\
&&+\mu^7\left(- \tilde{N}_{7,c}[o_1,o_1,o_1,z_1,z_1,\overline{z}_1,\overline{z}_1]\right) \nonumber \\
&& \vdots
\end{eqnarray}
We consider amplitude equations up to seventh order as this is the order where the 
nonlinearity of the higher order coupling energy enters first.
For Eq.~(\ref{eq:Expansion_z_coupled}) and Eq.~(\ref{eq:Expansion_o_coupled}) to be fulfilled
each individual order in $\mu$ has to be zero. 
At linear order in $\mu$ we get the two homogeneous equations
\begin{equation}
 \hat{L}^0_z z_1=0\, , \quad  \hat{L}^0_o o_1=0 \, .
\end{equation}
Thus $z_1$ and $o_1$ are elements of the kernel of $\hat{L}_z^0$ and $\hat{L}_o^0$.
Both kernels contain linear
combinations of modes with a wavevector on the critical circle i.e.
\begin{eqnarray}\label{eq_z1o1}
z_1(\mathbf{x},T)&=&\sum\limits_j^{n} \left( A_j^{(1)}(T)e^{\imath \vec{k}_j \cdot \vec{x}} +
 A_{j^-}^{(1)}(T)  e^{-\imath \vec{k}_j \cdot \vec{x}} \right)\nonumber \\
o_1(\mathbf{x},T)&=&\sum\limits_j^{n} \left( B_j^{(1)}(T)e^{\imath \vec{k}'_j \cdot \vec{x}} + 
\overline{B}_{j}^{(1)}(T) e^{-\imath  \vec{k}'_j \cdot \vec{x}} \right) ,
\end{eqnarray}
with the complex amplitudes $A_j^{(1)}=\mathcal{A}_j e^{\imath \phi_j}$,  
$B_j^{(1)}=\mathcal{B}_j e^{\imath \psi_j}$ and
$\vec{k}_j=k_{c,z} \left(\cos (j \pi/n), \sin (j \pi/n) \right)$, 
$\vec{k}'_j=k_{c,o} \left(\cos (j \pi/n), \sin (j \pi/n) \right)$. 
In view of  the hexagonal or stripe layout of the OD pattern shown in Fig.~\ref{fig:Numerics}, 
$n=3$ is an appropriate choice.
Since in cat visual cortex the typical wavelength for OD and OP maps are approximately the 
same \cite{Kaschube1,Kaschube2} i.e. $k_{c,o}=k_{c,z}=k_c$
the Fourier components of the emerging pattern are located on a common critical circle.
To account for species differences we also analyzed models with detuned OP and OD wavelengths
in part (II) of this study.\\
At second order in $\mu$ we get 
\begin{eqnarray}
\hat{L}^0 z_2+r_{z1} z_1-r_{z1} \partial_T z_1 &=& 0 \nonumber \\
\hat{L}^0 o_2+r_{o1} o_1- r_{z1} \partial_T o_1 &=&0 \, .
\end{eqnarray}
As $z_1$ and $o_1$ are elements of the kernel $r_{z1}=r_{o1}=0$.
At third order, when applying the solvability condition (see Methods), we get
\begin{eqnarray}
 r_{z2} \partial_T \, z_1 &=& r_{z2} z_1 - \hat{P}_c N_{3,u}[z_1,z_1,\overline{z}_1] \nonumber \\
r_{z2} \partial_T \, o_1&=& r_{o2} o_1-\sqrt{r_{o2}} \, \hat{P}_c \tilde{N}_{2,u}[o_1,o_1]-
\hat{P}_c \tilde{N}_{3,u}[o_1,o_1,o_1] \, . 
\end{eqnarray}
We insert the leading order fields Eq.~(\ref{eq_z1o1}) and obtain the amplitude equations
\begin{eqnarray}\label{eq:withkappa}
r_{z2} \partial_T A_i^{(1)} &=& r_{z2} A_i^{(1)}-\sum_j g_{ij}|A_j^{(1)}|^2A_i^{(1)} 
-\sum_j f_{ij}A_j^{(1)} A_{j^-}^{(1)}\overline{A}_{i^-}^{(1)} \nonumber \\
 r_{z2} \partial_T B_i^{(1)} &=& r_{o2} B_i^{(1)} -2\sqrt{r_{o2}}\,  
\overline{B}_{i+1}^{(1)}\overline{B}_{i+2}^{(1)}-\sum_{j} \widetilde{g}_{ij} |B_j^{(1)}|^2 B_i^{(1)} \, . 
\end{eqnarray}
%%%%
These uncoupled amplitude equations obtain corrections at higher order.
There are fifth order, seventh order and even higher order corrections to the uncoupled
amplitude equations. In addition, at seventh order enters the nonlinearity of the higher order
inter-map coupling energies.
The amplitude equations up to seventh order are thus derived from
\begin{eqnarray}
r_{z2} \partial_T \, z_1 &=& r_{z2} z_1- \hat{P}_c N_{3,u}[z_1,z_1,\overline{z}_1] \nonumber \\
r_{z2} \partial_T \, z_3 &=&  r_{z2} z_3- \dots - 
\hat{P} N_{3,u}[z_1,z_1,\overline{z_3}]  \\
r_{z2} \partial_T \, z_5 &=&  r_{z2} z_5-\dots -
\hat{P}_c N_{3,u}[z_3,z_1,\overline{z_3}]- 
\hat{P}_c N_{7,c}[z_1,z_1,\overline{z}_1,o_1,o_1,o_1,o_1] \, , \nonumber
\end{eqnarray}
and corresponding equations for the fields $o_1$, $o_3$, and $o_5$.
The field $z_1$ is given in Eq.~(\ref{eq_z1o1}) and its amplitudes $A^{(1)}$ and $B^{(1)}$ are determined 
at third order.
The field $z_3$ contains contributions from modes off the critical
circle $z_{3,off}$, $|\vec{k}_{off}| \neq k_c$ and
on the critical circle i.e. 
$z_3=z_{3,off}+ \sum\limits_j^{n} \left( A^{(3)}_j(T)e^{\imath \vec{k}_j \cdot \vec{x}} 
+ A^{(3)}_{j^-} (T)  e^{-\imath \vec{k}_j \cdot \vec{x}} \right)$. 
Its amplitude $A^{(3)}$ are determined at fifth order.
The field $z_5$ also contains contributions from modes off the critical circle $z_{5,off}$ and
on the critical circle i.e. $z_5=z_{5,off}+ 
\sum\limits_j^{n} \left( A^{(5)}_j(T)e^{\imath \vec{k}_j \cdot \vec{x}} + 
A^{(5)}_{j^-} (T)  e^{-\imath \vec{k}_j \cdot \vec{x}} \right)$. 
Its amplitude $A^{(5)}$ are determined at seventh order.
This leads to a series of amplitude equations 
\begin{eqnarray}
r_{z2} \partial_T A_i^{(1)} &=&  r_{z2} A_i^{(1)}-
\sum_j g_{ij} |A_j^{(1)}|^2 A_i^{(1)}- \sum_j f_{ij} A_j^{(1)} A_{j^-}^{(1)} \overline{A}_{i^-}^{(1)} \nonumber \\
 r_{z2} \partial_T A_i^{(3)} &=&  r_{z2} A_i^{(3)}-\dots - \sum_j g_{ij} |A_j^{(1)}|^2\overline{A}^{(3)}_i  \\
 r_{z2} \partial_T A_i^{(5)} &=&  r_{z2} A_i^{(5)}-\dots-  \sum_j g_{ij} |A_j^{(3)}|^2 A_i^{(1)}- \sum_{jlk} h_{ijlk} |A_j^{(1)}|^2|B_l^{(1)}|^2|B_k^{(1)}|^2A_i^{(1)}\, , \nonumber
\end{eqnarray}
which are solved order by order.
We set $r_{z2}=r_z$ and $r_{o2}=r_o$ and rescale to the fast time.
This leads to
\begin{eqnarray}
\partial_t A_i^{(1)} &=& r_z A_i^{(1)}-\sum_j g_{ij} |A_j^{(1)}|^2 A_i^{(1)}-
\sum_j f_{ij} A_j^{(1)} A_{j^-}^{(1)} \overline{A}_{i^-}^{(1)}  \nonumber \\
\partial_t A_i^{(3)} &=& r_z A_i^{(3)}-\dots -\sum_j g_{ij} |A_j^{(1)}|^2\overline{A}^{(3)}_i  \\
\partial_t A_i^{(5)} &=& r_z A_i^{(5)}-\dots- \sum_j g_{ij} |A_j^{(3)}|^2 A_i^{(1)}
- \sum_{jlk} h_{ijlk} |A_j^{(1)}|^2|B_l^{(1)}|^2|B_k^{(1)}|^2A_i^{(1)}\, . \nonumber
\end{eqnarray}
We can combine the amplitude equations up to seventh order by introducing
the amplitudes $A_j=A^{(1)}_j+ A^{(3)}_j+A^{(5)}_j$ and $B_j=B^{(1)}_j+ B^{(3)}_j+B^{(5)}_j$.
This leads to the amplitude equations
\begin{eqnarray}\label{eq:AmplEq_Coupled_seventhOrder}
\partial_t \, A_i &=& r_z A_i -\sum_j g_{ij}|A_j|^2A_i -\sum_j f_{ij}  A_j A_{j^-} \overline{A}_{i^-}\nonumber \\
&&- \sum_{jlk}  h_{ijlk} |A_j|^2|B_l|^2|B_k|^2 A_i -\dots \nonumber \\
\partial_t \, B_i &=& r_o B_i -2 \overline{B}_{i+1} \overline{B}_{i+2}-\sum_j \tilde{g}_{ij}|B_j|^2AB_i \nonumber \\
&&-\sum_{jlk}  h_{ijlk} |B_j|^2|A_l|^2|A_k|^2 B_i -\dots \, .
\end{eqnarray}
For simplicity we have written only the simplest inter-map coupling terms. 
Depending on the configuration of active modes additional contributions may enter the amplitude equations.
In addition, for the product-type coupling energy, there are coupling terms which contain the 
constant $\delta$, see Methods.
In case of $A\ll B \ll 1$ the inter-map coupling terms in dynamics of the modes $B$ are small.
In this limit the dynamics of the modes $B$ decouples from the modes $A$ and we can use the uncoupled
OD dynamics, see Methods.
In the following, we use the effective inter-map coupling strength $\epsilon B^4$ (and $\tau B^4$).
%%%%%%%%%%%%%%%%%%%%%%%%%%%%%%%%%%%%%
\section{Higher order inter-map coupling energies}
\subsection{Optima of particular optimization principles: Higher order coupling terms}\label{sec:Mapcoupling_higherorder}
In this article we demonstrated that the low order coupling terms can lead to
a complete suppression of OP selectivity i.e. vanishing magnitude of the order parameter $|z|$. 
As the coupling terms are effectively
linear they not only influence pattern selection but also whether there is a pattern at all.
This is in general not the case for higher order coupling energies using
the amplitude equations Eq.~(\ref{eq:AmplEq_Coupled_seventhOrder}).
In this case the coupling is an effective cubic interaction term and complete selectivity
suppression is impossible.
Moreover, as in the low-order energy case, we could identify the 
limit $r_z \ll r_o$ in which the backreaction onto the
OD map formally becomes negligible.
The potential is of the form
\begin{eqnarray}\label{eq:Potential_Septic_part3}
 V&=&V_A+V_B+\nonumber \\
&&+ \sum_{j,l,k}\, \sum_{u,v,w,p}\,  h_{uvwp}^{ijlk}\, A_jA_l \overline{A}_k \overline{A}_i
B_u B_v B_w B_p \, \delta_{\vec{k}_j+\vec{k}_l-\vec{k}_k-\vec{k}_i+\vec{k}_u+
\vec{k}_v+\vec{k}_w+\vec{k}_p,0} \nonumber \\
&&+\delta \sum_{j,l,k}\, \sum_{u,v,w}\,  h_{uvp}^{ijlk}\, A_jA_l \overline{A}_k \overline{A}_i
B_u B_v B_w  \, \delta_{\vec{k}_j+\vec{k}_l-\vec{k}_k-\vec{k}_i+\vec{k}_u+
\vec{k}_v+\vec{k}_w,0} \nonumber \\
&&+\delta^2 \sum_{j,l,k}\, \sum_{u,v}\,  h_{uvp}^{ijlk}\, A_jA_l \overline{A}_k \overline{A}_i
B_u B_v   \, \delta_{\vec{k}_j+\vec{k}_l-\vec{k}_k-\vec{k}_i+\vec{k}_u+
\vec{k}_v,0} \nonumber \\
&&+\delta^3 \sum_{j,l,k}\, \sum_{u}\,  h_{uvp}^{ijlk}\, A_jA_l \overline{A}_k \overline{A}_i
B_u    \, \delta_{\vec{k}_j+\vec{k}_l-\vec{k}_k-\vec{k}_i+\vec{k}_u,0} \nonumber \\
&&+\delta^4 \sum_{j,l,k}\,  h^{ijlk}\, A_jA_l \overline{A}_k \overline{A}_i
   \, \delta_{\vec{k}_j+\vec{k}_l-\vec{k}_k-\vec{k}_i,0} \, ,
\end{eqnarray}
where $\delta_{i,j}$ denotes the Kronecker delta and the uncoupled 
contributions 
\begin{eqnarray}\label{eq:Potential_uncoupled}
 V_A&=&-r_z \sum_j^3 |A_j|^2 
+\frac{1}{2} \sum_{i,j}^3 g_{ij} |A_i|^2|A_j|^2
+\frac{1}{2} \sum_{i,j}^3 f_{ij} A_iA_{i^-}\overline{A}_j\overline{A}_{j^-} \nonumber \\
V_B&=&-r_o \sum_j^3 |B_j|^2 +\frac{1}{2} \sum_{i,j}^3 \widetilde{g}_{ij} |B_i|^2|B_j|^2 \, . 
\end{eqnarray}
Amplitude equations can be derived from the potential by $\partial_t A_i=-\delta V/\delta \overline{A}_i$.
We have not written terms involving the modes $A_{j^-}$ or $\overline{B}_j$. The complete
amplitude equations involving all modes and the corresponding coupling coefficients are given in Text S2.
As for the low order coupling energies terms involving the constant $\delta$ depend only on
the coupling coefficient of the product-type energy $\tau$.
In the following we specify the amplitude equations for negligible backreaction where  
$\mathcal{B}=\mathcal{B}_{hex}$, $\mathcal{B}=\mathcal{B}_{st}$, or 
$\mathcal{B}=0$.\\
%%
%In the following we identify classes of stationary solutions of the amplitude equations Eq.~(\ref{eq:AmplEq})
%and provide their stability criteria for the two higher order pendants of the coupling energies.
\subsection{Product-type energy $U=\tau \, o^4 |z|^4$}
First, we studied the higher order product-type inter-map coupling energy $U=\tau \, o^4 |z|^4$. 
As for the lower order version of this coupling energy the shift $\delta(\gamma)$ explicitly
enters the amplitude equations resulting in a rather complex parameter 
dependence, see Eq.~(68) in the Methods section.
\subsubsection{Stationary solutions and their stability}
In the case of OD stripes the amplitude equations of OP modes read
\begin{eqnarray}\label{eq:Ampl_tau_ODstripes}
 \partial_t \, A_1 &=& r_z A_1 -\sum_j \left( g_{1j}^{(1)}|A_j|^2A_1+g_{1j}^{(2)} |A_j|^2A_{1^-} 
+g_{1j}^{(3)}A_jA_{j^-}\overline{A}_{1^-} +g_{1j}^{(4)} A_jA_{j^-}\overline{A}_1\right) \nonumber \\
&&-B^4 A_{1^-}^2 \overline{A}_1 
-\sum_{u\neq v \neq w} A_uA_v\overline{A}_w \left(
\left(8\delta^3 B+24\delta B^2\overline{B}\right) \delta_{\vec{k}_u+\vec{k}_v-\vec{k}_w,0} \right. \\
&&\left. +\left(8\delta^3 \overline{B}+24\delta B\overline{B}^2\right) 
\delta_{\vec{k}_u+\vec{k}_v-\vec{k}_w,2\vec{k}_1}
+8\delta B^3 \delta_{\vec{k}_u+\vec{k}_v-\vec{k}_w,-2\vec{k}_1} \right)\nonumber\\
\partial_t \, A_2 &=& r_z A_2 -\sum_j \left( g_{2j}^{(1)} |A_j|^2 A_2 
+g_{2j}^{(3)}A_jA_{j^-}\overline{A}_{2^-} \right)\nonumber \\
&& - g^{(2)}_{ii} A_2 A_{1^-}\overline{A}_1 - g^{(5)} A_2 A_1 \overline{A}_{1^-}
-1/2 g^{(2)}_{ii} A_{1^-}^2\overline{A}_{2^-} -1/2  g^{(5)} A_1^2 \overline{A}_{2^-} \nonumber \\
&&- \sum_{u, v, w} A_u A_v\overline{A}_w \left(
g^{(6)}_{uv} \delta_{\vec{k}_u+\vec{k}_v-\vec{k}_w,\vec{k}_2} +
g^{(7)}_{ij} \delta_{\vec{k}_u+\vec{k}_v-\vec{k}_w,\vec{k}_1+\vec{k}_2} \right. \nonumber \\
&& \left. +g^{(8)}_{ij} \delta_{\vec{k}_u+\vec{k}_v-\vec{k}_w,-\vec{k}_1+\vec{k}_2}+
g^{(9)}_{ij} \delta_{\vec{k}_u+\vec{k}_v-\vec{k}_w,2\vec{k}_1+\vec{k}_2}+
g^{(10)}_{ij} \delta_{\vec{k}_u+\vec{k}_v-\vec{k}_w,-2\vec{k}_1+\vec{k}_2} \right)\, , \nonumber
\end{eqnarray}
where $\delta_{i,j}$ denotes the Kronecker delta and
\begin{equation}
\begin{array}{ccc}
g_{ii}^{(1)}=&1+\delta^4+12\delta^2|B|^2+6|B|^4, & g_{ij\neq i}^{(1)}=2g_{ii}^{(1)}, \nonumber \\ 
g_{ii}^{(2)}=&g_{ij\neq i}^{(2)}=12\delta^2 B^2+8B^3 \overline{B}, & \nonumber \\ 
g_{ii}^{(3)}=&0, &  g_{ij\neq i}^{(3)}=2+12|B|^4+24\delta^2|B|^2+2\delta^4, \nonumber \\ 
g_{ii}^{(4)}=&0, & g_{ij\neq i}^{(4)}=12\delta^2B^2+8B^3\overline{B}, \nonumber \\ 
g^{(5)}=&12 \delta^2\overline{B}^2+8B\overline{B}^3, &  
g^{(6)}_{uu}=6|B|^4+6\delta|B|^2, \, \, g^{(6)}_{uv\neq u}=2g^{(6)}_{uu}, \nonumber \\ 
g^{(7)}_{uu}=&4\overline{B}\delta^3+1B\overline{B}^2\delta, &  g^{(7)}_{uv\neq u}=2g^{(7)}_{uu}, \nonumber \\ 
g^{(8)}_{uu}=&4B\delta^3+1B^2\overline{B}\delta, & g^{(8)}_{uv\neq u}=2g^{(8)}_{uu}, \nonumber \\ 
g^{(9)}_{uu}=&6\overline{B}^2\delta^2, & g^{(9)}_{uv\neq u}=2g^{(9)}_{uu}, \nonumber \\ 
g^{(10)}_{uu}=&6B^2\delta^2, & g^{(10)}_{uv\neq u}=2g^{(10)}_{uu}.
\end{array}
\end{equation}
The equation for the mode $A_3$ is given by interchanging the 
modes $A_2$ and $A_3$ in Eq.~(\ref{eq:Ampl_tau_ODstripes}).
The equations for the modes $A_{i^-}$ are given by interchanging the modes $A_i$ and $A_{i^-}$ and
interchanging the modes $B_i$ and $\overline{B}_i$.\\
% a)
In this case, at low inter-map coupling the OP stripes given by 
\begin{equation}\label{eq:OPstripes}
 z=\mathcal{A}_1 e^{\imath \left(\vec{k}_1 \cdot \vec{x} +\phi_1 \right)}-\mathcal{A}_{1^-} 
e^{-\imath \left(\vec{k}_1 \cdot \vec{x} +\phi_{1^-} \right)}\, ,
\end{equation}
with $\phi_1-\phi_{1^-}=2\psi_1+\pi$ run parallel to the OD stripes.
Their stationary amplitudes are given by
\begin{eqnarray}
A_1^2 &=& \left(u-v-\sqrt{u^2-2uv+v^2-16w^2}\right)^2 x / 32w \nonumber \\
A_{1^-} &=& x/2\, ,
\end{eqnarray}
with $x=r_z \left(u-v+\sqrt{u^2-2uv+v^2-16w^2}\right) / (uv-v^2-8w^2)$,
$u=2+13\mathcal{B}^4\tau +24\mathcal{B}^2\delta^2 \tau +2\delta^4 \tau$, 
$v= (6\mathcal{B}^4+12\mathcal{B}^2\delta^2+\delta^4)\tau$,
$w=(2\mathcal{B}^2+3\delta^2)\tau\mathcal{B}^2$.
The parameter dependence of these stripe solutions is shown 
in Fig.~\ref{fig:Amplitudes_tau}\textbf{\sffamily{A}}.\\
%The intersection angles are thus zero which contradicts the experimentally observed intersection angles. 
At large inter-map coupling the attractor states of the OP map consist of a 
stripe pattern containing only two preferred orientations, namely 
$\vartheta=\phi_1$ and $\vartheta=\phi_1+\pi/2$.
The zero contour lines of the OD map are along the maximum amplitude of 
orientation preference minimizing the energy term.\\
% b)
In addition there are rhombic solutions
\begin{equation}\label{eq:OPrhombs}
z=\mathcal{A}_1e^{\imath (\vec{k}_1 \cdot \vec{x}+\psi_1)}+
\mathcal{A}_{1^-}e^{-\imath (\vec{k}_1 \cdot \vec{x} -\psi_1+\pi)}+
\mathcal{A}_2e^{\imath (\vec{k}_2 \cdot \vec{x}+\psi_1)}+
\mathcal{A}_{2^-}e^{-\imath (\vec{k}_2 \cdot \vec{x}-\psi_1)} \, ,
\end{equation}
which exist also in the 
uncoupled case, see Fig.~\ref{fig:Amplitudes_tau}\textbf{\sffamily{B}}. 
However, these rhombic solutions are energetically not favored
compared to stripe solutions, see Fig.~\ref{fig:Amplitudes_tau}\textbf{\sffamily{C}}.
The inclusion of the inter-map coupling makes these rhombic solution even more stripe-like.\\ 
In case of a OD constant solution the amplitude equations read
\begin{equation}
\partial_t A_i=r_z A_i -\sum_j g_{ij} |A_j|^2A_i-\sum_j f_{ij} A_jA_{j^-}\overline{A}_{i^-} \, ,
\end{equation}
with $g_{ii}=1+\delta^4\tau$, $g_{ij}=2+2\delta^4 \tau$ and $f_{ij}=2+2\delta^4\tau$.
Inter-map coupling thus leads to a renormalization of the uncoupled interaction terms.
Stationary solutions are stripes with the amplitude
\begin{equation}
\mathcal{A}=\sqrt{\frac{r_z}{1+\delta^4\tau}} \, ,
\end{equation}
and rhombic solutions with the stationary phases $\phi_1+\phi_{1^-}-\phi_2-\phi_{2^-}=\pi$
and the stationary amplitudes
\begin{equation}
\mathcal{A}_1=\mathcal{A}_{1^-}=\mathcal{A}_2=\mathcal{A}_{2^-}=\sqrt{r_z / (5+5\delta^4\tau )} \, . 
\end{equation}
In the case of OD hexagons we identify,
% c,d,e)
in addition to stripe-like and rhombic solutions, uniform solutions $\mathcal{A}_i=\mathcal{A}$.
When solving the amplitude equations numerically we have seen that the phase relations
vary with the inter-map coupling strength $\tau$ for non-uniform solutions. But for the uniform solution the
phase relations are independent of the inter-map coupling strength.
We use the ansatz for uniform solutions 
\begin{eqnarray}\label{eq:UniformSol_part3}
\mathcal{A}_j &=& \mathcal{A}_{j^-}=\mathcal{A}, \quad  j=1,2,3  \nonumber \\
\phi_j &=& \psi_j+(j-1)2\pi/3+\Delta\, \delta_{j,2}  \nonumber \\
\phi_{j^-} &=& -\psi_j+(j-1)2\pi/3+\Delta\, \left( \delta_{j,1}+\delta_{j,3}\right)\, ,
\end{eqnarray}
where $\delta_{i,j}$ is the Kronecker delta and $\Delta$ a constant parameter.
This leads to the stationarity condition
\begin{equation}
6\mathcal{A}^2\mathcal{B}\left[4\left(-4B^3+7B^2\delta-B\delta^2+\delta^3 \right)
+\mathcal{B} \cos \Delta  \left(13B^2-8B\delta +6\delta^2 \right) \right] \sin \Delta =0\, .
\end{equation}
Four types of stationary solutions exist namely the $\Delta=0,\Delta=\pi$, which we already observed
in case of the low order energies, and the solutions
\begin{equation}\label{eq:Unif_d_gamma}
\Delta=\Delta(\gamma)=\pm \arccos \left(\frac{4(4\mathcal{B}^3-7\mathcal{B}^2\delta+
\mathcal{B}\delta^2-\delta^3)}{\mathcal{B}(13\mathcal{B}^2-8\mathcal{B}\delta +6\delta^2)} \right)\, ,
\end{equation}
which depends on $\mathcal{B}$ and $\delta$ and thus on the bias $\gamma$.
The course of Eq.~(\ref{eq:Unif_d_gamma}) as a function of $\gamma$ is shown 
in Fig.~\ref{fig:PD_tau}\textbf{\sffamily{B}}.
Stationary amplitudes for these solutions are given by
\begin{eqnarray}\label{eq:StatAmp_Unif_tau}
\mathcal{A}_{\Delta=0}^2&=&
\frac{r_z}{3\tau \left( 3/\tau+33\mathcal{B}^4+56\mathcal{B}^3\delta  
+50\mathcal{B}^2 \delta^2+16\mathcal{B}\delta^3+3\delta^4 \right)} \nonumber \\ 
\mathcal{A}_{\Delta=\pi}^2&=& 
\frac{r_z}{\tau \left( 9/\tau+483\mathcal{B}^4-504\mathcal{B}^3\delta  
+246\mathcal{B}^2 \delta^2-48\mathcal{B}\delta^3+9\delta^4 \right)} \\
\mathcal{A}_{\Delta(\gamma)}^2&=&\frac{r_z(13\mathcal{B}^2-8\mathcal{B}\delta +6\delta^2)}
{3\tau\left( 411 \mathcal{B}^6 \tau +704 \mathcal{B}^5 \delta \tau -376  \mathcal{B}^4 \delta^2 \tau
+32\mathcal{B}^3\delta^3 \tau+2\delta^2(9-7\delta^4\tau)-39\mathcal{B}^2(3\delta^4 \tau-1)
+8\mathcal{B}\delta(5\delta^4\tau-3) \right)}\nonumber
\end{eqnarray}
We study the stability properties of OP stripe-like, rhombic and uniform solutions using linear stability analysis.
The eigenvalues of the stability matrix , see Text S2, are calculated numerically.
Linear stability analysis shows that for $\tau \geq 0$ the $\Delta=0$ solution is unstable for all bias values.
The stability region of the $\Delta=\pi$ solution and the solution Eq.~(\ref{eq:Unif_d_gamma}) is bias dependent.
The bias dependent solution Eq.~(\ref{eq:Unif_d_gamma}) is stable for $\gamma>\gamma^*$
and $\gamma<\gamma_c$ for which $\Delta=\pi$, see Fig.~\ref{fig:PD_tau}\textbf{\sffamily{B}}. 
For larger bias $\gamma>\gamma_c$ only the $d=\pi$ uniform solution is stable.
%%
%%%%%%%%%%%%%%%%%%%%%
\subsubsection{Bifurcation diagram}
\begin{figure}[bt]
\begin{center}
\includegraphics[width=.9\linewidth]{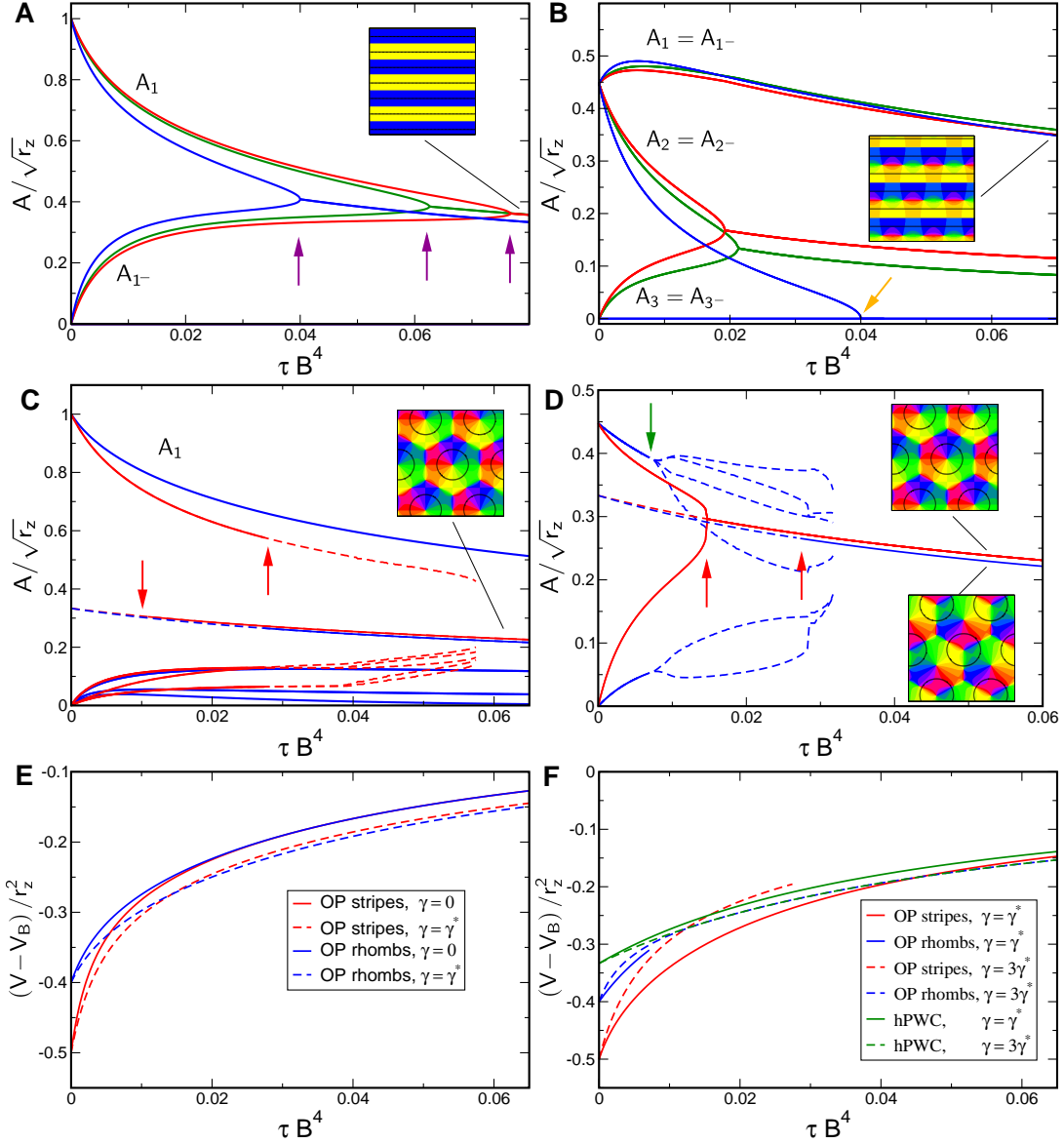} 
\end{center}
\caption{\textbf{Stationary amplitudes with coupling energy} $\mathbf{U= \tau |z|^4 o^4}$.
Solid (dashed) lines: stable (unstable) solutions.
\textbf{\sffamily{A,B}} OD stripes, $\gamma=0$ (blue), $\gamma=\gamma^*$ (green), $\gamma=1.4\gamma^*$ (red).
\textbf{\sffamily{C,D}} OD hexagons, $\gamma=\gamma^*$ (blue), $\gamma=3\gamma^*$ (red).
\textbf{\sffamily{A,C}} Transition from OP stripe solutions,
\textbf{\sffamily{B,D}} Transition from OP rhombic solutions.
\textbf{\sffamily{E}} Potential, Eq.~(\ref{eq:Potential_Septic_part3}), of OP stripes and OP rhombs 
interacting with OD stripes.
\textbf{\sffamily{F}} Potential, Eq.~(\ref{eq:Potential_Septic_part3}), of OP stripes, OP rhombs, and 
hPWC interacting with OD hexagons.
Arrows indicate corresponding lines in the phase diagram, Fig.~(\ref{fig:PD_tau}).
}
\label{fig:Amplitudes_tau}
\end{figure}

% OD stripes
The parameter dependence of OP solutions when interacting with OD stripes is 
shown in Fig.~\ref{fig:Amplitudes_tau}\textbf{\sffamily{A,B}}.
Similar to the low order variant of this coupling energy the amplitude of the stripes pattern $A_1$ is
suppressed while the amplitude of the opposite mode $A_{1^-}$ grows.
Finally both amplitudes collapse, leading to an orientation scotoma solution.
In contrast to the low order variant this stripe pattern is stable for arbitrary large inter-map
coupling.
In case of OP rhombic solutions inter-map coupling transforms this solution by reducing the amplitudes
$A_2=A_{2^-}$ while increasing the amplitudes $A_3=A_{3^-}$.
Without OD bias this solution is then transformed into the orientation scotoma stripe pattern, similar to the
low order variant of this energy. In contrast to the low order energy, 
for non-zero bias the amplitudes $A_2$ and $A_3$ stay small but non-zero.\\
% OD hexagons
The parameter dependence of OP solutions when interacting with OD hexagons 
is shown in Fig.~\ref{fig:Amplitudes_tau}\textbf{\sffamily{C,D}}.
For a small OD bias ($\gamma=\gamma^*$) OP rhombic solutions decay into OP stripe-like
patterns. These stripe-like patterns stay stable also for large-inter map coupling.
In case of a larger OD bias ($\gamma= 3\gamma^*$), both the OP stripe and the OP rhombic solutions
decay into the uniform PWC solution.
Thus for small bias there is a bistability between stripe-like and uniform PWC solutions while
for larger OD bias the uniform PWC solution is the only stable solution. 
% Potential
The potential of OP stripe and OP rhombic solutions is shown 
in Fig.~\ref{fig:Amplitudes_tau}\textbf{\sffamily{E,F}}.
In the uncoupled case as well as for small inter-map coupling strength OP stripe
solutions are for all bias values the energetic ground state.
For large inter-map coupling and a small bias ($\gamma \approx \gamma^*$)
rhombic solutions are unstable and the stripe-like solutions are energetically preferred compared
to PWC solutions.
For larger bias, however, PWC solutions are the only stable solutions for large inter-map coupling. 
\clearpage
%%%%%%%%%%%%%%%%%%%%%%%%%%%%
\subsubsection{Phase diagram}
\begin{figure}[bt]
\includegraphics[width=\linewidth]{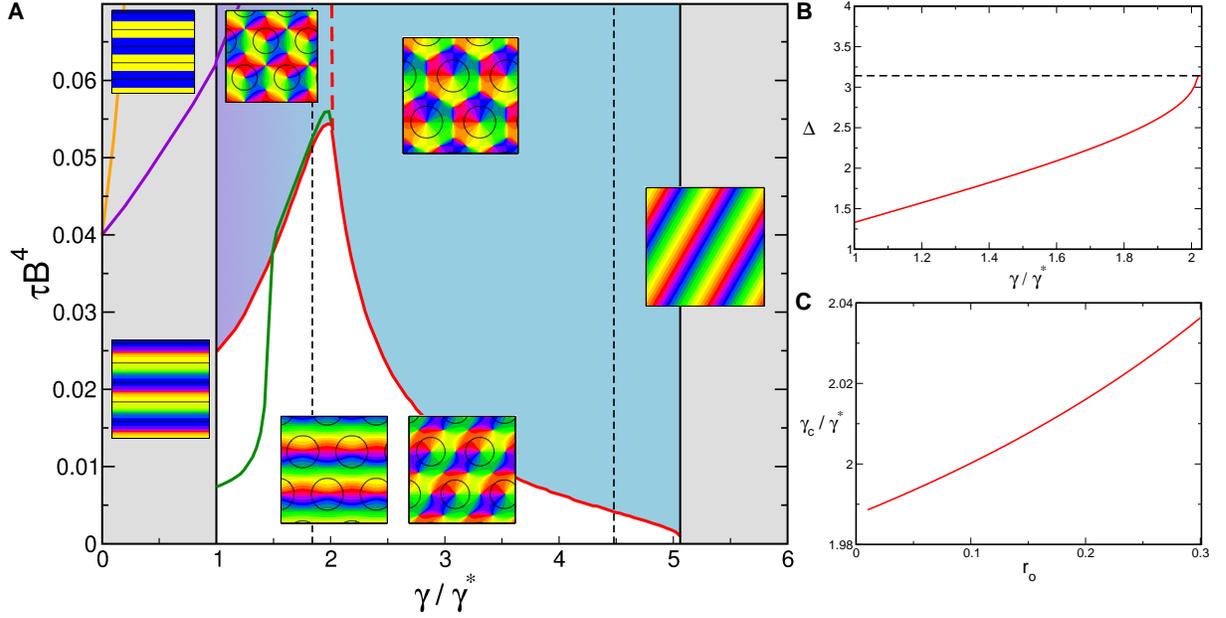} 
\caption{
\textbf{\sffamily{A}}
\textbf{Phase diagram with coupling energy} $\mathbf{U=\tau o^4|z|^4}$, $r_o=0.2, r_z \ll r_o$.
Vertical black lines: stability range of OD stripes, hexagons, and constant solutions.
Magenta (orange) line: Stability border of orientation scotoma stripes.
Green solid line: Stability border of rhombic solutions.
Red solid line: Stability border of PWC solutions, 
red dashed line: $\gamma_c$,
\textbf{\sffamily{B}} Course of Eq.~(\ref{eq:Unif_d_gamma}), dashed line: $\Delta=\pi$.
\textbf{\sffamily{C}} Stability border between Eq.~(\ref{eq:Unif_d_gamma}) solution and the $\Delta=\pi$ 
solution as a function of $r_o$ (vertical red line in \textbf{\sffamily{A}}).
}
\label{fig:PD_tau}
\end{figure}

The stability properties of all stationary solutions are summarized in the phase diagram Fig.~\ref{fig:PD_tau}.
%x-axis
Compared to the gradient-type interaction energy we cannot scale out the dependence on $r_o$. 
The phase diagram is thus plotted for $r_o=0.2$.
%y-axis
We rescale the inter-map coupling strength as $\tau \mathcal{B}^4$ where $\mathcal{B}$ is the
stationary amplitude of the OD hexagons.
% OD stripes
In the regime of stable OD stripes there is a transition from OP stripes towards the
orientation scotoma stripe solution.
% OD hexagons
In the regime of stable OD hexagons there is a transition from OP stripes 
towards PWC solutions (red line). The stability border of PWC solutions is strongly OD bias dependent
and has a peak at $\gamma \approx 2\gamma^*$.
For small OD bias $\gamma$ the uniform
solution Eq.~(\ref{eq:Unif_d_gamma}) is stable. With increasing bias
there is a smooth transition of this solution until at $\gamma=\gamma_c$ the $d=\pi$ uniform 
solution becomes stable.
In Fig.~\ref{fig:PD_tau}\textbf{\sffamily{C}} the stability border $\gamma_c$ between
the two types of uniform solutions is plotted as a function of $r_o$. We observe that there
is only a weak dependence on the control parameter and $\gamma_c \approx 2\gamma^*$.
%%%%
\subsubsection{Interaction induced pinwheel crystals}
\begin{figure}[th]
\includegraphics[width=\linewidth]{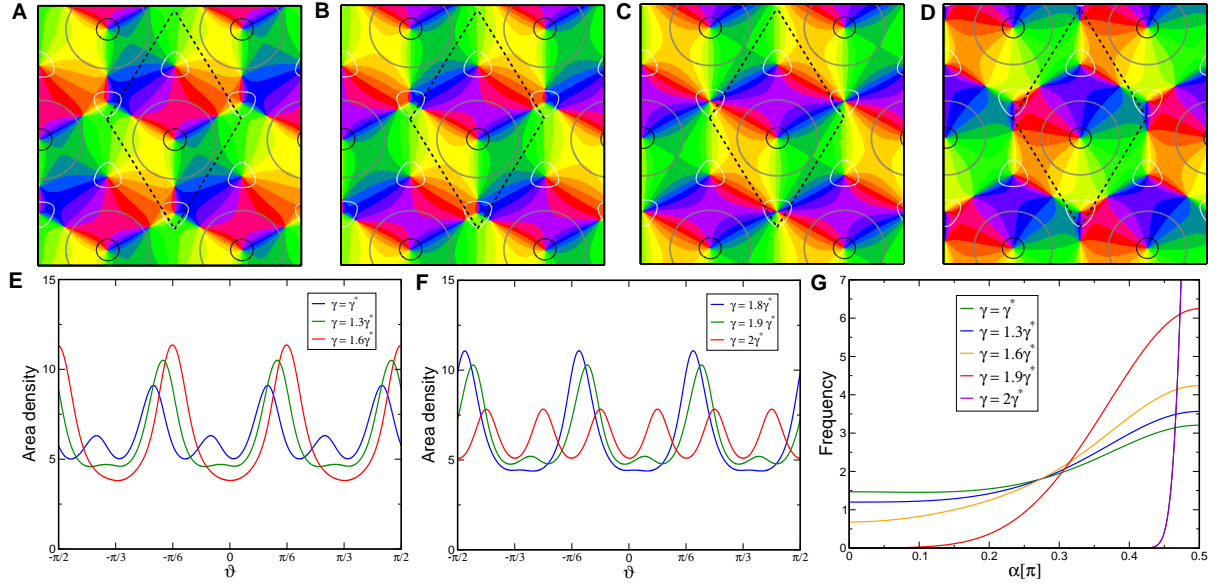} 
\caption{\textbf{Bias dependent pinwheel crystals}, Eq.~(\ref{eq:Unif_d_gamma}) 
\textbf{\sffamily{A}} $\gamma=\gamma^*$,
\textbf{\sffamily{B}} $\gamma=1.3\,\gamma^*$,
\textbf{\sffamily{C}} $\gamma=1.6\,\gamma^*$,
\textbf{\sffamily{D}} $\gamma=2\,\gamma^*$.
OP map, superimposed are the OD borders (gray),
90\% ipsilateral eye dominance (black), and 90\% contralateral eye dominance (white), $r_o=0.2$.
Dashed lines mark the unit cell of the regular pattern.
\textbf{\sffamily{E,F}} Distribution of orientation preference. 
\textbf{\sffamily{G}} Intersection angles between iso-orientation lines and OD borders.
}
\label{fig:Unif_gamma_depend}
\end{figure}

Figure~\ref{fig:Unif_gamma_depend} illustrates the uniform solutions Eq.~(\ref{eq:Unif_d_gamma})
for different values of the OD bias $\gamma$. 
For small bias, the OP pattern has six pinwheels per unit cell.
Two of them are located at OD maxima while one is located at an OD minimum.
The remaining three pinwheels are located near the OD border.
With increasing bias, these three pinwheels are pushed further away
from the OD border, being attracted to the OD maxima.
With further increasing bias three shifted pinwheels merge with the one at the OD maximum
building a single charge 1 pinwheel centered on a contralateral peak.
The remaining two pinwheels are located at an ispi and contra peak, respectively.
Note, compared to the Braitenberg PWC of the $\Delta=0$ uniform solution the charge 1 pinwheel
here is located at the contralateral OD peak. 
Finally, the charge 1 pinwheels split up again into four pinwheels. 
With increasing bias the solution more and more resembles the Ipsi-center PWC ($\Delta=\pi$ solution) 
which is stable also in the lower order version of the coupling energy.
Finally, at $\gamma/\gamma^* \approx 2$ the Ipsi-center PWC becomes stable and 
fixed for $\gamma>2\gamma^*$.
The distribution of preferred orientations for different values of the bias $\gamma$ is 
shown in Fig.~\ref{fig:Unif_gamma_depend}\textbf{\sffamily{E,F}}, reflecting the symmetry of each pattern.
%\begin{figure}[tb]
%\begin{center}
%\includegraphics[width=\linewidth]{/home/lars/Work/PhD_Thesis/pics/PWstab/Coverage/tau/Coverage_IS}
%\end{center}
%\caption{\textbf{(a),(b)} Distribution of orientation preference for bias dependent PWC, \refeqn{eq:Unif_d_gamma}. 
%\textbf{(c)} Intersection angles between iso-orientation lines and OD borders.}
%\label{fig:Coverage_tau}
%\end{figure}
The distribution of intersection angles is shown in Fig.~\ref{fig:Unif_gamma_depend}\textbf{\sffamily{G}}.
Remarkably, all solutions show a tendency towards perpendicular intersection angles.
This tendency is more pronounced with increasing OD bias. At about $\gamma/\gamma^* \approx 1.9$
parallel intersection angles are completely absent
and at $\gamma/\gamma^* \approx 2$
there are exclusively perpendicular intersection angles.
%%%%%%%%%%%%%%%%%%%%%%%%%%%%%%%%%%%%%%%%%%%%%%%
\subsection{Gradient-type energy $U= \epsilon \, |\nabla o \dotp \nabla z|^4$}
Finally, we examine the higher order version of the gradient-type inter-map coupling.
The interaction terms are independent of the 
OD shift $\delta$. In this case the coupling strength can be rescaled as $\beta \mathcal{B}^4$ and is
therefore independent of the bias $\gamma$. The bias in this case only determines the
stability of OD stripes, hexagons or the constant solution.
%%%%%%%%%%%%%%%%%%
\subsubsection{Stationary solutions and their stability}
As for its lower order pendant a coupling to OD stripes is relatively easy to analyze.
The energetic ground state corresponds to OP stripes with the direction perpendicular to the
OD stripes for which $U=0$. 
In addition, there are rhombic solutions with the stationary amplitudes 
$\mathcal{A}_1=\mathcal{A}_{1^-}=\mathcal{A}_2=\mathcal{A}_{2^-}=\sqrt{r_z / (5+80\epsilon \mathcal{B}^4)}$.
In case the OD map is a constant, $o(\mathbf{x})=\delta$, the gradient-type inter-map coupling leaves the 
OP unaffected. As for its lower order pendant the stationary states are 
therefore OP stripes running in an arbitrary direction and 
the uncoupled rhombic solutions.\\
%% OD hex non-uniform
In case of OD hexagons we identified three types of non-uniform solutions.
Besides stripe-like solutions of $z(\mathbf{x})$ with one dominant mode
we find rPWCs  
$\mathcal{A}_j=\mathcal{A}_{j^-}=\left(\mathcal{A},a,\mathcal{A}\right)$
with $a \ll \mathcal{A}$ and distorted rPWCs 
$\mathcal{A}_j=\left(\mathcal{A}_1,\mathcal{A}_2,\mathcal{A}_3\right), A_{j^-}=
\left(\mathcal{A}_3,\mathcal{A}_2,\mathcal{A}_1\right)$ with 
$\mathcal{A}_1 \neq \mathcal{A}_2 \neq \mathcal{A}_3$. 
Note, that distorted rPWCs are not stable in case of the product-type coupling energy
or the analyzed low-order coupling energies.
For these non-uniform solutions the stationary phases are inter-map coupling strength dependent.
We therefore calculate the stationary phases and amplitudes numerically using a Newton method
and initial conditions close to these solutions.\\
In case of OD hexagons there are further uniform solutions 
$A_j=A_{j^-}=\mathcal{A}, B_j=\mathcal{B}$ and $\psi_1=\psi_3=0,
\psi_2=\pi$.
The imaginary part of the amplitude equations, see Text S1, leads to equations for the phases $\phi_j$.
The ansatz Eq.~(\ref{eq:UniformSol_part3}) leads to the stationarity condition
\begin{equation}
\left( 13\cos \Delta-5 \right) \sin \Delta = 0 \, .
%-5  \sin \Delta+13 \cos \Delta \sin \Delta=0 \, .
\end{equation}
The solutions are $\Delta=0,\Delta=\pi$, and 
$\Delta=\pm \arccos \left(\frac{5}{13}\right) \approx \pm 1.176$ where
the stationary amplitude are given by 
%the equation
%\begin{equation}
%r_z \mathcal{A}-9\mathcal{A}^3+\epsilon \mathcal{B}^4 \mathcal{A}^3 \left(-61.875 +
%7.5 \cos (d)-4.875 \cos (2d)\right)  =0 \, ,
%\end{equation}
%which leads to
\begin{equation}\label{eq:StatAmp_Unif_eps}
\mathcal{A}=\sqrt{r_z / \left( 9+ \epsilon \mathcal{B}^4 \left( 61.875-7.5 \cos \Delta 
+4.875 \cos (2\Delta) \right)\right) } \, .
\end{equation}
%%%%
%%%% How to calculate stabilty properties
We calculated the stability properties of all stationary solutions by linear stability analysis
considering perturbations of the amplitudes $\mathcal{A}_ j \rightarrow \mathcal{A} +a_j$, 
$\mathcal{A}_ {j^-} \rightarrow \mathcal{A} +a_{j^-}$ and
of the phases $\phi_j \rightarrow \phi_j + \varphi_j$, $\phi_{j^-} \rightarrow \phi_{j^-} + \varphi_{j^-}$. 
This leads to a perturbation matrix $M$.
In general amplitude and phase perturbations do not decouple. We therefore calculate the eigenvalues 
of the perturbation $M$ matrix numerically.
It turns out that for this type of coupling energy only the uniform solutions 
with $\Delta=\pm \arccos \left(\frac{5}{13}\right)$ are stable while the $\Delta=0$ and
$\Delta=\pi$ solutions are unstable in general. 
%The corresponding eigenvalues are all negative above $\epsilon \mathcal{B}^4 \approx 0.42$
%(red solid line in \reffigP{fig:PDcoupled}{a})
%For $\Delta=0,\pi$ there is for all $\epsilon>0$ a positive eigenvalue.\\
%%%%
\subsubsection{Bifurcation diagram}
\begin{figure}[tb]
\centering
\includegraphics[width=\linewidth]{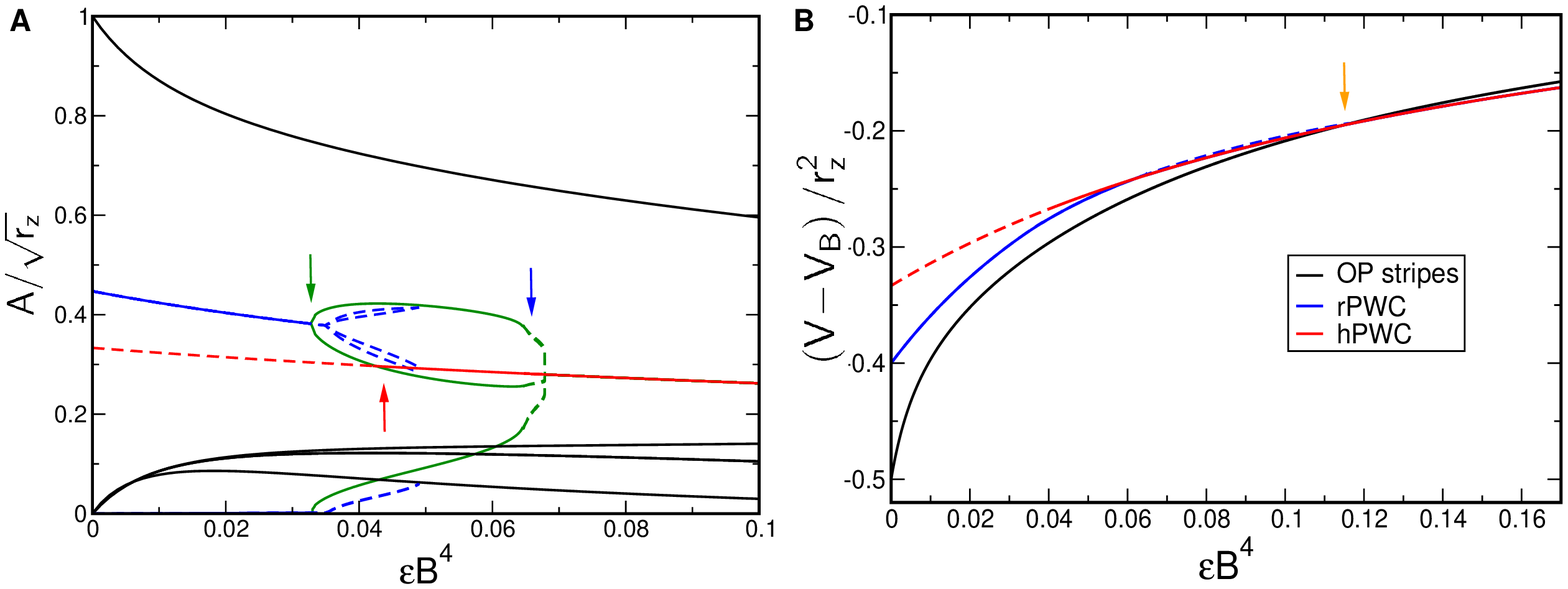}
\caption{\textbf{Stationary amplitudes with coupling energy} $\mathbf{U=\epsilon \, |\nabla z \dotp \nabla o|^4}$, 
\textbf{\sffamily{A}} 
Solid (dashed) lines: Stable (unstable) solutions.
Blue: rPWC, green: distorted rPWC, red: hPWC. Black lines: stripe-like solutions.
\textbf{\sffamily{B}} Potential, Eq.~(\ref{eq:Potential_Septic_part3}), of OP stripes (black), OP rhombs (blue), 
and hPWC solutions (red). 
Arrows indicate corresponding lines in the phase diagram, Fig.~(S\ref{fig:PDcoupled}).
}
\label{fig:Bif_eps}
\end{figure}

For increasing inter-map coupling strength the amplitudes of the OP stripe and OP rhombic solutions 
are shown in Fig.~\ref{fig:Bif_eps}\textbf{\sffamily{A}}.
%@ /home/reichl/PinwheelStabilization/PD_Section/new/ 
In case of stable OD hexagons there is a transition from rPWC (blue) towards 
distorted rPWC (green). The distorted rPWCs then decay into the hPWC (red).
In case of OP stripes (black dashed lines) inter-map coupling leads to a slight suppression of the dominant mode
and a growth of the remaining modes. This growth saturates at small amplitudes and thus the
solution stays stripe-like.
This stripe-like solution remains stable for arbitrary large inter-map coupling. 
Therefore there is a bistability between hPWC solutions and stripe-like solutions for large
inter-map coupling. \\
%%%%
The stability borders for the rPWC and distorted rPWC solutions were obtained by
calculating their bifurcation diagram numerically from the amplitude equations, see Text S1.
With increasing map coupling we observe a transition from a rPWC towards a 
distorted rPWC at $\epsilon \mathcal{B}^4 \approx 0.033$ (blue dashed 
line in Fig.~\ref{fig:PDcoupled}\textbf{\sffamily{A}}), see also 
Fig.~\ref{fig:PWwandering}\textbf{\sffamily{A}}.
The distorted rPWC loses its stability at $\epsilon \mathcal{B}^4 \approx 0.065$ 
(blue solid line in Fig.~\ref{fig:PDcoupled}\textbf{\sffamily{A}})
and from thereon all amplitudes are equal corresponding to the hPWC. 
There is a bistability between hPWC, rPWC, and stripe-like solutions.
To calculate the inter-map coupling needed for the hexagonal solution to become the energetic ground state 
we calculated the potential Eq.~(\ref{eq:Potential_Septic_part3}) for the three solutions.
In case of the uniform solution Eq.~(\ref{eq:UniformSol_part3}) the potential is given by
\begin{eqnarray}
V&=&-6\mathcal{A}^2r_z-3\mathcal{B}^2r_o +27\mathcal{A}^4+\frac{45}{2}\mathcal{B}^4 \nonumber \\
&&+\frac{1}{16}\mathcal{A}^4\mathcal{B}^4\epsilon \left(
3210-456 \cos \Delta +90 \cos(2\Delta) \right) \, . 
\end{eqnarray}
The potential in case of the rhombic and stripe-like solutions was obtained by numerically
solving the amplitude equations using Newtons method and initial conditions close to these solutions.
Above $\epsilon \mathcal{B}^4\approx 0.12$ the hPWC is energetically preferred compared
to stripe-like solutions (red dashed line in in Fig.~\ref{fig:PDcoupled}) and thus corresponds
to the energetic ground state for large inter-map coupling, see Fig.~(\ref{fig:Bif_eps})\textbf{\sffamily{B}}.
%%%%%
\subsubsection{Phase diagram}
\begin{figure}
%\centering
\begin{center}
\includegraphics[width=.66\linewidth]{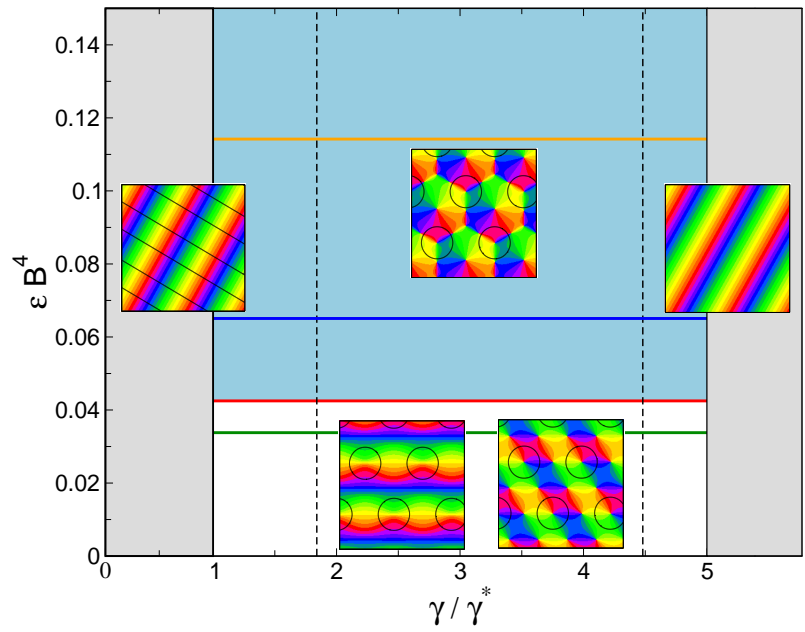}
\end{center}
\caption{\textbf{Phase diagram with coupling energy} $\mathbf{U=\epsilon \, |\nabla z \dotp \nabla o|^4}$, 
for $r_z \ll r_o$.
Vertical lines: stability range of OD hexagons,
green line: transition from rPWC to distorted rPWC,
red line: stability border of hPWC,
blue line: stability border of distorted rPWC. 
Above orange line: hPWC corresponds to ground state of energy.
}
\label{fig:PDcoupled}
\end{figure}

We calculated the phase diagram of the coupled system in the limit 
$r_z \ll r_o$, shown in Fig.~\ref{fig:PDcoupled}.
The phase diagram contains the stability borders of the uncoupled OD solutions
$\gamma^*,\gamma_2^*,\gamma_3^*,\gamma_4^*$. They correspond to vertical lines,
as they are independent of the inter-map coupling in the limit $r_z \ll r_o$. At $\gamma=\gamma^*$
hexagons become stable. Stripe solutions become unstable at $\gamma=\gamma_2^*$.
At $\gamma=\gamma_3^*$ the homogeneous solution becomes stable while at
$\gamma=\gamma_4^*$ hexagons loose their stability. In units $\gamma/\gamma^*$
the borders $\gamma_2^*, \gamma_3^*, \gamma_4^*$ vary slightly with $r_o$ , see 
Fig.~\ref{fig:constTerm}, and are drawn here for $r_o=0.2$.
We rescale the inter-map coupling strength as $\epsilon \mathcal{B}^4$ where $\mathcal{B}$ is the
stationary amplitude of the OD hexagons. The stability borders of OP solutions are then horizontal lines.
For $\gamma<\gamma^*$ or for $\gamma>\gamma_4^*$ pinwheel free orientation stripes
are dynamically selected. For $\gamma^*<\gamma<\gamma_4^*$ and above a critical effective coupling
strength $\epsilon \mathcal{B}^4\approx 0.042$ hPWC solutions are stable and become 
the energetic ground state of Eq.~(\ref{eq:Potential_Septic_part3}) above $\epsilon \mathcal{B}^4 \approx 0.117$.
Below $\epsilon \mathcal{B}^4 \approx 0.065$, rPWC solutions are stable leading to a bistability
region between rPWC and hPWC solutions.
We find in this region that rhombic solutions transform into distorted rhombic solutions above an effective
coupling strength of $\epsilon \mathcal{B}^4 \approx 0.033$. 
%%%%
\subsubsection{Interaction induced pinwheel crystals}
\begin{figure}[tb]
\begin{center}
\includegraphics[width=.6\linewidth]{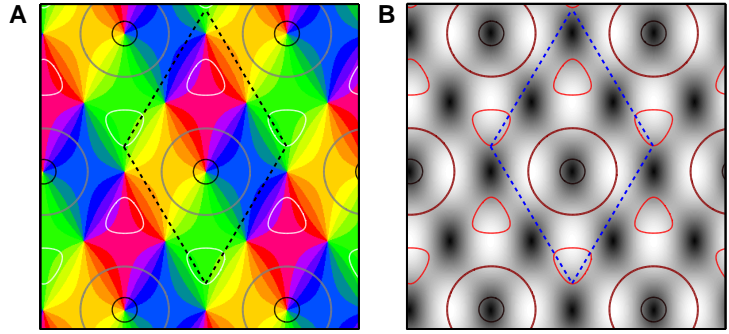}
\end{center}
\caption{\textbf{Rhombic pinwheel crystals.}
\textbf{\sffamily{A}} OP map with superimposed 
OD borders (gray), 90\% ipsilateral eye dominance (black), 
and 90\% contralateral eye dominance (white), $\gamma=3\gamma^*, r_o=0.2$.
\textbf{\sffamily{B}}
Selectivity $|z(\mathbf{x})|$, white: high selectivity, black: low selectivity.}
\label{fig:rPWC}
\end{figure}

\begin{figure}[tb]
\begin{center}
\includegraphics[width=\linewidth]{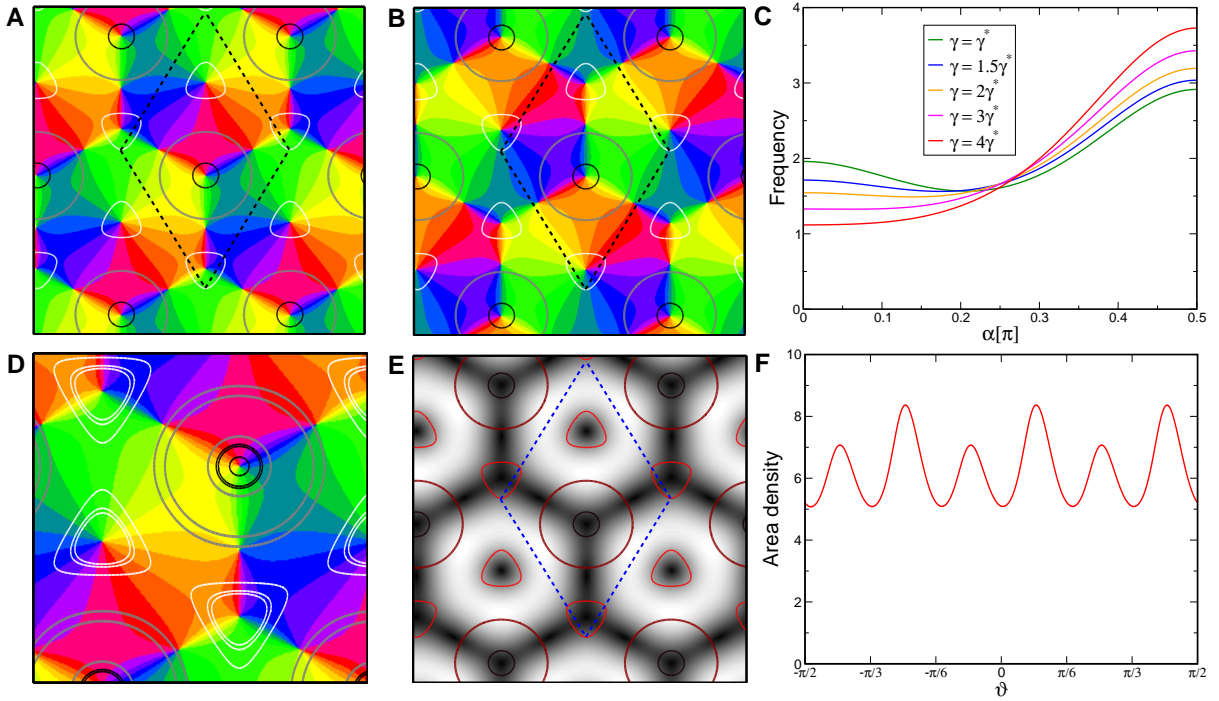}
\end{center}
\caption{\textbf{Contra-center pinwheel crystals.} 
\textbf{\sffamily{A,B}}
OP map, superimposed are the OD borders (gray),
90\% ipsilateral eye dominance (black), and 90\% contralateral eye dominance (white), 
$r_o=0.2, \gamma=3\gamma^*$.
\textbf{\sffamily{A}} $\Delta=\arccos (5/13)$,
\textbf{\sffamily{B}} $\Delta=-\arccos (5/13)$.
\textbf{\sffamily{C}} Distribution of orientation preference.
\textbf{\sffamily{D}} OP map
with superimposed OD map for three
different values \mbox{($\gamma=\gamma^*,\gamma=
\left( \gamma_4^* -\gamma^* \right) /2 +\gamma^*, \gamma=\gamma_4^*$)} of the OD bias.
\textbf{\sffamily{E}} Selectivity $|z(\mathbf{x})|$, white: high selectivity, black: low selectivity.
\textbf{\sffamily{F}} Distribution of intersection angles.
}
\label{fig:hPWC}
\end{figure}

\begin{figure}[tb]
\centering
\includegraphics[width=.8\linewidth]{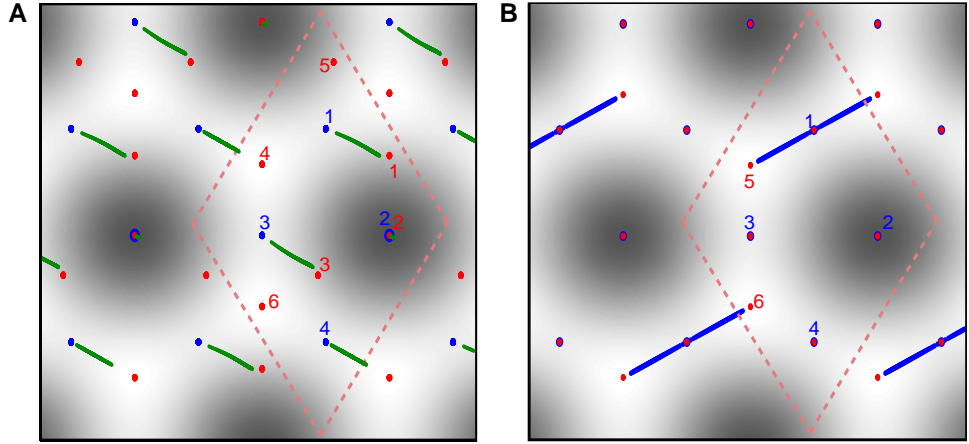}
\caption{\textbf{Inter-map coupling strength dependent pinwheel positions.}  
OD map, superimposed pinwheel positions (points) for different inter-map coupling 
strengths, $\gamma/\gamma^{*}=3$.
Numbers label pinwheels within the unit cell (dashed lines). 
Blue (green, red) points: pinwheel positions for rPWC (distorted rPWC, hPWC) solutions.
\textbf{\sffamily{A}} $\mathbf{U=\epsilon \, |\nabla z \dotp \nabla o|^4}$, using stationary amplitudes 
from Fig.~(\ref{fig:Bif_eps})(a). Positions of distorted rPWCs move continuously (pinwheel 1,3,4).
\textbf{\sffamily{B}} $\mathbf{U=\tau \, |z|^4  o^4}$, using stationary amplitudes 
from Fig.~(\ref{fig:Amplitudes_tau})\textbf{\sffamily{D}}. Positions of rPWCs move continuously (pinwheel 5,6).
}
\label{fig:PWwandering}
\end{figure}

%1) Description of rPWC
First, we studied the spatial layout of the rhombic solutions which is illustrated in Fig.~\ref{fig:rPWC}.
The rPWC solutions are symmetric under rotation by 180 degree.
The rhombic solution has 4 pinwheels per unit cell and its pinwheel density is thus
$\rho=4 \, \cos (\pi/6) \approx  3.5$.
One may expect that the energy term Eq.~(\ref{eq:Energy_HO}) favors pinwheels to co-localize with OD extrema.
In case of the rhombic layout there is only one pinwheel at an OD extremum while the
other three pinwheels are located at OD saddle-points which are also energetically favorable
positions with respect to $U$.
The orientation selectivity $|z(\mathbf{x})|$ for the rPWC is shown in 
Fig.~\ref{fig:rPWC}\textbf{\sffamily{B}}. The pattern of selectivity is arranged in small patches of
highly selective regions.\\
%2) Description of Contra-Center PWC
The hexagonal layout of the two stable uniform solutions is shown in Fig.~\ref{fig:hPWC}.
The $\Delta=\pm \arccos (5/13)$ solutions have six pinwheels per unit cell.
Their pinwheel density is therefore $\rho=6\cos \pi/6 \approx 5.2$.
Three pinwheels of the same topological charge are located at the extrema of the OD map. 
Two of these are located at the OD maximum while one is located at the
OD minimum. The remaining three pinwheels are not at an OD extremum but near
the OD border. The distance to the OD border depends on the OD bias, see Fig.~\ref{fig:hPWC}\textbf{\sffamily{D}}. 
For a small bias ($\gamma \approx \gamma^*$)
these three pinwheels are close to the OD borders and with increasing bias the OD border
moves away from the pinwheels. 
The pinwheel in the center of the OP hexagon is at the contralateral OD peak.
Because these pinwheels organize most of the map while the others essentially only match one OP 
hexagon to its neighbors we refer to this pinwheel crystal as the \textit{Contra-center pinwheel crystal}.
Note, that some pinwheels are located at the vertices of the hexagonal pattern. Pinwheels located 
between these vertices (on the edge) are not in the middle of this edge.
Solutions with $\Delta=\pm\arccos (5/13)$ are therefore not symmetric under a rotation by 60 degree
but symmetric under a rotation by 120 degree.
Therefore the solution $\Delta=+ \arccos (5/13)$ cannot be transformed into the
solution  $\Delta=-\arccos (5/13)$ by a rotation of the OD and OP pattern by 180 degrees.
This symmetry is also reflected by the distribution of preferred 
orientations, see Fig.~\ref{fig:hPWC}\textbf{\sffamily{F}}.
%The pattern of the Contra-center PWC is selective to all orientations but the 
%distribution of represented orientations is not flat, see \reffigP{fig:hPWC}{f}. 
Six orientations are slightly overrepresented. Compared to the Ipsi-center PWC, which 
have a $60^\circ$ symmetry, this distribution illustrates the $120^\circ$ symmetry of the pattern. 
%%%%
%4) Coverage, IS
%\begin{figure}[tb]
%\begin{center}
%\includegraphics[width=.8\linewidth]{/home/lars/Work/PhD_Thesis/pics/PWstab/Coverage/eps/Coverage_IS}
%\end{center}
%\caption{\textbf{(a)} Distribution of orientation preference for contra-center PWC.
%\textbf{(b)} Statistics of their intersection angles.}
%\label{fig:Coverage_d=acos}
%\end{figure}
The distribution of intersection angles is continuous, see Fig.~\ref{fig:hPWC}\textbf{\sffamily{C}}.
Although there is a fixed uniform solution with varying OD bias the distribution
of intersection angles changes. 
The reason for this is the bias dependent change in the OD borders, see Fig.~\ref{fig:hPWC}\textbf{\sffamily{D}}.
For all bias values there is a tendency towards perpendicular intersection angles, although
for low OD bias there is an additional small peak at parallel intersection angles. 
%%%%%
The orientation selectivity $|z(\mathbf{x})|$ for the hPWC is 
shown in Fig.~\ref{fig:hPWC}\textbf{\sffamily{E}}. The pattern shows hexagonal bands of high selectivity.\\
%3) Transition rhombic -> Hexagonal
%%%% Track pinwheel positions
Finally, we study changes in pinwheel positions during the transition from a rPWC towards 
a hPWC i.e. with increasing inter-map coupling strength.
In case of the higher order gradient-type coupling energy there is a transition towards
a contra-center PWC, see Fig.~\ref{fig:PWwandering}\textbf{\sffamily{A}}.
In the regime where the distorted rPWC is stable, three of the four pinwheels of the rPWC
are moving either from an OD saddle-point to a position near an OD border (pinwheel 1 and 3) or
from an OD saddle-point to an OD extremum (pinwheel 4). One pinwheel
(pinwheel 2) is fixed in space.
At the transition to the hPWC two additional pinwheels are created, one near an OD border
(pinwheel 5) and one at an OD extremum (pinwheel 6).
%%%% Tau
We compare the inter-map coupling strength dependent pinwheel positions of the gradient-type
coupling energy with those of the product type coupling 
energy, see Fig.~(\ref{fig:PWwandering})\textbf{\sffamily{B}}.
In this case three (pinwheel 2,3,4) of the four rPWC pinwheels have a inter-map coupling 
strength independent position.
The remaining pinwheel (pinwheel 1) with increasing inter-map coupling strength splits up into three 
pinwheels.
While one of these three pinwheels (pinwheel 1) is fixed in space the remaining two 
pinwheels (pinwheel 5,6) move towards
the extrema of OD. 
Thus for large inter-map coupling, where hPWC solutions are stable, all six pinwheels are located at 
OD extrema.
%%%%%%%%%%%%%%%%%%%%%%%%%%%%%%%%
\section{Summary}
We derived amplitude equations and analyzed ground states of the higher order inter-map coupling energies.
We calculated local and global optima and derived corresponding phase diagrams.
A main difference between phase diagrams for low order and high order coupling energies
consists in the collapse of orientation selectivity above a critical coupling strength that
occurs only in the low order models.
In contrast, for the high order versions, orientation selectivity is preserved
for arbitrarily strong inter-map coupling.
In order to neglect the backreaction on the dynamics of the modes $B$ we assumed
$A\ll B \ll 1$.
Our results, however, show that for the stability of pinwheel crystals a finite amplitude
$B$ is necessary. A decrease in $B$ cannot be compensated by another parameter
(as it would be $r_z$ in case of the low order inter-map coupling energies).
For a finite $B$ even higher order corrections to the amplitude equations than those 
presented here can thus become significant. Such terms are we neglected in the present treatment.
In part (II) of this study we numerically confirm our main results 
for the higher order inter-map coupling energies.\\
From a practical point of view, the analyzed phase diagrams and 
pattern properties indicate that the higher order gradient-type coupling
energy is the simplest and most convenient choice for constructing models that 
reflect the correlations of map layouts in the visual cortex.
For this coupling, intersection angle statistics are reproduced well, pinwheels can
be stabilized, and pattern collapse cannot occur.
%%%%
\newpage
\begin{flushleft}
{\Large
\textbf{Text S2: Supporting Information for\\
Coordinated optimization of visual cortical maps\\
(I) Symmetry-based analysis}
}
\end{flushleft}
%\section*{Text S12}
\section{Amplitude equations for higher order coupling energies}
Here, we list the amplitude equations for the OP dynamics in case of the high order 
inter-map coupling energies $U=\epsilon |\nabla z \cdot \nabla o|^4$ and $U=\tau |z|^4o^4$.
{\allowdisplaybreaks % only local!!!
\begin{eqnarray*}
% g_ij
\partial_t \, A_i &=& r_z A_i -\sum_j |A_j|^2 A_i \left( g_{ij}+\delta^4 g'_{ij} 
+\delta^2 \sum_u g_{iju} |B_u|^2  \right. \\
&& \left. +\delta g^{\prime\prime}_{ij} \left(B_1B_2B_3+\overline{B}_1\overline{B}_2\overline{B}_3\right)
+\sum_{u,v} g_{ijuv} |B_u|^2|B_v|^2 \right)\\
&&-\sum_j |A_{j^-}|^2 A_i \left( g_{ij^-}+\delta^4 g'_{ij^-} 
+\delta^2 \sum_u g_{ij^-u} |B_u|^2  \right. \\
&& \left. +\delta g^{\prime\prime}_{ij^-} \left(B_1B_2B_3+\overline{B}_1\overline{B}_2\overline{B}_3\right)
+\sum_{u,v}^3 g_{ij^-uv} |B_u|^2|B_v|^2 \right)\\
% f_ij
&&-\sum_j A_jA_{j^-} \overline{A}_{i^-} \left( f_{ij}+\delta^4 f'_{ij} 
+\delta^2 \sum_u f_{iju} |B_u|^2  \right. \\
&& \left. +\delta f^{\prime\prime}_{ij} \left(B_1B_2B_3+\overline{B}_1\overline{B}_2\overline{B}_3\right)
+\sum_{u,v} f_{ijuv} |B_u|^2|B_v|^2 \right)\\
% \propto A_i (ohne Triads)
&&-\sum_{j\neq l} A_i A_j \overline{A}_l \left(\sum_u h_{ijlu}^{(1)} \overline{B}_j B_l |B_u|^2
+\delta^2 h_{ijl}^{(1)} \overline{B}_j B_l\right)\\
&&-\sum_{j, l} A_i A_{j^-} \overline{A}_l \left(\sum_u h_{ij^-lu}^{(1)} B_j B_l |B_u|^2
+\delta^2 h_{ij^-l}^{(1)} B_j B_l\right)\\
&&-\sum_{j, l} A_i A_j \overline{A}_{l^-} \left(\sum_u h_{ijl^-u}^{(1)} \overline{B}_j \overline{B}_l |B_u|^2
+\delta^2 h_{ijl^-}^{(1)} \overline{B}_j \overline{B}_l\right)\\
&&-\sum_{j \neq l} A_i A_{j^-} \overline{A}_{l^-} \left(\sum_u h_{ij^-l^-u}^{(1)} B_j \overline{B}_l |B_u|^2
+\delta^2 h_{ij^-l^-}^{(1)} B_j \overline{B}_l\right)\\
% \propto B_i (ohne Triads)
&& -\sum_{l,j\neq k} A_j A_l \overline{A}_k h_{ijlk}^{(2)} \overline{B}_j \overline{B}_l B_k B_i 
 -\sum_{l,j,k} A_{j^-} A_l \overline{A}_k h_{ij^-lk}^{(2)} B_j \overline{B}_l B_k B_i \\
&& -\sum_{l,j\neq k} A_j A_l \overline{A}_{k^-} h_{ijlk^-}^{(2)} \overline{B}_j \overline{B}_l \overline{B}_k B_i 
 -\sum_{l,j\neq k} A_{j^-} A_{l^-} \overline{A}_k h_{ij^-l^-k}^{(2)} B_j B_l B_k B_i \\
&& -\sum_{l,j\neq k} A_{j^-} A_l \overline{A}_{k^-} h_{ij^-lk^-}^{(2)} B_j \overline{B}_l \overline{B}_k B_i 
 -\sum_{l,j\neq k} A_{j^-} A_{l^-} \overline{A}_{k^-} h_{ij^-l^-k^-}^{(2)} B_j B_l \overline{B}_k B_i \\
&&-\sum_{j,l} |A_j|^2 A_l \left( \sum_u h_{ijlu}^{(3)} |B_u|^2 \overline{B}_l B_i + \delta^2 h_{ijl}^{(3)} \overline{B}_l B_i \right) \\
&&-\sum_{j,l} |A_{j^-}|^2 A_l \left( \sum_u h_{ij^-lu}^{(3)} |B_u|^2 \overline{B}_l B_i + \delta^2 h_{ij^-l}^{(3)} \overline{B}_l B_i \right) \\
&&-\sum_{j,l} |A_j|^2 A_{l^-} \left( \sum_u h_{ijl^-u}^{(3)} |B_u|^2 B_l B_i + \delta^2 h_{ijl^-}^{(3)} B_l B_i \right) \\
&&-\sum_{j,l} |A_{j^-}|^2 A_{l^-} \left( \sum_u h_{ij^-l^-u}^{(3)} |B_u|^2 B_l B_i + \delta^2 h_{ij^-l^-}^{(3)} B_l B_i \right) \\
% \propto A_i (mit Triads)
% [ABB]
&& -\sum_{j,u} \delta h_{iju}^{(4)} A_i \overline{A}_j A_u B_{u+1} B_{u+2} B_j -\sum_{j,u} \delta h_{iuj}^{(4)} A_i A_j \overline{A}_u \overline{B}_{u+1} \overline{B}_{u+2} \overline{B}_j \\
&& -\sum_{j,u} \delta h_{ij^-u}^{(4)} A_i \overline{A}_{j^-} A_u B_{u+1} B_{u+2} \overline{B}_j -\sum_{j,u}\delta h_{ij^-u}^{(4)} A_i A_{j^-} \overline{A}_u \overline{B}_{u+1} \overline{B}_{u+2} \overline{B}_j \\
&& -\sum_{j,u} \delta h_{iju^-}^{(4)} A_i \overline{A}_j A_{u^-} \overline{B}_{u+1} \overline{B}_{u+2} B_j -\sum_{j,u} \delta h_{iju^-}^{(4)} A_i A_j \overline{A}_{u^-} B_{u+1} B_{u+2} B_j \\
&& -\sum_{j,u} \delta h_{ij^-u^-}^{(4)} A_i \overline{A}_{j^-} A_{u^-} \overline{B}_{u+1} \overline{B}_{u+2} \overline{B}_j -\sum_{j,u}\delta  h_{ij^-u^-}^{(4)} A_i A_{j^-} \overline{A}_{u^-} B_{u+1} B_{u+2} \overline{B}_j \\
%[AAB]
&& 
-\sum_{u}  h^{(5)}_{iu} A_i A_u \overline{A}_{(u+1)^-} B_{u+2} B_1B_2B_3
-\sum_{u}  h^{(5)}_{iu^-} A_i A_{u^-} \overline{A}_{(u+1)} \overline{B}_{u+2} B_1B_2B_3\\
&&
-\sum_{u}  \tilde{h}^{(5)}_{iu} A_i A_u \overline{A}_{(u+1)^-} B_{u+2} \overline{B}_1\overline{B}_2\overline{B}_3
-\sum_{u}  \tilde{h}^{(5)}_{iu^-} A_i A_{u^-} \overline{A}_{(u+1)} \overline{B}_{u+2} \overline{B}_1\overline{B}_2\overline{B}_3\\
&& 
-\sum_{u} \delta^3 h^{(5)} A_i A_u \overline{A}_{(u+1)^-} B_{u+2}-\sum_{u} \delta^3 h^{(5)} A_i A_{u^-} \overline{A}_{(u+1)} \overline{B}_{u+2}\\
%\propto B_i (mit Triads)
%[AAB]
&&-\sum_{j,u} \delta h^{(6)}_{iju} \overline{A}_j A_u A_{u+1} B_{u+2} B_j B_i
-\sum_{j,u} \delta h^{(6)}_{ij^-u} \overline{A}_{j^-} A_u A_{u+1} B_{u+2} \overline{B}_j B_i\\
&&-\sum_{j,u} \delta h^{(6)}_{iju^-} \overline{A}_{j} A_{u^-} A_{(u+1)^-} \overline{B}_{u+2} B_j B_i
-\sum_{j,u} \delta h^{(6)}_{ij^-u^-} \overline{A}_{j^-} A_{u^-} A_{(u+1)^-} \overline{B}_{u+2} \overline{B}_j B_i\\
%[ABB]
&&-\sum \delta h_{iju}^{(7)} |A_j|^2 A_u B_{u+1} B_{u+2} B_i
-\sum \delta h_{ij^-u}^{(7)} |A_{j^-}|^2 A_u B_{u+1} B_{u+2} B_i \\
&& -\sum \delta h_{iju^-}^{(7)} |A_j|^2 A_{u^-} \overline{B}_{u+1} \overline{B}_{u+2} B_i
 -\sum \delta h_{ij^-u^-}^{(7)} |A_{j^-}|^2 A_{u^-} \overline{B}_{u+1} \overline{B}_{u+2} B_i\\
% kein \propto A_i order \propto B_i
&&-\sum \delta h_{ij}^{(8)} A_j^2 \overline{A}_{i+2} \overline{B}_j^2 \overline{B}_{i+1}
-\sum \delta h_{ij^-}^{(8)} A_{j^-}^2 \overline{A}_{i+2} B_j^2 \overline{B}_{i+1}\\
&& -\sum \delta^3 h_{ij}^{(9)} |A_j|^2  A_{(i+2)^-} \overline{B}_{i+1}
-\sum \delta^3 h_{ij^-}^{(9)} |A_{j^-}|^2  A_{(i+2)^-} \overline{B}_{i+1}
\end{eqnarray*}
}
where the indices are considered to be cyclic i.e. $j+3=j$. All sums are considered to run from 1 to 3.
In the article, these amplitude equations are specified in case of OD stripes, Eq.~(109), OD 
hexagons, Eq.~(110), or a constant OD solution, Eq~(111).
\section{Coupling coefficients}
In the following, we list the non-zero elements of the coupling coefficients.
{\allowdisplaybreaks % only local!!!
\begin{equation*}
\begin{array}{cccc}
g_{ii}=1 & g_{ij}=2 & g'_{ii}=\tau & g'_{ij}=2\tau \\
g_{iji}=g_{ijj}=g_{iju}=24\tau & g_{iii}=g_{iij}=12\tau &&\\
g^{\prime\prime}_{ij}=48\tau & g^{\prime\prime}_{ii}=24\tau &&\\
g_{ijii}=6\epsilon+12\tau & g_{ijij}=33\epsilon+48\tau & g_{ijjj}=6\epsilon+12\tau & g_{ijiu}=3\epsilon+48\tau \\
g_{ijju}=-3\epsilon +48\tau & g_{ijuu}=1.5\epsilon+12\tau && \\
f_{ij}=2 & f'_{ij}=2\tau &&\\
f_{iju}=f_{ijj}=f_{iji}=24\tau & f^{\prime\prime}_{ij}=48\tau &&\\
f_{ijii}=12\tau & f_{ijij}=48\tau+33\epsilon & f_{ijjj}=12\tau+6\epsilon & f_{ijiu}=48\tau+3\epsilon\\ 
f_{ijju}=48\tau-3\epsilon & f_{ijuu}=12\tau+1.5\epsilon &&\\
h^{(1)}_{ijli}=15\epsilon+48\tau & h^{(1)}_{ijlj}=0.75\epsilon+24\tau & h^{(1)}_{ijll}=0.75\epsilon+24\tau &\\
h^{(1)}_{ijii}=21\epsilon+24\tau & h^{(1)}_{ijij}=9.75\epsilon+24\tau & h^{(1)}_{ijil}=1.5\epsilon+48\tau &\\
h^{(1)}_{iiji}=10.5\epsilon+12\tau & h^{(1)}_{iijj}=4.875\epsilon+12\tau & h^{(1)}_{iiju}=0.75\epsilon+24\tau &\\
h^{(1)}_{ij^-li}=15\epsilon+48\tau & h^{(1)}_{ij^-lj}=h^{(1)}_{ij^-ll}=0.75\epsilon+24\tau & &\\
h^{(1)}_{ijl^-i}=15\epsilon+48\tau & h^{(1)}_{ijl^-j}=h^{(1)}_{ijl^-l}=0.75\epsilon+24\tau & &\\
h^{(1)}_{ij^-l^-i}=15\epsilon+48\tau & h^{(1)}_{ij^-l^-j}=h^{(1)}_{ij^-l^-l}=0.75\epsilon+24\tau & &\\
h^{(1)}_{ijl}=24\tau & h^{(1)}_{iij}=12\tau & h^{(1)}_{ij^-l}=h^{(1)}_{ii^-j}=24\tau & h^{(1)}_{ij^-j}=12\tau\\
h^{(1)}_{ijl^-}=24\tau & h^{(1)}_{iij^-}=12\tau & h^{(1)}_{iii^-}=6\tau & h^{(1)}_{iji^-}=24\tau\\
h^{(1)}_{ij^-l^-}=24\tau & h^{(1)}_{ii^-j^-}=24\tau & h^{(1)}_{ij^-i^-}=24\tau & \\
h^{(2)}_{ijli}=7.5\epsilon+24\tau & h^{(2)}_{iiij}=10.5\epsilon+12\tau &h^{(2)}_{ij^-li}=7.5\epsilon+24\tau &h^{(2)}_{ii^-ij}=21\epsilon+24\tau\\
h^{(2)}_{ijli^-}=15\epsilon+48\tau & h^{(2)}_{ijji^-}=8.25\epsilon+12\tau & h^{(2)}_{ijjl^-}=3.75\epsilon+12\tau & \\
h^{(2)}_{ij^-l^-i}=7.5\epsilon+24\tau & h^{(2)}_{ij^-l^-j}=7.5\epsilon+24\tau & h^{(2)}_{ij^-l^-l}=7.5\epsilon+24\tau & \\
h^{(2)}_{ij^-ll^-}=7.5\epsilon+24\tau & h^{(2)}_{ij^-li^-}=15\epsilon+48\tau &&\\
h^{(2)}_{ij^-l^-i^-}=15\epsilon+48\tau & h^{(2)}_{ij^-j^-i^-}=8.25\epsilon+12\tau & h^{(2)}_{ij^-j^-l^-}=3.75\epsilon+12\tau & \\
h^{(3)}_{ijli}=0.75\epsilon+24\tau & h^{(3)}_{ijlj}=15\epsilon+48\tau & h^{(3)}_{ijll}=0.75\epsilon+24\tau & h^{(3)}_{ijji}=4.875\epsilon+12\tau \\
h^{(3)}_{ijjj}=10.5\epsilon+12\tau & h^{(3)}_{ijjl}=0.75\epsilon+24\tau & h^{(3)}_{ijl}=24\tau & h^{(3)}_{ijj}=12\tau \\  
h^{(3)}_{iiji}=12\epsilon+24\tau & h^{(3)}_{iijj}=9.75\epsilon+24\tau & h^{(3)}_{iijl}=1.5\epsilon+48\tau & h^{(3)}_{iij}=24\tau \\  
h^{(3)}_{ii^-ji}=21\epsilon+24\tau &  h^{(3)}_{ii^-jj}=9.75\epsilon+24\tau  & h^{(3)}_{ii^-jl}=1.5\epsilon+48\tau & \\
h^{(3)}_{ij^-ji}=9.75\epsilon+24\tau &  h^{(3)}_{ij^-jj}=21\epsilon+24\tau  & h^{(3)}_{ij^-jl}=15\epsilon+48\tau & \\ 
h^{(3)}_{ij^-li}=0.75\epsilon+24\tau &  h^{(3)}_{ij^-lj}=0.75\epsilon+24\tau  & h^{(3)}_{ij^-ll}=15\epsilon+48\tau & \\
\end{array}
\end{equation*}
\begin{equation*}
\begin{array}{cccc}
h^{(3)}_{ii^-j}=24\tau & h^{(3)}_{ij^-j}=24\tau & h^{(3)}_{ij^-l}=24\tau &\\
h^{(3)}_{iii^-i}=16\epsilon+8\tau & h^{(3)}_{iii^-j}=12\epsilon+24\tau & h^{(3)}_{iji^-i}=4\epsilon+8\tau & h^{(3)}_{iji^-j}=16.5\epsilon+24\tau \\
h^{(3)}_{iji^-l}=1.5\epsilon+24\tau & h^{(3)}_{iij^-i}=21\epsilon+24\tau & h^{(3)}_{iij^-j}=9.75\epsilon+24\tau & h^{(3)}_{iij^-l}=1.5\epsilon+48\tau \\
h^{(3)}_{ijj^-i}=9.75\epsilon+24\tau & h^{(3)}_{ijj^-j}=21\epsilon+24\tau & h^{(3)}_{ijj^-l}=1.5\epsilon+48\tau & h^{(3)}_{ijl^-i}=0.75\epsilon+24\tau \\
h^{(3)}_{ijl^-j}=15\epsilon+48\tau & h^{(3)}_{ijl^-l}=0.75\epsilon+24\tau & h^{(3)}_{iii^-}=12\tau & h^{(3)}_{iji^-}=12\tau \\
h^{(3)}_{iij^-}=24\tau & h^{(3)}_{ijj^-}=24\tau & h^{(3)}_{ijl^-}=24\tau &\\
h^{(3)}_{ii^-i^-i}=8\epsilon+4\tau & h^{(3)}_{ii^-i^-j}=6\epsilon+12\tau & h^{(3)}_{ii^-j^-i}=21\epsilon+24\tau & h^{(3)}_{ii^-j^-j}=9.75\epsilon+24\tau \\
h^{(3)}_{ii^-j^-l}=1.5\epsilon+48\tau & h^{(3)}_{ij^-i^-i}=4\epsilon+8\tau & h^{(3)}_{ij^-i^-j}=16.5\epsilon+24\tau & h^{(3)}_{ij^-i^-l}=1.5\epsilon+24\tau \\
h^{(3)}_{ij^-j^-i}=4.875\epsilon+24\tau & h^{(3)}_{ij^-j^-j}=10.5\epsilon+12\tau & h^{(3)}_{ij^-j^-l}=0.75\epsilon+24\tau & h^{(3)}_{ij^-l^-i}=0.75\epsilon+24\tau \\
h^{(3)}_{ij^-l^-j}=15\epsilon+48\tau & h^{(3)}_{ij^-l^-l}=0.75\epsilon+24\tau & h^{(3)}_{ii^-i^-}=6\tau & h^{(3)}_{ij^-j^-}=12\tau \\
h^{(3)}_{ii^-j^-}=24\tau & h^{(3)}_{ij^-l^-}=24\tau & h^{(3)}_{ii^-i^-}=12\tau &  \\
h^{(4)}_{iji}=12\tau & h^{(4)}_{ijl}=24\tau &h^{(4)}_{ii^-i}=h^{(4)}_{ij^-i}=24\tau &h^{(4)}_{ii^-j}=48\tau\\
h^{(4)}_{ijj^-j}=h^{(4)}_{ij^-l}=48\tau &h^{(4)}_{iji^-}=h^{(4)}_{ijj^-}=h^{(4)}_{ijl^-}=48\tau&h^{(4)}_{ij^-i^-}=h^{(4)}_{ij^-l^-}=24\tau&\\
h^{(5)}_{ii}=0.375\epsilon+12\tau & \tilde{h}^{(5)}_{ii}=2h^{(5)}_{ii} & h^{(5)}_{ii+1}=7.5\epsilon+48\tau & \tilde{h}^{(5)}_{ii+1}=2h^{(5)}_{ii+1}\\ 
h^{(5)}_{ii+2}=0.75\epsilon+24\tau & \tilde{h}^{(5)}_{ii+2}=2h^{(5)}_{ii+2} & h^{(5)}_{ii^-}=1.5\epsilon+48\tau &  \tilde{h}^{(5)}_{ii^-}=1/2h^{(5)}_{ii^-}\\
h^{(5)}_{i(i+1)^-}=15\epsilon+48\tau &  \tilde{h}^{(5)}_{i(i+1)^-}=1/2h^{(5)}_{i(i+1)^-} &&\\
h^{(5)}=8\tau &&&\\
h^{(6)}_{ijj}=h^{(6)}_{iju}=24\tau & h^{(6)}_{iij}=8\tau & h^{(6)}_{ii^-i}=h^{(6)}_{ij^-i}=48\tau & h^{(6)}_{ii^-j}=h^{(6)}_{ij^-j}=24\tau\\
h^{(6)}_{iii^-}=h^{(6)}_{iij^-}=24\tau & h^{(6)}_{iji^-}=h^{(6)}_{ijj^-}=h^{(6)}_{iju^-}=48\tau &&\\
h^{(6)}_{ij^-j^-}=h^{(6)}_{ii^-i^-}=48\tau  &h^{(6)}_{ij^-u^-}=h^{(6)}_{ii^-j^-}=24\tau&h^{(6)}_{ij^-i^-}=48\tau & \\
h^{(7)}_{ijj}=12\tau & h^{(7)}_{iij}=h^{(7)}_{ijl}=24\tau &h^{(7)}_{ii^-j}=h^{(7)}_{ij^-j}=h^{(7)}_{ij^-l}=24\tau&\\
h^{(7)}_{iii^-}=h^{(7)}_{iij^-}=h^{(7)}_{iji^-}=48\tau & h^{(7)}_{ijj^-}=h^{(7)}_{ijl^-}=48\tau &\\
h^{(7)}_{ii^-j^-}=h^{(7)}_{ij^-i^-}=48\tau & h^{(7)}_{ii^-i^-}=h^{(7)}_{ij^-j^-}=24\tau &h^{(7)}_{ij^-l^-}=48\tau&\\
h^{(8)}_{ii}=h^{(8)}_{ij}=12\tau & h^{(8)}_{ii^-}=h^{(8)}_{ij^-}=12\tau &&\\
h^{(9)}_{ij}=h^{(9)}_{ii}=8\tau &&&\\
%g_{ij} = 2(1+\tau \delta^4), &
%g_{ii} = 1+\tau \delta^4, &
%f_{ij} = 2(1+\tau \delta^4), &
%f_{ii} = 0 
\end{array}
\end{equation*}
}
Note, that coupling coefficients involving the constant shift $\delta$ only occur in case of the
product-type inter-map coupling energy.
\newpage
\begin{flushleft}
{\Large
\textbf{Text S3: Supporting Information for\\
Coordinated optimization of visual cortical maps\\
(I) Symmetry-based analysis}
}
\end{flushleft}
% \section*{Text S2}
\section*{Stability matrix}
Here, we state the stability matrices $M$ used in the linear stability analysis of the coupled 
amplitude equations.
The stability matrix is defined by \\
\begin{equation*}
 \partial_t \left( \begin{array}{c} \text{Re} A_i \\ \text{Re} A_{i^-} \\ \text{Im} A_i \\ \text{Im}A_{i^-}
                   \end{array}\right) = M \left( \begin{array}{c} \text{Re} A_i \\ \text{Re} A_{i^-} \\ \text{Im} A_i \\ \text{Im}A_{i^-}
                   \end{array}\right) = \left( \begin{array}{cc} M1 & M2 \\M3 & M4 \end{array} \right) 
\left( \begin{array}{c} \text{Re} A_i \\ \text{Re} A_{i^-} \\ \text{Im} A_i \\ \text{Im}A_{i^-}
                   \end{array}\right)
\end{equation*}
for the uniform solutions Eq.~(60). % Eq.~(\ref{eq:UniformSol_part3}).
We separate the uncoupled contributions from the inter-map coupling contributions i.e.
$M1=M1^{(u)}+M1^{(\alpha)}$.
The uncoupled contributions are given by
\begin{equation*}
 M1^{(u)}=r_z \mathbf{I}  
+ \mathcal{A}^2 \left( \begin{array}{cccccc} 
-13 & -3 & 3 & -2 c&
- 3c -\sqrt{3} s &
3c -\sqrt{3} s \\ 
 -3 & -23/2 & 0 &- 3c -\sqrt{3} s &
-2 c & 0\\
3 & 0 &-23/2 & 3 c -\sqrt{3} s & 0 & -2c \\
-2c & -3c -\sqrt{3} s & 3c -\sqrt{3} s &
-12 -\cos 2\Delta & -2-2u & 2 +2v \\
-3d+\sqrt{3}s & -2d & 0&  -2-2u & -12-\cos(2\Delta+10\pi/3)  & -4d^2\\
3c-\sqrt{3}s & 0 & -2c & -2-2v & -4s^2 & -12+u
\end{array} \right)
\end{equation*}

\begin{equation*}
 M2^{(u)}=  \mathcal{A}^2 \left( \begin{array}{cccccc}
0 & \sqrt{3} & \sqrt{3} & 2s & 2\cos(\Delta-\pi/6) & 2\cos(\Delta +\pi/6) \\
\sqrt{3} & \sqrt{3}/2 & 0 & 2 \cos(\Delta-\pi/6) & 2s & 4s\\
\sqrt{3} & 0 & -\sqrt{3}/2 & 2\cos(\Delta+\pi/6) & 4s & 2s\\                 
-2s & \sqrt{3}d-3s & \sqrt{3}d+3s & -\sin 2\Delta & 2x & 2y \\
\sqrt{3} d-2s & -2d & 0 & 2x & y & 2\sin 2\Delta \\
\sqrt{3}d+3s & 0 & -2s & 2y & 2\sin 2\Delta & -x                  
\end{array} \right)
\end{equation*}
  % check equations!!!!!!!
\begin{equation*}
 M3^{(u)}=  \mathcal{A}^2 \left( \begin{array}{cccccc}                      
0 & \sqrt{3} & \sqrt{3} & 2s & 2\cos(\Delta -\pi/6 ) & 2\cos(\Delta+\pi/6) \\
\sqrt{3} & \sqrt{3}/2 &  0 & 2\cos(\Delta-\pi/6) & 2s & 4s\\
\sqrt{3} & 0 & -\sqrt{3}/2 & 2\cos(\Delta+\pi/6) & 4s & 2s\\
-2s & \sqrt{3}c-3s & \sqrt{3} c+3s & -\sin 2\Delta & 2x & 2y\\
\sqrt{3}d-3s & -2s & 0 & 2x &y & 2\sin 2\Delta \\
\sqrt{3} c+3s & 0 & -2s & 2y & 2\sin 2\Delta & -x
 \end{array} \right)
\end{equation*}

\begin{equation*}
 M4^{(u)}=r_z \mathbf{I}  
+ \mathcal{A}^2 \left( \begin{array}{cccccc} 
-11 & -1 & 1 & -2 c& 2w & 2z \\ 
 -1 & -25/2 & -4 & 2w & -2c & -4c\\
1 & -4 & -25/2 & 2z & -4c & -2c\\
-2c & 2w & 2z &  -12+\cos 2\Delta & 2(u-1) &2(1+v) \\
2w & -2c & -4c & 2(u-1) & -12+\cos(2\Delta+10\pi/3) & -4c^2 \\
2z & -4c & -2c & 2(v+1) & -4c^2 & -12+\cos(2\Delta +8\pi/3)\, ,
\end{array} \right)
\end{equation*}
where $\mathbf{I}$ denotes the $6\times 6$ identity matrix and 
$s=\sin \Delta, c=\cos \Delta, u=\sin(2\Delta +, \pi/6), v=\sin(2\Delta -\pi/6), 
x=\cos(2\Delta +\pi/6), y=\cos(2\Delta -\pi/6), w=\sin(\Delta-\pi/6), z=\sin(\Delta+\pi/6)$.\\
The coupling part in case of the low order product-type inter-map coupling energy is given by
\begin{equation*}
 M1^{(\alpha)}=\alpha  \left( \begin{array}{cccccc} 
-(6\mathcal{B}^2+\delta^2) & 2\mathcal{B}^2 & -2\mathcal{B}^2 & -\mathcal{B}^2 &
2\mathcal{B} (\mathcal{B}-\delta) & -2\mathcal{B} (\mathcal{B}-\delta) \\
2\mathcal{B}^2 &   -(6\mathcal{B}^2+\delta^2) & 2\mathcal{B}^2 & 2\mathcal{B}(\mathcal{B}-\delta) &
-\mathcal{B}^2 & 2\mathcal{B} (\mathcal{B}-\delta) \\
-2\mathcal{B}^2 & 2\mathcal{B}^2 & -(6\mathcal{B}^2+\delta^2) & 2\mathcal{B}(-\mathcal{B}+\delta) &
2\mathcal{B}(\mathcal{B}-\delta) & -\mathcal{B}^2 \\
-\mathcal{B}^2 & 2\mathcal{B}(\mathcal{B}-\delta) & -2\mathcal{B}(\mathcal{B}-\delta) & 
-(6\mathcal{B}^2+\delta^2) & 2\mathcal{B}^2 & -2\mathcal{B}^2 \\
2\mathcal{B}^2(\mathcal{B}-\delta) & -\mathcal{B}^2 & 2\mathcal{B}(\mathcal{B}-\delta) &
2\mathcal{B}^2 & -(6\mathcal{B}^2+\delta^2) & 2\mathcal{B}^2  \\
2\mathcal{B}(-\mathcal{B}+\delta) & 2\mathcal{B}(\mathcal{B}-\delta) & -\mathcal{B}^2 & 
-2\mathcal{B}^2 & 2\mathcal{B}^2 & -(6\mathcal{B}^2+\delta^2)
\end{array}\right)
\end{equation*}
$M4^{(\alpha)}=M1^{(\alpha)}, M2^{(\alpha)}=M3^{(\alpha)}=0$.\\
The coupling part in case of the low order gradient-type inter-map coupling energy is given by
\begin{equation*}
 M1^{(\beta)}=\beta \mathcal{B}^2 \left( \begin{array}{cccccc} 
 3 & -5/4 & 5/4 & 1& -5/4 & 5/4 \\
-5/4 & 3 & -5/4 & -5/4 & 1 & -5/4 \\
5/4 & -5/4 & 3 & 5/4 & -5/4 & 1 \\
1 & -5/4 & 5/4 & 3 & -5/4 & 5/4 \\
-5/4 & 1 & -5/4 & -5/4 & 3 & -5/4 \\
5/4 & -5/4 & 1 & 5/4 & -5/4 & 3                     
                     \end{array}\right)
\end{equation*}
$M4^{(\beta)}=M1^{(\beta)}, M2^{(\beta)}=M3^{(\beta)}=0$.\\
The stationary amplitudes $\mathcal{A}$ are given in Eq.~(63), %Eq.~(\ref{eq:StatAmp_unif_alp}), 
Eq.~(68), Eq.~(81), and Eq.~(83).
%Eq.~(\ref{eq:Unif_d_beta}), Eq.~(\ref{eq:StatAmp_Unif_tau}), and Eq.~(\ref{eq:StatAmp_Unif_eps}).
The stationary amplitudes $\mathcal{B}$ and the constant $\delta$ are given 
by Eq.~(94) and Eq.~(116), respectively.

% Do NOT remove this, even if you are not including acknowledgments
\section*{Acknowledgments}
We thank Ghazaleh Afshar, Eberhard Bodenschatz, Theo Geisel, Min Huang, Wolfgang Keil, 
Michael Schnabel, Dmitry Tsigankov, and Juan Daniel Fl\'{o}rez Weidinger for discussions.
%This work was supported by the HFSP, BMBF,  and the MPG.

%\section*{Appendix}\label{sec:Appendix}

%%%%%%%%%%%%%%%%%%%%%%%%%%%%

%%%%%%%%%%%%%%%%%%%%%%%%%%%%%%%%%%%%%%%%%%%%%%%%%%%
%\section*{References}
% The bibtex filename
%\bibliography{biblio}

\begin{thebibliography}{100}
\providecommand{\url}[1]{\texttt{#1}}
\providecommand{\urlprefix}{URL }
\expandafter\ifx\csname urlstyle\endcsname\relax
  \providecommand{\doi}[1]{doi:\discretionary{}{}{}#1}\else
  \providecommand{\doi}{doi:\discretionary{}{}{}\begingroup
  \urlstyle{rm}\Url}\fi
\providecommand{\bibAnnoteFile}[1]{%
  \IfFileExists{#1}{\begin{quotation}\noindent\textsc{Key:} #1\\
  \textsc{Annotation:}\ \input{#1}\end{quotation}}{}}
\providecommand{\bibAnnote}[2]{%
  \begin{quotation}\noindent\textsc{Key:} #1\\
  \textsc{Annotation:}\ #2\end{quotation}}
\providecommand{\eprint}[2][]{\url{#2}}

\bibitem{Kaschube6}
Kaschube M, Schnabel M, L{\"o}wel S, Coppola DM, White LE, et~al. (2010)
  Universality in the evolution of orientation columns in the visual cortex.
\newblock Science 330: 1113--1116.
\input{Kaschube6}

\bibitem{HubelWiesel2}
Hubel DH, Wiesel TN (1977) Functional architecture of macaque monkey visual
  cortex.
\newblock Proc R Soc London 198: 1--59.
\input{HubelWiesel2}

\bibitem{LoewelBuch}
Payne BR, Peters A (2002) The Cat Primary Visual Cortex.
\newblock Academic Press.
\input{LoewelBuch}

\bibitem{Grinwald}
Grinwald A, Lieke E, Frostig RD, Gilbert CD, Wiesel TN (1986) Functional
  architecture of cortex revealed by optical imaging of intrinsic signals.
\newblock Nature 324: 361--364.
\input{Grinwald}

\bibitem{Blasdel3}
Blasdel GG, Salama G (1986) Voltage-sensitive dyes reveal a modular
  organization in monkey striate cortex.
\newblock Nature 321: 579--585.
\input{Blasdel3}

\bibitem{Swindale3}
Swindale NV, Matsubara J, Cynader M (1987) Surface organization of orientation
  and direction selectivity in cat area 18.
\newblock J Neurosci 7: 1414--1427.
\input{Swindale3}

\bibitem{Grinwald1}
Bonhoeffer T, Grinwald A (1991) Iso-orientation domains in cat visual cortex
  are arranged in pinwheel-like patterns.
\newblock Nature 353: 429--431.
\input{Grinwald1}

\bibitem{Bartfeld}
Bartfeld E, Grinwald A (1992) Relationships between orientation-preference
  pinwheels, cytochrome oxidase blobs, and ocular-dominance columns in primate
  striate cortex.
\newblock PNAS 89: 11905--11909.
\input{Bartfeld}

\bibitem{Blasdel4}
Blasdel GG (1992) Diferential imaging of ocular dominance and orientation
  selectivity in monkey striate cortex.
\newblock J Neurosci 12: 3115--3138.
\input{Blasdel4}

\bibitem{Blasdel}
Obermayer K, Blasdel GG (1993) Geometry of orientation and ocular dominance
  columns in monkey striate cortex.
\newblock J Neurosci 13: 4114--4129.
\input{Blasdel}

\bibitem{Grinwald5}
Bonhoeffer T, Grinwald A (1993) The layout of iso-orientation domains in area
  18 of cat visual cortex: optical imaging reveals a pinwheel-like
  organization.
\newblock J Neurosci 13: 4157-80.
\input{Grinwald5}

\bibitem{Bosking3}
Weliky M, Bosking WH, Fitzpatrick D (1996) A systematic map of direction
  preference in primary visual cortex.
\newblock Nature 379: 725-728.
\input{Bosking3}

\bibitem{Grinwald_Dir}
Shmuel A, Grinwald A (1996) Functional organization for direction of motion and
  its relationship to orientation maps in cat area 18.
\newblock J Neurosci 16: 6945--6964.
\input{Grinwald_Dir}

\bibitem{Chapman}
Chapman B, Stryker MP, Bonhoeffer T (1996) Development of orientation
  preference maps in ferret primary visual cortex.
\newblock J Neurosci 16: 6443--6453.
\input{Chapman}

\bibitem{Rao}
Rao SC, Toth LJ, Sur M (1997) Optically imaged maps of orientation preference
  in primary visual cortex of cats and ferrets.
\newblock J Comp Neurol 387: 358--370.
\input{Rao}

\bibitem{Bosking}
Bosking WH, Zhang Y, Schofield B, Fitzpatrick D (1997) Orientation selectivity
  and the arrangement of horizontal connections in tree shrew striate cortex.
\newblock J Neurosci 17: 2112--2127.
\input{Bosking}

\bibitem{Das2}
Das A, Gilbert CD (1997) Distortions of visuotopic map match orientation
  singularities in primary visual cortex.
\newblock Nature 387: 594--598.
\input{Das2}

\bibitem{Loewel}
L{\"o}wel S, Schmidt KE, Kim DS, Wolf F, Hoffs{\"u}mmer F, et~al. (1998) The
  layout of orientation and ocular dominance domains in area 17 of strabismic
  cats.
\newblock Eur J Neurosci 10: 2629--2643.
\input{Loewel}

\bibitem{Stryker2}
Issa NP, Trepel C, Stryker MP (2000) Spatial frequency maps in cat visual
  cortex.
\newblock J Neurosci 15: 8504--8514.
\input{Stryker2}

\bibitem{White}
White LE, Coppola DM, Fitzpatrick D (2001) The contribution of sensory
  experience to the maturation of orientation selectivity in ferret visual
  cortex.
\newblock Nature 411: 1049.
\input{White}

\bibitem{Galuske}
Galuske RAW, Schmidt KE, Goebel R, Lomber SG, Payne BR (2002) The role of
  feedback in shaping neural representations in cat visual cortex.
\newblock PNAS 99: 17083-88.
\input{Galuske}

\bibitem{Casagrande}
Xu X, Bosking WH, White LE, Fitzpatrick D, Casagrande VA (2005) Functional
  organization of visual cortex in the prosimian bush baby revealed by optical
  imaging of intrinsic signals.
\newblock J Neurophysiol 94: 2748--2762.
\input{Casagrande}

\bibitem{White_Dir}
Li Y, Fitzpatrick D, White LE (2006) The development of direction selectivity
  in ferret visual cortex requires early visual experience.
\newblock Nature Neuroscience 9: 676--681.
\input{White_Dir}

\bibitem{White3}
White LE, Fitzpatrick D (2007) Vision and cortical map development.
\newblock Neuron 56: 327--338.
\input{White3}

\bibitem{Horton4}
Adams DL, Horton JC (2003) Capricious expression of cortical columns in the
  primate brain.
\newblock Nature Neuroscience 6.
\input{Horton4}

\bibitem{Koulakov2}
Chklovskii DB, Koulakov AA (2004) Maps in the brain: What can we learn from
  them?
\newblock Annu Rev Neurosci 27: 369--392.
\input{Koulakov2}

\bibitem{Horton5}
Horton JC, Adams DL (2005) The cortical column: a structure without a function.
\newblock Phil Trans R Soc B 360: 837--862.
\input{Horton5}

\bibitem{Koulakov}
Koulakov AA, Chklovskii DB (2001) Orientation preference patterns in mammalian
  visual cortex: A wire length minimization approach.
\newblock Neuron 29: 519--527.
\input{Koulakov}

\bibitem{Koulakov4}
Koulakov AA, Chklovskii DB (2002) Direction of motion maps in the visual
  cortex: a wire length minimization approach.
\newblock Neurocomputing 44-46: 489--494.
\input{Koulakov4}

\bibitem{Durbin}
Durbin R, Mitchison GJ (1990) A dimension reduction framework for understanding
  cortical maps.
\newblock Nature 343: 644--647.
\input{Durbin}

\bibitem{Obermayer2}
Obermayer K, Ritter H, Schulten K (1990) A principle for the formation of the
  spatial structure of cortical feature maps.
\newblock PNAS 87: 8345--8349.
\input{Obermayer2}

\bibitem{Swindale11}
Swindale NV (1991) Coverage and the design of striate cortex.
\newblock Biol Cybern 65: 415--424.
\input{Swindale11}

\bibitem{Schulten}
Erwin E, Obermayer K, Schulten K (1995) Models of orientation and ocular
  dominance columns in the visual cortex: A critical comparison.
\newblock Neural Computation 7: 425--468.
\input{Schulten}

\bibitem{Swindale6}
Swindale NV (1996) The development of topography in the visual cortex: {A}
  review of models.
\newblock Network 7.
\input{Swindale6}

\bibitem{Goodhill7}
Cimponeriu A, Goodhill GJ (2000) Dynamics of cortical map development in the
  elastic net model.
\newblock Neurocomputing 32-33: 83--90.
\input{Goodhill7}

\bibitem{Goodhill}
Goodhill GJ, Cimponeriu A (2000) Analysis of the elastic net model applied to
  the formation of ocular dominance and orientation columns.
\newblock Comput Neural Syst 11: 153--168.
\input{Goodhill}

\bibitem{Swindale5}
Swindale NV (2000) How many maps are there in visual cortex?
\newblock Cereb Cortex 10: 633--634.
\input{Swindale5}

\bibitem{Goodhill5}
Carreira-Perpinan MA, Goodhill GJ (2004) Influence of lateral connections on
  the structure of cortical maps.
\newblock J Neurophysiol 92: 2947--2959.
\input{Goodhill5}

\bibitem{Swindale1}
Swindale NV (2004) How different feature spaces may be represented in cortical
  maps.
\newblock Network: Comput Neural Syst 15: 217--242.
\input{Swindale1}

\bibitem{Goodhill2}
Carreira-Perpinan MA, Lister RJ, Goodhill GJ (2005) A computational model for
  the development of multiple maps in primary visual cortex.
\newblock Cereb Cortex 15: 1222--1233.
\input{Goodhill2}

\bibitem{Sur5}
Yu H, Farley BJ, Jin DZ, Sur M (2005) The coordinated mapping of visual space
  and response features in visual cortex.
\newblock Neuron 47: 267--280.
\input{Sur5}

\bibitem{Goodhill9}
Goodhill GJ (2007) Contributions of theoretical modeling to the understanding
  of neural map development.
\newblock Neuron 56: 301-311.
\input{Goodhill9}

\bibitem{Sur2}
Farley BJ, Yu H, Jin DZ, Sur M (2007) Alteration of visual input results in a
  coordinated reorganization of multiple visual cortex maps.
\newblock J Neurosci 27: 10299--10310.
\input{Sur2}

\bibitem{Bonhoeffer}
H{\"u}bener M, Shoham D, Grinwald A, Bonhoeffer T (1997) Spatial relationships
  among three columnar systems in cat area 17.
\newblock J Neurosci 17: 9270--9284.
\input{Bonhoeffer}

\bibitem{Crair2}
Crair MC, Ruthazer ES, Gillespie DC, Stryker MP (1997) Ocular dominance peaks
  at pinwheel center singularities of the orientation map in cat visual cortex.
\newblock J Neurophysiol 77: 3381--3385.
\input{Crair2}

\bibitem{Loewel7}
Engelmann R, Crook JM, L{\"o}wel S (2002) Optical imaging of orientation and
  ocular dominance maps in area 17 of cats with convergent strabismus.
\newblock Visual Neuroscience 19: 39--49.
\input{Loewel7}

\bibitem{Matsuda}
Matsuda Y, Ohki K, Saito T, Ajima A, Kim DS (2000) Coincidence of ipsilateral
  ocular dominance peaks with orientation pinwheel centers in cat visual
  cortex.
\newblock NeuroReport 11: 3337--3343.
\input{Matsuda}

\bibitem{Obermayer}
Obermayer K, Blasdel GG, Schulten K (1992) Statistical-mechanical analysis of
  self-organization and pattern formation during the development of visual
  maps.
\newblock Phys Rev A 45: 7568--7589.
\input{Obermayer}

\bibitem{Bednar}
Bednar JA, Miikkulainen R (2006) Joint maps for orientation, eye, and direction
  preference in a self-organizing model of {V}1.
\newblock Neurocomputing 69: 1272--1276.
\input{Bednar}

\bibitem{Wolf3}
Wolf F, Geisel T (1998) Spontaneous pinwheel annihilation during visual
  development.
\newblock Nature 395: 73--78.
\input{Wolf3}

\bibitem{Goodhill4}
Carreira-Perpinan MA, Goodhill GJ (2002) Are visual cortex maps optimized for
  coverage?
\newblock Neural Computation 14: 1545--1560.
\input{Goodhill4}

\bibitem{Swindale2}
Swindale NV (1982) A model for the formation of orientation columns.
\newblock Proc R Soc Lond B 215: 211--230.
\input{Swindale2}

\bibitem{Grinwald3}
Shmuel A, Grinwald A (2000) Coexistence of linear zones and pinwheels within
  orientation maps in cat visual cortex.
\newblock PNAS 97: 5568--5573.
\input{Grinwald3}

\bibitem{Ohki}
Ohki K, Chung S, Kara P, H{\"u}bener M, Bonhoeffer T, et~al. (2006) Highly
  ordered arrangement of single neurons in orientation pinwheels.
\newblock Nature 442: 925.
\input{Ohki}

\bibitem{Ohki2}
Ohki K, Ch'ng YH, Kara P, Reid C (2005) Functional imaging with cellular
  resolution reveals precise micro-architecture in visual cortex.
\newblock Nature 433: 597-603.
\input{Ohki2}

\bibitem{Kaschube1}
Kaschube M, Wolf F, Geisel T, L{\"o}wel S (2002) Genetic influence on
  quantitative features of neocortical architecture.
\newblock J Neurosci 22: 7206--7217.
\input{Kaschube1}

\bibitem{Cho2}
Cho MW, Kim S (2004) Understanding visual map formation through vortex dynamics
  of spin hamiltonian models.
\newblock Phys Rev Lett 92: 018101.
\input{Cho2}

\bibitem{Hoffsummer}
Hoffs{\"u}mmer F, Wolf F, Geisel T, L{\"o}wel S, Schmidt KE (1996) Sequential
  bifurcation and dynamic rearrangement of columnar patterns during cortical
  development.
\newblock In: Bower J, editor, Computation and Neural Systems.
\input{Hoffsummer}

\bibitem{Hoffsummer2}
Hoffs{\"u}mmer F, Wolf F, Geisel T, L{\"o}wel S, Schmidt KE (1995) Sequential
  bifurcation of orientation-- and ocular dominance maps.
\newblock In: Proceedings of the International Conference on Artificial Neural
  Networks. Paris: EC2 \& Cie, volume~I, p. 535.
\input{Hoffsummer2}

\bibitem{Hoffsummer3}
Hoffs{\"u}mmer F, Wolf F, Geisel T, Schmidt KE, L{\"o}wel S (1997) Sequential
  emergence of orientation-- and ocular dominance maps.
\newblock In: Elsner N, Menzel R, editors, Learning and Memory, Proceedings of
  the 23rd G{\"o}ttingen Neurobiology Conference 1995. Stutttgart: Thieme
  Verlag, p.~97.
\input{Hoffsummer3}

\bibitem{Kadar}
Lee HY, Yahyanejad M, Kardar M (2003) Symmetry considerations and development
  of pinwheels in visual maps.
\newblock PNAS 100: 16036--16040.
\input{Kadar}

\bibitem{Reichl4}
Reichl L, L{\"o}wel S, Wolf F (2009) Pinwheel stabilization by ocular dominance
  segregation.
\newblock Phys Rev Lett 102: 208101.
\input{Reichl4}

\bibitem{Manneville}
Mannevile P (1990) Dissipative Structures and Weak Turbulence.
\newblock Academic Press.
\input{Manneville}

\bibitem{Cross}
Cross MC, Hohenberg PC (1993) Pattern formation outside of equilibrium.
\newblock Rev Mod Phys 65: 851--1112.
\input{Cross}

\bibitem{Greenside}
Cross MC, Greenside H (2009) Pattern Formation and Dynamics in Nonequilibrium
  Systems.
\newblock Cambride University Press.
\input{Greenside}

\bibitem{Braitenberg2}
Braitenberg V, Sch{\"u}tz A (1998) Cortex: Statistics and Geometry of Neuronal
  Connectivity.
\newblock Berlin: Springer.
\input{Braitenberg2}

\bibitem{Cowan3}
Bressloff PC, Cowan JD, Golubitsky M, Thomas PJ, Wiener MC (2002) What
  geometric visual hallucinations tell us about the visual cortex.
\newblock Neural Computation 14: 473--491.
\input{Cowan3}

\bibitem{Cowan}
Thomas PJ, Cowan JD (2004) Symmetry induced coupling of cortical feature maps.
\newblock Phys Rev Lett 92: 188101.
\input{Cowan}

\bibitem{Wolf1}
Wolf F (2005) Symmetry, multistability and long-range interactions in brain
  development.
\newblock Phys Rev Lett 95: 208701.
\input{Wolf1}

\bibitem{Wolf2}
Wolf F (2005) Symmetry breaking and pattern selection in visual cortical
  development.
\newblock Les houches 2003 lecture notes. Methods and Models in neurophysics.
  Amsterdam: Elsevier.
\input{Wolf2}

\bibitem{Kaschube3}
Kaschube M, Schnabel M, Wolf F (2008) Self-organization and the selection of
  pinwheel density in visual cortical development.
\newblock New J Phys 10: 015009.
\input{Kaschube3}

\bibitem{Cowan2}
Bressloff PC, Cowan JD (2002) The visual cortex as a crystal.
\newblock Physica D 173: 226--258.
\input{Cowan2}

\bibitem{Chapman3}
Chapman B, G{\"o}deke I (2002) No {On}-{Off} maps in supragranular layers of
  ferret visual cortex.
\newblock J Neurophysiol 88: 2163--2166.
\input{Chapman3}

\bibitem{Busse}
Busse FH (1978) Non-linear properties of thermal convection.
\newblock Rep Prog Phys 41: 1929--1967.
\input{Busse}

\bibitem{Bodenschatz}
Bodenschatz E, Pesch W, Ahlers G (2000) Recent developments in
  {R}ayleigh-{B}{\'e}nard convection.
\newblock Annu Rev Fluid Mech 32: 709--778.
\input{Bodenschatz}

\bibitem{Soward}
Soward A (1985) Bifurcation and stability of finite amplitude convection in a
  rotating layer.
\newblock Physica D 14: 227--241.
\input{Soward}

\bibitem{Vinals}
Zhang W, Vinals J (1997) Pattern formation in weakly damped parametric surface
  waves.
\newblock J Fluid Mech 336: 301--330.
\input{Vinals}

\bibitem{Miller4}
Miller KD (2010) $\pi$ = visual cortex.
\newblock Science 330: 1059--1060.
\input{Miller4}

\bibitem{Stevens}
Stevens CF (2011) A universal design principle for visual system pinwheels.
\newblock Brain Behav Evol 77: 132-135.
\input{Stevens}

\bibitem{Blasdel5}
Obermayer K, Blasdel GG (1997) Singularities in primate orientation maps.
\newblock Neural Computation 9: 555--575.
\input{Blasdel5}

\bibitem{Bonhoeffer2}
Bonhoeffer T, Kim DS, Malonek D, Shoham D, Grinwald A (1995) Optical imaging of
  the layout of functional domains in area 17 and across the area 17/18 border
  in cat visual cortex.
\newblock Eur J Neurosci 7: 1973--1988.
\input{Bonhoeffer2}

\bibitem{Swindale7}
Swindale NV (1992) A model for the coordinated development of columnar systems
  in primate striate cortex.
\newblock Biol Cybern 66: 217--230.
\input{Swindale7}

\bibitem{Mermin}
Mermin ND (1979) The topological theory of defects in ordered media.
\newblock Rev Mod Phys 51: 591--648.
\input{Mermin}

\bibitem{Scherf}
Wolf F, Pawelzik K, Scherf O, Geisel T, L{\"o}wel S (2000) How can squint
  change the spacing of ocular dominance columns?
\newblock J Physiol 94: 525--537.
\input{Scherf}

\bibitem{SwiftHohenberg}
Swift JB, Hohenberg PC (1977) Hydrodynamic fluctuations at the convective
  instability.
\newblock Phys Rev A 15: 319--328.
\input{SwiftHohenberg}

\bibitem{Cowan5}
Bressloff PC, Cowan JD, Golubitsky M, Thomas PJ, Wiener MC (2001) Geometric
  visual hallucinations euclidean symmetry, and the functional archtitecture of
  striate cortex.
\newblock Phil Trans R Soc B 356: 1-32.
\input{Cowan5}

\bibitem{Herrmann2}
Mayer N, Herrmann MJ, Asada M, Geisel T (2007) Pinwheel stability in a
  non-{E}uclidean model of pattern fomation in the visual cortex.
\newblock Journal of the Korean Physical Society 50: 150--157.
\input{Herrmann2}

\bibitem{Schnabel2}
Schnabel M (2008) A Symmetry of the Visual World in the Architecture of the
  Visual Cortex.
\newblock Ph.D. thesis, G{\"o}ttingen University.
\input{Schnabel2}

\bibitem{Schnabel3}
Schnabel M, Kaschube M, White LE, Wolf F (2008) Signatures of shift-twist
  symmetry in the layout of orientation preference maps.
\newblock Society for Neuroscience Abstracts 30 (866).
\input{Schnabel3}

\bibitem{Reichl6}
Reichl L, Heide D, L\"owel S, Crowley JC, Kaschube M, et~al. Coordinated
  optimization of visual cortical maps ({II}) {N}umerical studies.
\newblock \quad .
\input{Reichl6}

\bibitem{Kaschube2}
Kaschube M, Wolf F, Puhlmann M, Rathjen S, Schmidt KE, et~al. (2003.) The
  pattern of ocular dominance columns in cat primary visual cortex: Intra- and
  interindividual variability of column spacing and its dependence on genetic
  background.
\newblock Eur J Neurosci 18: 3251--3266.
\input{Kaschube2}

\bibitem{Braitenberg}
Braitenberg V, Braitenberg C (1979) Geometry of orientation columns in the
  visual cortex.
\newblock Biol Cybern : 179--186.
\input{Braitenberg}

\bibitem{Cho1}
Cho MW, Kim S (2005) Different ocular dominance map formation influenced by
  orientation preference columns in visual cortices.
\newblock Phys Rev Lett 94: 068701.
\input{Cho1}

\bibitem{Grossberg}
Grossberg S, Olson SJ (1994) Rules for the cortical map of ocular dominance and
  orientation columns.
\newblock Neural Networks 7: 883--894.
\input{Grossberg}

\bibitem{Tanaka}
Nakagama H, Tani T, Tanaka S (2006) Theoretical and experimental studies of
  relationship between pinwheel centers and ocular dominance columns in the
  visual cortex.
\newblock Neuroscience Research 55: 370--382.
\input{Tanaka}

\bibitem{Miller}
Erwin E, Miller KD (1998) Correlation-based development of ocularly matched
  orientation and ocular dominance maps: Determination of required input
  activities.
\newblock J Neurosci 18: 9870--9895.
\input{Miller}

\bibitem{Pierre}
Pierre DM (1997) Modeling Orientation and Ocular Dominance Columns in the
  Visual Cortex.
\newblock Ph.D. thesis, MIT.
\input{Pierre}

\bibitem{Swindale8}
Mitchison GJ, Swindale NV (1999) Can hebbian volume learning explain
  discontinuities in cortical maps?
\newblock Neural Computation 11: 1519--1526.
\input{Swindale8}

\bibitem{Crair}
Crair MC, Gillespie DC, Stryker MP (1998) The role of visual experience in the
  development of columns in cat visual cortex.
\newblock Science 279: 556--570.
\input{Crair}

\bibitem{Stryker3}
Issa NP, Trachtenberg JT, Chapman B, Zahs KR, Stryker MP (1999) The critical
  period for ocular dominance plasticity in the ferret's visual cortex.
\newblock J Neurosci 19: 6965--6978.
\input{Stryker3}

\bibitem{Horton}
Horton JC, Hocking DR (1996) Intrinsic variability of ocular dominance column
  periodicity in normal macaque monkeys.
\newblock J Neurosci 16: 7228--7339.
\input{Horton}

\bibitem{Swindale4}
Swindale NV, Shoham D, Grinwald A, Bonhoeffer T, H{\"u}bener M (2000) Visual
  cortex maps are optimized for uniform coverage.
\newblock Nature Neuroscience 3: 822--826.
\input{Swindale4}

\bibitem{Swindale9}
Swindale NV, Shoham D, Grinwald A, Bonhoeffer T, H{\"u}bener M (2002) Reply to
  carreira-perpinan and goodhill. are visual cortex maps optimized for
  coverage?
\newblock Neural Computation 14: 2053--2056.
\input{Swindale9}

\end{thebibliography}

%%%%%%%%%%%%%%%%%%%%%%%%%%%%%%%%%%%
\listoffigures
%\section*{Figure Legends}
%\begin{figure}[!ht]
%\begin{center}
%%\includegraphics[width=4in]{figure_name.2.eps}
%\end{center}
%\caption{
%{\bf Bold the first sentence.}  Rest of figure 2  caption.  Caption 
%should be left justified, as specified by the options to the caption 
%package.
%}
%\label{Figure_label}
%\end{figure}

%\section*{Tables}
%\begin{table}[!ht]
%\caption{
%\bf{Table title}}
%\begin{tabular}{|c|c|c|}
%table information
%\end{tabular}
%\begin{flushleft}Table caption
%\end{flushleft}
%\label{tab:label}
% \end{table}

\end{document}